\documentclass[aps,prx,showpacs,floatfix,twocolumn,superscriptaddress]{revtex4-1}
\usepackage{graphicx}
\usepackage{bm,amsmath}
\usepackage{dcolumn}
\usepackage{amsfonts,amssymb}
\usepackage{chemfig}
\usepackage{diagbox}

\begin{document}
\title{Multiscale Excitations in the Diluted Two-dimensional \\ $S=1/2$ Heisenberg Antiferromagnet}

\author{Liuyun Dao}
\affiliation{Center for Advanced Quantum Studies, School of Physics and Astronomy, Beijing Normal University, Beijing 100875, China}

\author{Hui Shao}
\email{huishao@bnu.edu.cn}
\affiliation{Center for Advanced Quantum Studies, School of Physics and Astronomy, Beijing Normal University, Beijing 100875, China}
\affiliation{Key Laboratory of Multiscale Spin Physics, Ministry of Education, Beijing 100875, China}

\author{Anders W. Sandvik}
\email{sandvik@bu.edu}
\affiliation{Department of Physics, Boston University, 590 Commonwealth Avenue, Boston, Massachusetts 02215, USA}
\affiliation{Beijing National Laboratory for Condensed Matter Physics and Institute of Physics, Chinese Academy of Sciences, Beijing, 100190, China}
\date{\today}

\begin{abstract}
We study the excitation spectrum of the $S=1/2$ Heisenberg model on the randomly diluted square lattice by stochastic analytic continuation of imaginary-time
correlations obtained by quantum Monte Carlo simulations in the low-temperature limit. Focusing on relatively low dilution fractions, $p=1/16$ and $p=1/8$, the
dynamic structure factor $S({\bf q},\omega)$ exhibits a strongly damped magnon peak with anomalous dispersion near ${\bf q}=(0,0)$ and $(\pi,\pi)$, a
non-dispersive low-energy peak from localized excitations, and a second peak within the continuum connecting these two features. A magnon with 
anomalous logarithmic dispersion, close to our result, was predicted in spin wave and $T$-matrix theory [A. Chernyshev et al., Phys.~Rev.~B {\bf 65},
  104407 (2002)], with the quasiparticle description breaking down at the low localization energy. However, no intermediate mode was
predicted. Analyzing spectral functions in real space for individual vacancy realizations by energy tomography, we find that these excitations are concentrated on 
a subset of the spins adjacent to vacancies, with fewer spins involved as the energy is lowered. We argue that the low-energy excitations are those of
a sparse random network of effective moments at a fraction of the vacancies, leading to a damped diffusive-like behavior with localization at the lowest
energies (the localized mode). In the case of the magnon, there is a shift in the real-space spectral weight distribution,
from the spins away from vacancies at high energy to those adjacent to vacancies at lower energy. We also analyze the Anderson quantum rotor excitation at
$\omega \propto N^{-1}$ (with $N=L^2$ the system size), which in the clean system is visible in
$S({\bf q},\omega)$ only at ${\bf q}=(\pi,\pi)$ but spreads through the entire Brillouin zone due to the lack of translational invariance when $p>0$. Beyond
weight close to ${\bf q}=(0,0)$ and $(\pi,\pi)$, which we explain by local sublattice imbalance within a classical dimer-monomer model, there is also
intricate structure arising from correlated singlet fluctuations, which we demonstrate by enhancing said fluctuations with four-spin couplings in the
Hamiltonian. The features in momentum space reflect spatially nonuniform breaking of the spin rotation symmetry and should be experimentally
observable with elastic neutron scattering on layered quantum antiferromagnets doped with nonmagnetic impurities. All the $\omega > 0$ excitations should
be observable by inelastic neutron scattering. Technically, our work demonstrates that surprisingly complex spectral information can be obtained from
quantum Monte Carlo data, despite the ``ill-posed'' numerical analytic continuation problem.

\end{abstract}
\maketitle

\section{Introduction}
\label{sec:intro}

Quantum systems with quenched disorder may exhibit rich behaviors that cannot be realized in uniform systems. Experimentally, some amount of disorder is
always present, and it is also often possible to control the amount of disorder, thus enabling systematic studies of disorder induced phenomena. An understanding
of weak disorder effects is often necessary in order to correctly interpret experiments intended to probe the physics of uniform systems
\cite{Kimchi18,Magnaterra23}, while in other cases new phases of matter can be induced by strong or weak disorder \cite{Hong21,Xiang23}.
One of the simplest examples from the area of quantum magnetism is two-dimensional (2D) Heisenberg antiferromagnets doped with nonmagnetic impurities
(e.g., replacing Cu by Zn in copper oxides), which exhibit interesting
phenomena in experiments \cite{Cheong91,Ting92,Corti95,Carretta97,Vajk02,Papinutto05} that have inspired extensive analytical
\cite{Harris77,Bulut89,Brenig91,Wan93,Yasuda97,Yasuda99,Chen00,Chernyshev01,Chernyshev02,Mucciolo04,Liu09,Liu13} and
computational \cite{Manousakis92,Kato00,Sandvik02,Hoglund04,Yu05,Lucher05,Anfuso06,Yu06,Wang06,Liu09,Wang10,Changlani13a,Weinberg17} research on
diluted $S=1/2$ Heisenberg models with nearest neighbor interactions $J{\bf S}_i \cdot {\bf S}_j$.
Many works have focused on systems at and close to the classical percolation point (up to which the system remains
long-range ordered \cite{Sandvik02}), where interplay of geometrical and quantum fluctuations lead to unusual low-energy excitations
\cite{Wang06,Wang10,Changlani13a,Weinberg17}. An important aspect here is the local sublattice imbalance caused by random dilution of a bipartite
lattice. It was recently argued \cite{Bhola22} based on classical dimer-monomer models (which can explain some aspects of the excitations
at the percolation point \cite{Wang06,Wang10}) that sublattice imbalance may possibly lead to profound changes in the excitations of the N\'eel state
even at low dilution. A classical critical point of the dimer-monomer model was demonstrated where the size of isolated regions of confined monomers
diverges, thus leading to monomer percolation. This new type of percolation point may strictly be located at zero dilution in 2D, with an associated
length scale that diverges as the dilution fraction $p \to 0$. In 3D, the percolation of monomer region likely takes place at a nonzero value of $p$.

The intriguing possibility of the monomer percolation phenomenon having a counterpart in the antiferromagnet is one of the motivations of our study presented here,
along with previous detailed, but so far untested predictions from spin-wave and T-matrix theory \cite{Brenig91,Wan93,Chernyshev01,Chernyshev02}. Focusing on
the ground state, we carry out extensive quantum Monte Carlo (QMC) studies of the 2D Heisenberg model at ultralow temperatures in systems with relatively low
dilution fractions, $p=1/16$ and $1/8$. We compute imaginary-time correlation functions and perform analytic continuation to the real-frequency ($\omega$) and
momentum (${\bf q}$) resolved dynamic structure factor $S({\bf q},\omega)$, using the stochastic analytic continuation (SAC) \cite{Shao23} method. 
In addition to studying the ${\bf q}$ and $\omega$ dependence of the disorder averaged $S({\bf q},\omega)$ for system sizes $N=L^2$ with $L$ up to $64$,
we also investigate the real-space distribution $S({\bf r},\omega)$ of spectral weight in individual vacancy realizations. Tomographic analysis of $S({\bf r},\omega)$
in different energy windows provides direct insights into the nature of different types of excitations.

\begin{figure}[t]
\includegraphics[width=75mm]{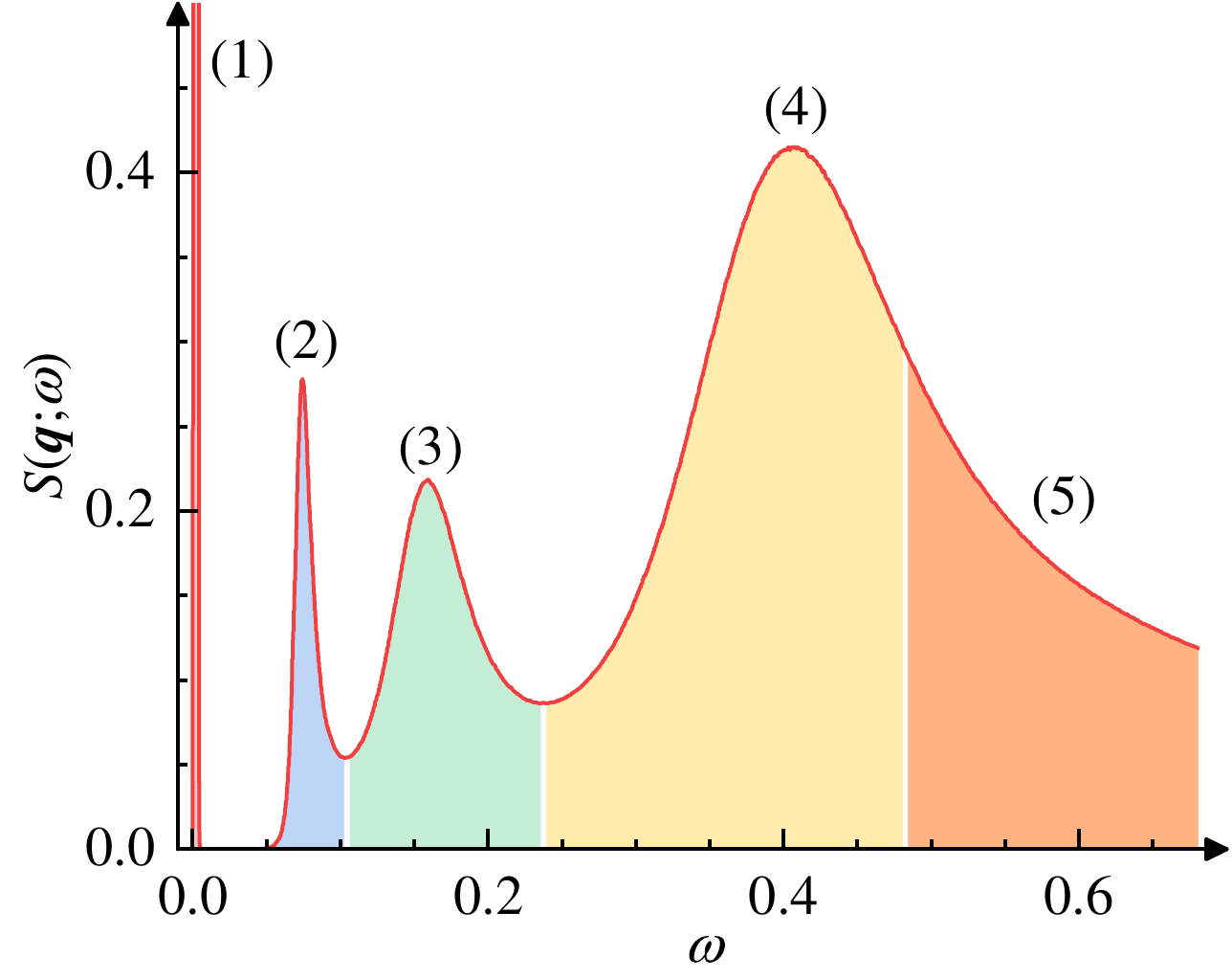}
\caption{Example of a disorder averaged dynamic structure factor, obtained for a system with $L=64$ at dilution fraction $p=1/8$. The momentum here is
${\bf q}=2\pi(3,3)/L$ but the overall peak structure is typical for a wide range of ${\bf q}$. The marked spectral features arise from excitations identified as: (1)
quantum rotor states with energy scaling as $\omega_{\rm rot} \propto L^{-2}$, (2) localized excitations with the peak energy $\omega_{\rm loc}$ decreasing
with $L$ but remaining at $\omega_{\rm loc} > 0$, (3) a weakly dispersive diffusive-like mode proposed here to arise from a random network of spins adjacent to
vacancies, (4) damped magnon with anomalous dispersion close to ${\bf q}=(0,0)$ and $(\pi,\pi)$, and (5) a multi-magnon continuum with a tail extending up to
$\omega \approx 5$.}
\label{schematic}
\end{figure}

Our numerical calculations, in synergy with previous analytical results, lead to an intriguing picture of the excitations of the diluted antiferromagnet,
which apart from its direct experimental relevance also presents a key paradigmatic ``strong disorder'' problem in quantum many-body theory (as elaborated
in Ref.~\onlinecite{Chernyshev02}). Unbiased numerical calculations of dynamic spectral functions pose formidable challenges in their own right. We have
achieved unprecedented resolution of  the dynamic structure factor at multiple energy scales, identifying five different spectral attributes in
$S({\bf q},\omega)$ that can be assigned to specific types of excitations. The observed behaviors are intricate and multifaceted. As a preview and
orientation for the reader, we here first briefly summarize the different excitation modes uncovered by our study, aided by an example of $S({\bf q},\omega)$
displayed in Fig.~\ref{schematic}.

Figure \ref{schematic} is based on calculations that will be discussed in detail later on but here serves to illustrate the different excitations identified
through a variety of probes in both momentum and real space. For all ${\bf q}$, a very sharp peak is present at $\omega_{\rm rot} \propto 1/L^2$ that originates
from the lowest Anderson quantum rotor mode \cite{Anderson52,Manousakis91,Sandvik10}
related to the breaking of the spin-rotation symmetry in the N\'eel state when $L \to \infty$. There is another
dispersionless peak that shifts down in energy with increasing $L$ but remains at $\omega_{\rm loc} > 0$. It predominantly arises from a very small number of
excited spins adjacent to
vacancies. This localized mode is followed by a broad and only weakly dispersing peak that we identify as a band of damped excitations, likewise involving
mainly spins next to vacancies but a much larger subset as compared to the localized excitations. While our study does not provide a definite conclusion, it
is possible that these excitations are diffusive-like and smoothly connect to the localized mode, in the sense of a diffusion constant that approaches zero at
the localization energy. A band of damped magnons follows at higher energy, with the character of the magnons changing from the lowest (always above the diffusive
peak) to highest energy. At the lower energies, the dispersion is anomalous (close to logarithmic \cite{Chernyshev02}) and in real space the spectral weight
in the corresponding energy window is large on all the spins next to vacancies. At high energy, with ${\bf q}$ close to the magnetic zone boundary (on which
also the magnons in the uniform system have the highest energy), there is much less spectral weight on those spins and more within the uniform bulk patches between
the vacancies. Finally, the magnon peak is always followed by a very large multimagnon tail that extends far below the $\omega$ cutoff in Fig.~\ref{schematic}.

In the reminder of this Introduction, we briefly review the previous analytical results in Sec.~\ref{sec:predictions}, and in Sec.~\ref{sec:summary} we
present more details on our results, their interpretations and how they compare with the previous predictions. We outline the rest of the paper
in Sec.~\ref{sec:outline}.

\subsection{Previously predicted excitations}
\label{sec:predictions}

According to the most sophisticated spin-wave and $T$-matrix analysis \cite{Chernyshev02}, the excitation spectrum should indeed exhibit a multiscale structure,
with strongly damped spin waves renormalized down in energy by the disorder at low energy, close to ${\bf q}=(0,0)$ and $(\pi,\pi)$. There is a
dispersionless mode at a very low $p$ dependent energy, $\omega_{\rm loc} \sim {\rm e}^{-\pi/4p}$, reflecting a localization length of order $l \sim {\rm e}^{\pi/4p}$
from the impurity scattering of the spin waves becoming dominant at long wavelengths. On approaching the energy $\omega_{\rm loc}$ from above, the assumption of
the spin waves propagating in a disorder averaged medium also becomes invalid, and the dynamics is then completely incoherent, with no well defined quasi particle.
The resulting excitations were instead described as diffusive. Our results partially confirm this picture but also demonstrate an even richer excitation spectrum
with an additional peak related to the diffusive-like modes, marked (3) in Fig.~\ref{schematic}, before the excitations become fully localized at $\omega_{\rm loc}$.

The available theoretical predictions are detailed \cite{Chernyshev02}, with the anomalous dispersion at small $p$ and momenta close to $(0,0)$ and $(\pi,\pi)$
given in units of $J$ by
\begin{equation} 
\omega_{\rm ano}(k)= c_0k[1+2p\ln(k)/\pi],
\label{wklog}  
\end{equation}  
where $k$ is the momentum relative to one of the above two ${\bf q}$ points, $c_0$ is the spin-wave velocity of the clean system, and the lattice
constant is set to unity. This dispersion relation should break down at the momenta where $\omega_{\rm ano}(k)$ approaches the localization energy $\omega_{\rm loc}$,
which is the minimum excitation energy for all ${\bf q}$. The localized states produce a sharp peak in $S({\bf q},\omega)$, and an essentially featureless spectral
function was predicted between it and the magnon peak when the two excitations are individually discernible. The two peaks were found to merge,
$\omega_{\rm ano}(k) \to \omega_{\rm loc}$, when the magnons localize.

The anomalous magnon is also significantly damped, with the peak in $S({\bf q},\omega)$ having a width of order $pk$. Further, the anomalous dispersion
should only apply at energies below  $\omega \approx J$, and a separate band of excitations, similar to the spin waves in the undiluted system, remains
for ${\bf q}$ close to the magnetic zone boundary (where $\omega \approx 2J$). In some parts of the Brillouin zone (BZ), this mode coexists with the anomalous
magnons at lower energy. The exact nature of the cross-over or coexistence between the two types of magnons is not clear, because crucial interactions
when the spin waves resonate with the local impurity modes were not taken into account in the $T$-matrix approach.

Technically, the calculation leading to these predictions \cite{Chernyshev02} is based on mutually noninteracting spin waves (linear spin-wave theory)
within which the single-vacancy system is first solved exactly. The multi-vacancy system is then treated perturbatively within the $T$-matrix formalism,
summing all diagrams involving single impurity scattering. Disorder averaging is carried out before evaluating $S({\bf q},\omega)$, which effectively
implies that the excitations are those of an impurity-averaged bulk medium.

\subsection{Computational results and their interpretation}
\label{sec:summary}

The predictions outlined above should be valid at low dilution and low energy; specifically, when single-impurity scattering dominates. However, it is a priori
not clear whether collective impurity effects washed out by the approximations may possibly leave out some excitations or create artificial features, as indeed
demonstrated by our numerical results previewed in Fig.~\ref{schematic} and summarized in some more detail next. Our numerical approach is complementary to the
theory, in the sense that it is well suited for accessing excitations at small to moderate length scales and there are no approximations beyond the limitations
of numerical analytic continuation (which we will argue to be minor here). Because of practical restrictions in the system size in QMC simulations
at extremely low temperatures (realizing the ground state), here $N=L^2$ spins with $L \le 64$, we cannot reach momenta in the very close
neighborhood of $(0,0)$ and $(\pi,\pi)$. We nevertheless identify long-wavelength features similar to those predicted, though not always with full
quantitative agreement. As already mentioned, we also demonstrate some qualitative discrepancies between theory and numerics.

We observe the dispersionless localized mode and extrapolate its energy to the thermodynamic limit for $p=1/8$. The result, $\omega_{\rm loc} \approx 0.1$,
would require the factor $A$ of the predicted peak energy $\omega_{\rm loc} = A{\rm e}^{-\pi/4p}$ derived in Ref.~\onlinecite{Chernyshev02} (to which we refer
for explanations) to be as large as $50$, but the form may only strictly apply for much smaller $p$. At $p=1/16$, we are not able to distinguish our extrapolated
value from $0$ within numerical uncertainties. At higher energy we observe the magnon peak, with dispersion relation close to
$\omega_0({\bf q})$ of the clean 2D Heisenberg model for ${\bf q}$ at and close to the zone boundary, then dropping sharply in a way closely matching
Eq.~(\ref{wklog}) when approaching $(0,0)$ and $(\pi,\pi)$.

The dispersion of the high-energy magnons on the zone boundary includes the characteristic drop relative to low-order spin-wave theory at ${\bf q}=(\pi,0)$
\cite{Singh95,Sandvik01} also in the diluted systems. This $(\pi,0)$ anomaly in the clean 2D Heisenberg model has been identified as a precursor to spinon
deconfinement \cite{Piazza15,Shao17,Powalski18}. Thus, our results indicate that this precursor survives also in the diluted system, which we explore further
by adding the same four-spin interaction to the Heisenberg Hamiltonian as in Ref.~\onlinecite{Shao17}.

In the spin-wave theory the magnon in the diluted system is split into two
branches at energies close to $J$ \cite{Brenig91,Chernyshev02}, but here we always find a single peak in this energy range. The double peak may reflect
an inability of the analytic theory to fully take into account interactions between the spin waves and the impurity modes \cite{Chernyshev02}. Low-order
spin-wave theory in general also cannot quantitatively describe high-energy magnons and multi-magnon continua, which in the uniform system is most apparent
in the $(\pi,0)$ anomaly, which in addition to the energy suppression also involves a significant multimagnon continuum at higher energy \cite{Sandvik01,Shao17}.
In the diluted systems we find an even more prominent continuum, extending up to $\omega \sim 5J$ over much of the BZ. Apart from the
multimagnon tail, the magnon peak is always very broad, in qualitative agreement with the prediction of the width being proportional to $pk$ at
small $k$. 

In contrast to the spin-wave theory, as exemplified by the peak marked (3) in Fig.~\ref{schematic}, we find a weakly dispersive and relatively broad peak
in $S({\bf q},\omega)$ at energies between the magnon and the localized mode marked (1)---at about 20\% of the magnon energy close to the zone boundary and
smaller relative suppression close to ${\bf q}=(0,0)$ and $(\pi,\pi)$. The peak emerges systematically with increasing system size in the energy range where
the analytic calculation \cite{Chernyshev02} only produced a rather flat continuum. We gain further insights into the nature of this new peak as well as the
localized mode below it by investigating the real-space (${\bf r}$) resolved local dynamic structure factor $S({\bf r},\omega)$ for individual vacancy
realizations. By conducting tomography of such functions in $\omega$ space, i.e., separating contributions from different energy windows, we can clearly see
that the second low-energy mode is concentrated on some of the spins adjacent to the vacancies. These spins should interact with each other through the bulk
medium, with an effective coupling that necessarily must be much weaker than the bare Heisenberg coupling $J$, thus qualitatively explaining the reduced energy
scale as compared to the magnon. Based on our real-space insights, we conjecture that the excitations are those of a random network of a subset of the spins
adjacent to vacancies. Because of the smoothly evolving spectral weight distribution in real space as the energy approaches the localization energy, it is
likely that the peaks (2) and (3) in Fig.~\ref{schematic} both arise from the random spin network, with full localization taking place only at the lowest
energies, peak (2), and with diffusive-like behavior at higher energy up to peak (3). This connection may in principle arise with a diffusion constant
approaching zero at the localization energy. As mentioned above, the previous $T$-matrix calculations also \cite{Chernyshev02} produced significant spectral
weight above the localized mode, but not a second peak below the magnon.

The magnons at the zone boundary occupy mainly the uniform patches of space between the vacancies. At the dilution fractions $p$ considered here, the
patches can typically accommodate several wave lengths at the highest energies, explaining the closeness of their (peak) energy to $\omega_0({\bf q})$ of
the uniform system, though the damping is very significant. At longer wave lengths, where the magnon drops below $\omega_0({\bf q})$,
to approximately $\omega_{\rm ano}(k)$ in Eq.~(\ref{wklog}) with a slightly renormalized velocity, the tomography informs us that both the spins at and
between the vacancies are involved in the excitations, now with all the spins adjacent to vacancies carrying elevated spectral weight and the other
spins having almost uniform but lower weight.

Thus, starting from the high energy magnons with energy close to $\omega_0$, the excitations mainly affect the spins within the patches
between vacancies. Spectral weight is gradually transferred to all the spins at distance one lattice spacing from the vacancies as the magnons drop to
the anomalous energy $\omega_{\rm ano}$. When the energy is further reduced toward the diffusive-like mode, the spectral weight away
from the vacancies is further depleted and also removed from many of the spins neighboring vacancies, leaving a rather small subset of spins that
carry large spectral weight. There is a sharp further reduction in the number of participant spins when the energy drops toward
$\omega_{\rm loc}$.

All the above excitations should be observable by inelastic neutron scattering experiments in 2D quantum antiferromagnets doped with nonmagnetic
impurities, e.g., cuprate oxides with Cu substituted by Zn or Mg. A potentially even better starting point may be Cu(DCOO)$_2\cdot$4D$_2$O, if it also
can be successfully doped with Zn of Mg---we are not aware of any attempts so far. The pristine system is currently regarded as the best realization
of the 2D Heisenberg model, with the measured inelastic neutron scattering cross section being in very close agreement with that of the model
\cite{Piazza15,Shao17}.

In addition to all the $\omega>0$ excitations, in the finite systems studied here, the spin-rotation symmetry is not broken and collective excitations are
expected at energy scaling as $1/N$---the Anderson tower of quantum rotor states that become degenerate in the thermodynamic limit and facilitate the broken
symmetry \cite{Anderson52,Manousakis91,Sandvik10}.
These excitations are not included in spin wave and related calculations starting from a state with broken symmetry, but they are important as a probe of
the nature of the ground state. In the case of a pristine lattice, the rotor states exist at wave-vectors ${\bf q}=(0,0)$ (even total spin $S$ modes)
and $(\pi,\pi)$ (odd $S$ modes), but only the $S=1$ state at $(\pi,\pi)$ contributes to $S({\bf q},\omega)$.
In the diluted system, where translational symmetry is broken, the weight of the rotor state in $S({\bf q},\omega)$ spreads out over momentum space in a
way reflecting specific large-scale inhomogeneities in the zero-energy fluctuations of the N\'eel order.  We find an interesting structure of the ${\bf q}$
dependent rotor weight that should be experimentally visible in elastic neutron scattering through broadening of the magnetic $(\pi,\pi)$ Bragg peak and
spectral weight moving to other parts of the BZ, including a broad maximum forming around $(0,0)$.

Peaks of rotor weight close to ${\bf q}=(0,0)$ and $(\pi,\pi)$ can be qualitatively explained by a classical dimer-monomer model, which was previously
used to study effects of local sublattice imbalance in $S=1/2$ Heisenberg antiferromagnets at the percolation point
\cite{Wang10,Changlani13a,Changlani13b,Ghosh15}, and also in other fractal clusters \cite{Changlani13b,Ghosh15}.
Beyond the two peaks in the BZ, there are other weak variations of the rotor weight as a function of
${\bf q}$ that are not explained by classical sublattice imbalance. We will show that a minimum at ${\bf q}=(\pi/2,\pi/2)$ originates from a tendency of
resonating dimer singlets to form local columnar correlations, as a precursor to a long-range ordered valence-bond solid state existing in an extended
model space. We find support for this scenario by including a four-spin interaction $Q$ in the Hamiltonian (the $J$-$Q$ model \cite{Sandvik07}),
which further depletes the rotor modes close to ${\bf q}=(\pi/2,\pi/2)$ as a consequence of correlated singlet fluctuations---the precursor
to spinon deconfinement.

A recent comprehensive mathematical and computational study of the dimer-monomer model at low dilution found a novel type of monomer percolation phenomenon
in the limit of low dilution \cite{Bhola22}. It was proposed, without details, that some aspects of the low-energy excitations of the diluted Heisenberg model
could reflect this percolation point $p_c$, especially in 3D systems where $p_c >0$ but perhaps also in 2D, where $p_c = 0$. Though,
as mentioned above, in the rotor excitation we find signatures of sublattice imbalance as modeled by the dimer-monomer model, a possible relationship between
monomer percolation and the $p \to 0$ delocalization predicted in the $T$-matrix theory is less clear. We do find compelling evidence from our
tomographic approach of the localized excitations also being tied to spins in the monomer rich regions. We will discuss how the length scale
$l \propto {\rm e}^{\pi/4p}$ in the spin wave theory \cite{Chernyshev02} may be compatible with power law divergence of a monomer correlation length
in the effective dimer-monomer description \cite{Bhola22}.

\subsection{Outline}
\label{sec:outline}

The rest of the paper is organized as follows: In Sec.~\ref{sec:methods} we introduce the quantum spin models and also the classical dimer-monomer model that
we use to qualitatively understand some of our results for the rotor excitations. We also provide tests of the SAC method by studying small systems that
are amenable to exact Lanczos diagonalization. In Sec.~\ref{sec:sqw} we present our results for the disorder averaged ${\bf q}$ dependent and local dynamic
structure factor, and also for the real-space resolved spectral function of individual vacancy realizations. In Sec.~\ref{sec:rotor} we further analyze the
way the quantum rotor contributions spread out in the BZ by the disorder. In Sec.~\ref{sec:jq} we present complementary results for the $J$-$Q$ model that
are helpful for interpreting the rotor weight distribution. We conclude in In Sec.~\ref{sec:conc} with a brief summary and further discussion of our
results. Auxiliary results supporting our conclusions are presented in three appendices.

\section{Models and methods}
\label{sec:methods}

\subsection{Diluted quantum antiferromagnets}

The $S=1/2$ Heisenberg model diluted by static vacancies is defined by the Hamiltonian
\begin{equation}
H_J = J \sum_{\langle ij\rangle} n_in_j {\bf S}_i \cdot {\bf S}_j ,
\label{ham}
\end{equation}  
where $\langle ij\rangle$ denotes nearest neighbors on the 2D square lattice and $n_i=1$ and $n_i=0$ for spins and vacancies, respectively.
We set the exchange constant to $J=1$ and define the doping fraction $p=N_0/N$, where $N_0$ is the number of randomly distributed vacancies
in the canonical ensemble and $N=L^2$ is the number of lattice sites. The classical percolation point above which there is no connected cluster of
size $\propto N$ is $p_{\rm c} \approx 0.407$, and the system remains N\'eel ordered up to this point (with the fractal percolating cluster being ordered)
\cite{Sandvik02}. We here consider $p \ll p_{\rm c}$ and mainly focus on the case of $N_0/2$ vacancies on each of the two sublattices. In Appendix
\ref{sec:nanb} we provide results showing no substantial differences for systems without this restriction.

We will also discuss some aspects of a system with an added four spin interaction, which when strong leads to a quantum phase transition from the
N\'eel ground state to a columnar valence-bond solid \cite{Sandvik07}. In the presence of vacancies, this $Q$ interaction is
\begin{equation}
H_Q = -Q \sum_{\langle ijkl\rangle} n_in_jn_kn_l (\hbox{$\frac{1}{4}$}-{\bf S}_i \cdot {\bf S}_j)(\hbox{$\frac{1}{4}$}-{\bf S}_k \cdot {\bf S}_l),
\label{jqham}
\end{equation}  
where $\langle ijkl\rangle$ corresponds to sites on a $2\times 2$ plaquette with $ij$ and $kl$ forming opposite edges (with both vertical and horizontal
orientation). The $J$-$Q$ Hamiltonian is $H=H_J+H_Q$ and we here use $J+Q=1$ as the energy scale. In addition to the Heisenberg model with $Q=0$, we will
consider $Q/J=1$ and $Q/J=2$, where the system is still in the N\'eel state, far from the transition point at $Q/J\approx 22$, but already exhibits large
changes in the excitation spectrum (as found in the case of the clean system in Ref.~\onlinecite{Shao17}). The purpose of $Q>0$ here is to amplify
effects that we argue arise from local columnar dimer fluctuations already at $Q=0$.

\subsection{Spectral functions}
\label{sec:spectral}
           
Our main aim here is to investigate the excitations of the system through the dynamic spin structure factor $S({\bf q},\omega)$, where $\omega$ is the
excitation energy in units of $J=1$ for the Heisenberg model and $J+Q=1$ in the $J$-$Q$ model. The momentum (wave vector) is ${\bf q}=(q_x,q_y)$,
with $(q_x,q_y)=2\pi(k_x,k_y)/L$ and $k_x,k_y=0,\ldots,L-1$.

For any bosonic operator $O$, the spectral function at temperature $T=0$ can be written in the basis of eigenstates $\vert n\rangle$ and
eigenvalues $E_n$ of the Hamiltonian as:
\begin{eqnarray}
S(\omega)=\pi\sum_{n}{\vert \langle n \vert O \vert 0 \rangle \vert}^2 \delta(\omega-E_n+E_0).
\label{sqwexact}
\end{eqnarray}
For the dynamic structure factor $S({\bf q},\omega)$ of the spin-rotation invariant Heisenberg model
the operator can be taken as the Fourier transform of the spin-$z$ operator:
\begin{eqnarray}
O_{\bf q}=\frac{1}{\sqrt{N}}\sum_{\boldsymbol{r}}e^{-i\boldsymbol{q}\cdot\boldsymbol{r}}S_{\boldsymbol{r}}^{z},
\label{opq}
\end{eqnarray}
where $\boldsymbol{r}=(x,y)$ is the site coordinate with unit lattice constant. This operator produces only $S=1$ excitations when acting on the $S=0$ ground state of
a clean $L \times L$ system with even $L$. However, the diluted system can contain isolated clusters with unequal occupation of the two sublattices, which have ground
states with $S>0$. The excitations then can have spin $S + 1$ or $S-1$ relative to the ground state $S$. Our methods automatically perform the proper spin rotational
averaging for ground states of any $S$. Since we are working at low dilution, the spectral functions are completely dominated by one large connected cluster.

While $S(\boldsymbol{q},\omega)$ can be computed for small systems using Lanczos exact diagonalization \cite{Dagotto96} (as we will also do),
for larger systems it is not possible to obtain exact results. Here we compute the corresponding imaginary-time dependent correlation functions using QMC
simulations on large lattices and perform numerical analytic continuation to real frequency.

The imaginary-time correlation function is
\begin{eqnarray}
G({\bf q},\tau)=\langle O_{-{\bf q}}(\tau)O_{\bf q}(0)\rangle,
\end{eqnarray}
where $O(\tau)=e^{\tau H}O e^{-\tau H}$ and its relationship to the dynamic structure factor is
\begin{eqnarray}
	G(\boldsymbol{q},\tau)=\frac{1}{\pi}\int_{0}^{\infty}S(\boldsymbol{q},\omega)e^{-\tau \omega}.
	\label{Gtau}
\end{eqnarray}
It is customary to define $S({\bf q},\omega)$ with the full spin operator ${\bf S}_{\bf r}$ instead of just the $z$ component in Eq.~(\ref{opq}), which we
accomplish here by multiplying our QMC computed $G({\bf q},\tau)$ by $3$. In most cases the overall normalization will not be of importance.

We perform the QMC calculations of $G({\bf q},\tau)$ with the Stochastic Series Expansion (SSE) method \cite{Sandvik99,Sandvik10} at sufficiently
low $T$ to converge to the ground state. For efficient evaluation of correlations in imaginary time, we use the method of exact ``time slicing''
\cite{Sandvik19} (see Ref.~\onlinecite{Sandvik97} for other approaches and the relationship to conventional path integrals), with each slice of width
$\Delta_\tau$ expanded to all contributing orders. The space-time correlations can then be easily computed with the states separated by the operators
corresponding to a multiple $n$ of slices, whence $\tau \in  \{n\Delta_\tau\}$. Since there are, as we will see, excitations with energy scaling as
$L^{-2}$ and our goal is to resolve these, we take the inverse temperature as high as $\beta =4L^2$ for $L \le 64$. Disorder averages are computed over at least
1000 random vacancy samples.

\subsection{Stochastic analytic continuation}
\label{sub:sac}

We employ the SAC method \cite{Shao23} to invert Eq.~(\ref{Gtau}) for $S(\boldsymbol{q},\omega)$. We here briefly describe how we apply the
method to the particular model discussed here and also provide test comparing results for small systems with exact Lanczos calculations.

For a set of imaginary times $\tau_i=1,\dots,N_{\tau}$, an SSE simulation provides an unbiased statistical estimate $\bar{G}_i=\bar{G}(\tau_i)$
of the correlation function, where here we suppress the momentum label ${\bf q}$ for simplicity and include it below only when presenting results.
Importantly, the statistical errors of different data points $i$ are correlated \cite{Jarrell96}, which has to be properly taken into account for the SAC
method to be statistically sound. With the SSE data divided into bins indexed by $b=1,2,\ldots,N_B$, the covariance matrix is given by: 
\begin{eqnarray}
	C_{ij}=\frac{1}{N_B(N_B-1)}\sum_{b=1}^{N_B}(G_i^b-\bar{G}_i)(G_j^b-\bar{G}_j),
	\label{cijdef}
\end{eqnarray}
where $G_i^b$ denotes the mean of the data in bin $b$ and $\bar{G}_i$ is the average over all $b$.
The diagonal elements of the covariance matrix $C$ are the conventional variances (squares of
the standard statistical errors), $\sigma_i^2 \equiv C_{ii}$.

With the expected presence of excitations at very low energy, we need correspondingly long imaginary times.
It is then neither feasible nor necessary to use a dense linear grid of $\tau$ points, and we instead use a linear grid only for $\tau \in [0,2]$
at spacing $\Delta_\tau=0.1$, thereafter switching to values of the form $\tau = an^2\Delta_\tau$ for $\tau > 2$, with integer $n$ and $a$
chosen such that 100-200 time points are typically computed. As discussed further below, not all of these data points may be used as input in the SAC,
because of an error-determined cut-off imposed on the maximum time and also because the set may be pruned to ensure a stable covariance matrix.

Given a proposal for $S(\omega)$, a corresponding set of values $\{G_i\}$ can be computed by Eq.~(\ref{Gtau}). The deviation of this set from the
SSE output set $\{\bar{G}_i\}$ is quantified in the standard way by the "goodness-of-fit"
\begin{eqnarray}
\chi^2=\sum_{i=1}^{N_{\tau}}\sum_{j=1}^{N_{\tau}}(G_i-\bar{G}_i)[C^{-1}]_{ij}(G_j-\bar{G}_j),
\end{eqnarray}
which after diagonalizing and transforming everything to the basis of eigenvectors of the covariance matrix takes the conventional uncorrelated form
\begin{eqnarray}
\chi^2=\sum_{i=1}^{N_{\tau}}(G_i-\bar{G}_i)^2/\epsilon_i^2,
\label{chi2}
\end{eqnarray}
where $\epsilon^2_i$ are the eigenvalues of $C$ (i.e., $\epsilon_i$ are the standard deviations of the independently fluctuating stochastic modes)
and the transformed $G_i$ and $\bar{G}_i$ are implied.

In SAC, the spectral function is normally parametrized using a large number of $\delta$ functions located at frequencies $\omega_i$ and with
positive semi-definite amplitudes $a_i$. The parameter space $\{a_i,\omega_i\}$ is treated as that of a statistical mechanics problem with $\chi^2$
corresponding to the energy, sampling according to the Boltzmann-like distribution,
\begin{eqnarray}
P(S)\propto \exp \left (-\frac{\chi^2}{2\Theta} \right),
\end{eqnarray}
at a fictitious temperature $\Theta$. For $\Theta \to 0$, this procedure reaches a minimum goodness-of-fit $\chi^2_{\rm min} > 0$ ($\chi^2_{\rm min} =0$
being possible only if negative amplitudes are allowed), which is taken as a proxy for an effective number of degrees of freedom $N_{\rm dof}$ of the fit,
motivated by the fact that the expected value of $\chi^2_{\rm min}$ (in a fit to a valid function) equals $N_{\rm dof}$ in the $\chi^2$ distribution.
The temperature $\Theta$ is then adjusted so that
\begin{eqnarray}
\langle \chi^2(\Theta)\rangle = \chi^2_{\rm min} + a\sqrt{2\chi^2_{\rm min}},
\end{eqnarray}
where $2\chi^2_{\rm min}$ acts as a proxy for the variance of $\chi^2_{\rm min}$ according to the $\chi^2$ distribution. Here the factor $a$ should be of order
$1$ (normally we take $a\approx 0.25 \sim 0.5$), which corresponds to a statistically good average fit, with $\langle \chi^2\rangle$ falling within a standard
deviation of the smallest goodness-of-fit expected, while avoiding overfitting for essentially the same reason. This $\Theta$ criterion is discussed in
detail in Ref.~\onlinecite{Shao23} and was recently further motivated in Ref.~\onlinecite{Schumm24}. The sampled spectrum is collected in the form of
a histogram with bin width adjusted according to the energy scales appearing in the spectrum.

The SAC method is equivalent to the conventional Maximum-Entropy \cite{Jarrell96} method (MEM) under certain conditions (essentially for a very large number
of $\delta$ functions and with the entropic weighting in the MEM adjusted so that $\chi^2_{\rm MEM} = \langle \chi^2_{\rm SAC}\rangle$) \cite{Shao23}, but 
effectively with a different form of the entropy when sampling the spectrum in the space of both amplitudes and frequencies of the $\delta$ functions (while
the conventionally used Shannon entropy in the MEM is realized when sampling only the frequencies of equal-weight $\delta$ functions) \cite{Shao23,Ghanem23}.
It has been shown that the fidelity of the method is often (but now always \cite{Schumm24}) better than the conventional MEM when also the amplitudes
are sampled \cite{Shao23} (a SAC parametrization first suggested by Beach \cite{Beach04}), as we do in all cases here.

\begin{figure}[t]
\includegraphics[width=75mm]{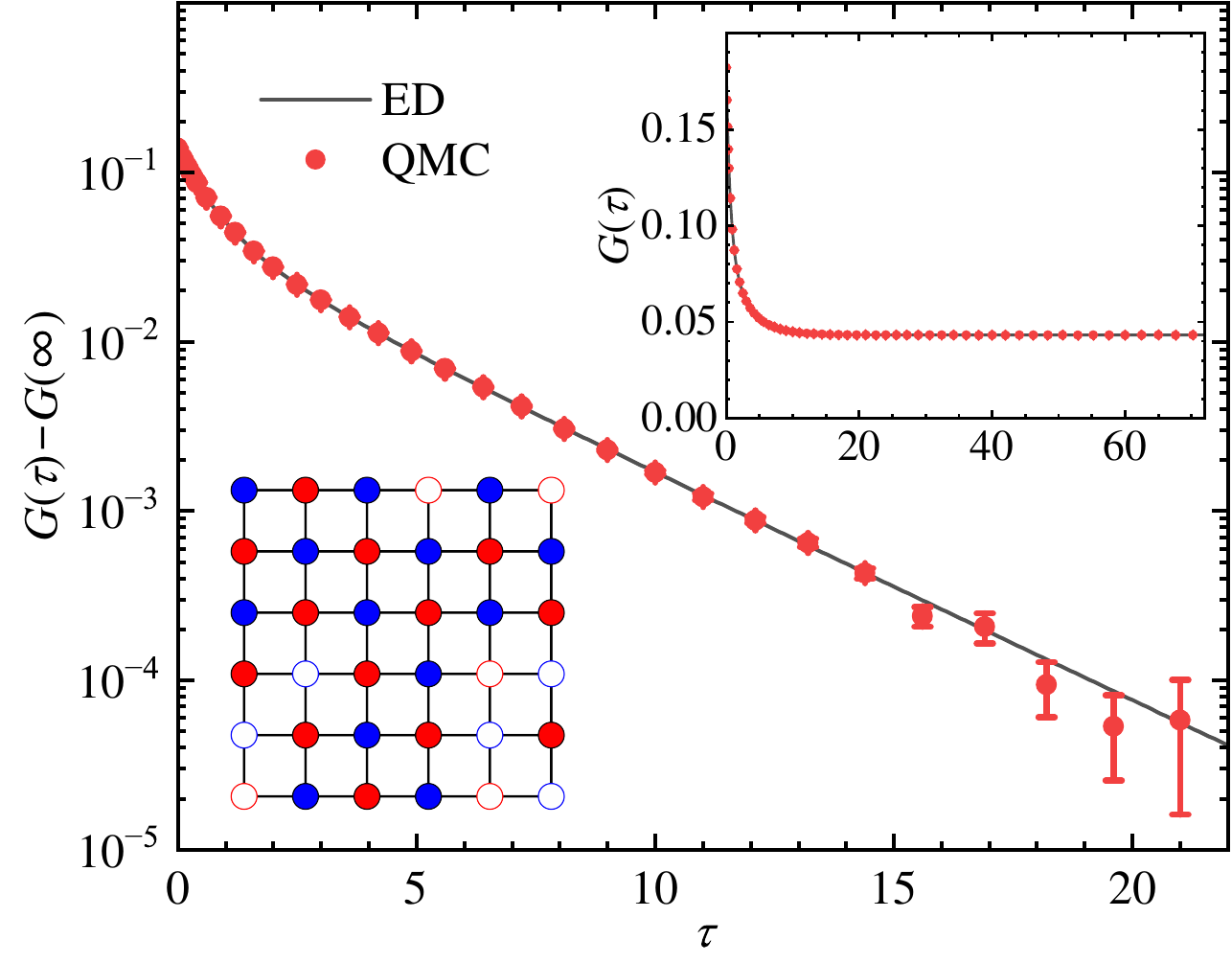}
\caption{SSE computed imaginary-time correlation function at $\boldsymbol{q}=(\pi/3,0)$ for the $6\times 6$ lattice 
in the inset, where there are five vacancies on each sublattice (the sublattices being indicated by red and blue, with spins and vacancies
shown as solid and open circles, respectively). One of the spins (the rightmost one on the second row from the bottom) of the periodic
lattice is isolated from the rest of the system, which forms a connected cluster. Because of the ground state degeneracy of this system,
$G({\bf q},\tau)$ does not decay to zero at large $\tau$, as seen in the inset. After subtracting the large-$\tau$ constant, the decay is
exponential, as shown in the main figure, with the asymptotic exponential decay reflecting the gap, here $\Delta=0.309$, of the large
connected cluster. The curves show the corresponding exact results from Lanczos diagonalization.}
\label{gtau1}
\end{figure}

For a connected spin cluster with equal number of sites on the two sublattices, the ground state is a singlet and $G(\tau)$ decays to zero as
$\tau \to \infty$. However, systems randomly diluted with vacancies will inevitably contain some clusters with unequal sublattice occupation; the
most common case being a single isolated spin, which also causes an opposite imbalance in the remaining large cluster. Such imbalanced clusters
lead to ground state degeneracies (finite total spin of the ground state of odd-sized clusters, as already discussed above in Sec.~\ref{sec:spectral})
and, accordingly, $G(\tau)$ approaches a nonzero constant for large $\tau$. This behavior corresponds to a $\delta$ function at $\omega=0$ in frequency
space, which we omit by removing the large-$\tau$ constant from $G(\tau)$ before performing the analytic continuation.

We illustrate this subtraction scheme for a $6\times 6$ system in Fig.~\ref{gtau1}, where there is a single isolated spin and the remaining large
cluster also has an odd number of spins. The correlation function $G({\bf q},\tau)$ at ${\bf q}=(\pi/3,0)$ is shown for $\tau$ up to $\beta/2$ in the
inset, where the decay toward a nonzero constant is clear. After subtracting this constant, the asymptotic decay is exponential,
$G_{\bf q}(\tau) \propto e^{-\tau\Delta}$, where in this case the gap is $\Delta=0.3090$. The SSE results agree with the exact $G(\tau)$ for all ${\bf q}$,
both before and after subtraction (with the same gap for all ${\bf q}$ because of the lack of translation symmetry). The $\omega=0$ contribution can be
added to the spectrum after the analytic continuation, but in most cases, at the dilution fractions considered here, it is very small and not clearly
visible in the graphs presented below.

Though the maximum imaginary time value, $\tau=\beta/2$ [when taking into account that $G(\tau)=G(\beta-\tau)$ for a bosonic operator], is in principle
as large as $2N$ in our calculations, $G(\tau)$ is typically dominated by statistical noise at these large time separations. In practice, we only use
the $\tau$ points at which the computed relative statistical error (conventional standard deviation), after subtraction of the possible constant, is less
than 10\%. This cut-off is reflected in data in the main part of Fig.~\ref{gtau1}, where the largest $\tau$ value is about $22$ even though $\beta/2=72$,
following our criterion $\beta=4N$ to ensure sufficient convergence to the ground state. As $L$ increases, spectral features appear at lower energies and
the largest $\tau$ value increases accordingly. As an example, for $L=64$ the 10\% criterion for the relative statistical error can correspond to time
separations as large as $\tau \approx 4000$ and a total of $N_\tau \approx 70$ elements in the SAC input set $\{G(\tau_i)\}$.

It should be pointed out here that the statistical error in $G(\tau)$ has two sources: from the finite sampling in the SSE simulation and from
sample-to-sample fluctuations. The latter are typically larger, and, therefore, our SSE runs are rather short, which is possible because the
equilibration and autocorrelation times are short even for randomly diluted Heisenberg systems \cite{Sandvik02}. We can then generate a large
number of vacancy samples for each $L$ and $p$. We have not investigated the details of the error correlations from the two sources of fluctuations,
but all covariance effects are taken properly into account collectively by the covariance matrix, which in our SAC approach is diagonalized
for use in Eq.~(\ref{chi2}).

\subsection{Tests on small systems}
\label{sec:test}

For small systems, e.g., the diluted $6\times 6$ lattice considered in Fig.~\ref{gtau1}, the dynamic structure factor can be directly computed in the form
of Eq.~(\ref{sqwexact}) using Lanczos exact diagonalization. The number of contributing $\delta$ functions is then small (at temperature $T=0$ in our study)
and cannot be well approximated by any continuous function. A direct comparison with results obtained with the SAC method, where an essentially continuous
average spectrum is produced, is then typically not very meaningful. However, the locations and amplitudes of the $\delta$ functions depend significantly on the
details of the vacancy realization, and $S({\bf q},\omega)$ averaged over a large number of samples can therefore be well approximated by a smooth
continuum (which, as we will see, still can contain very sharp peaks). We here use this approach to test the SAC method.

\begin{figure}[t]
\includegraphics[width=64mm]{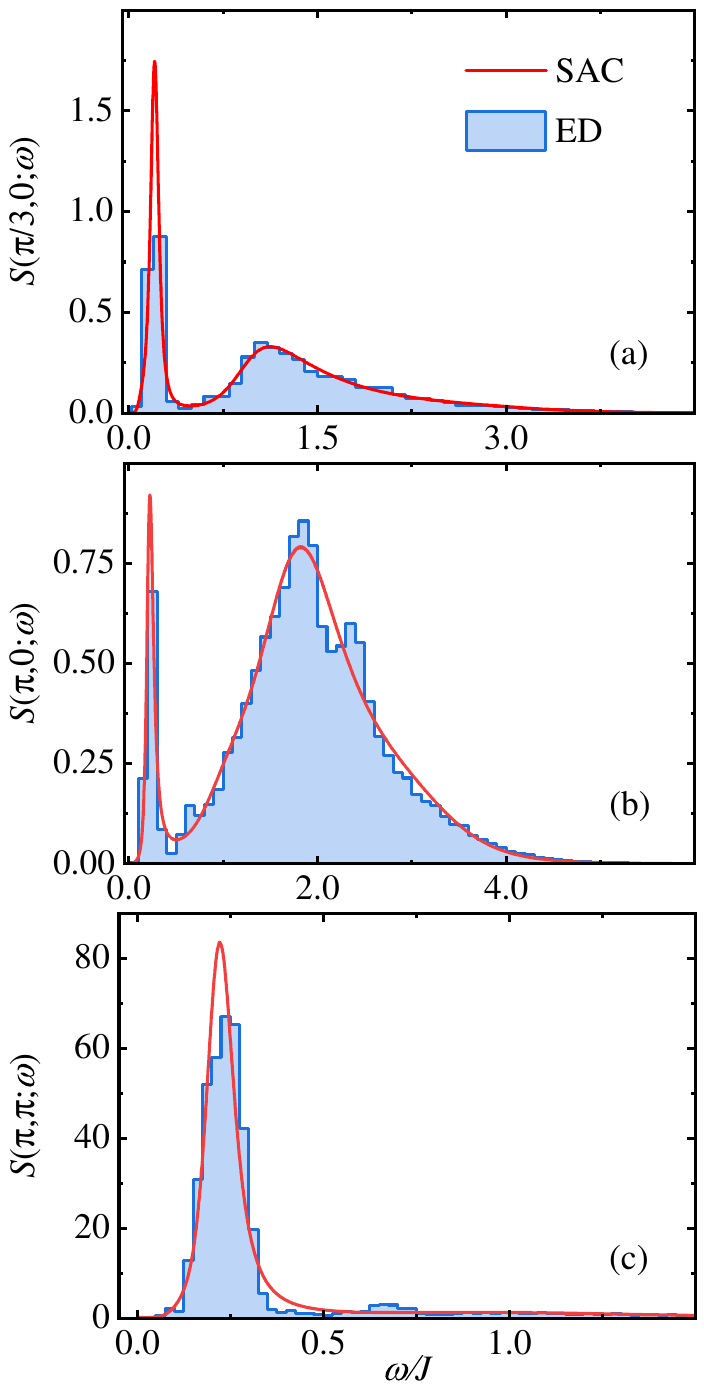}
\caption{Disorder averaged (2000 samples) $S(\boldsymbol{q},\omega)$ for system size $L=6$ diluted at $p=5/18$. Results are shown at
(a) ${\bf q}=(\pi/3,0)$, (b) ${\bf q}=(\pi,0)$, and (c) ${\bf q}=(\pi,\pi)$. The histograms represent the distribution of $\delta$ functions in the
exact expression Eq.~(\ref{sqwexact}) and the red curves are the results of the SAC method applied to SSE data.}
\label{sw_ed1}
\end{figure}

\begin{figure}[t]
\includegraphics[width=64mm]{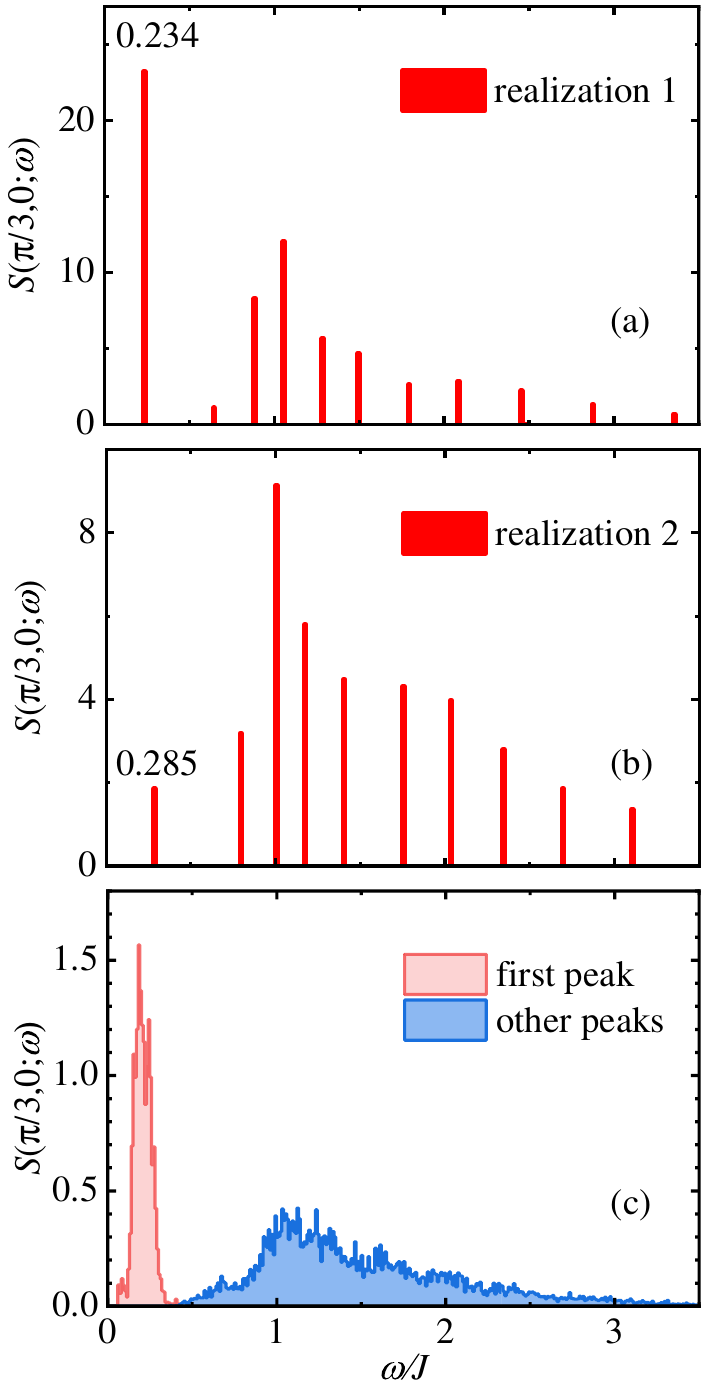}
\caption{Separation of low-energy peak and continuum. The exact form of $S(\boldsymbol{q},\omega)$ at ${\bf q}=(\pi/3,0)$ obtained with two different
vacancy realizations from the same set as in Fig.~\ref{sw_ed1} is shown in (a) and (b). The height of each bar corresponds to the squared matrix element
in Eq.~(\ref{sqwexact}). The location of the lowest $\delta$ function is indicated. The distribution of the energy of the lowest peak over 2000 samples
is shown in red in (c), while the other contributions form the distribution shown in blue. There is practically no overlap between these two parts of the 
distribution. Here the data is the same as in Fig.~\ref{sw_ed1}(a) but the bin width in the histogram is smaller.}
\label{first_peak}
\end{figure}

Results for $L=6$ obtained with both the Lanczos method and SSE combined with SAC are displayed in Fig.~\ref{sw_ed1} at three different momenta.
The overall shapes of the Lanczos spectra, represented as histograms with suitable bin width, are remarkably well reproduced by the SAC method,
except for small-scale structure that cannot realistically be reproduced by any numerical analytic continuation method applied to QMC data. In all
cases, there is a sharp Gaussian like low-energy peak followed by a broad profile with non-trivial shape that depends on the momentum. For
${\bf q}=(\pi/3,0)$ and $(\pi,0)$ in Figs.~\ref{sw_ed1}(a) and \ref{sw_ed1}(b), even the high-energy tails are almost flawlessly reconstructed
by the SAC, while the much smaller second peak at ${\bf q}=(\pi,\pi)$ in Fig.~\ref{sw_ed1}(c) cannot be well replicated because of its small relative
spectral weight. The frequency resolution of peaks in analytic continuation of QMC data generally is better at low frequency, but the ability to resolve
a specific detail also depends on its relative weight and other features present.

Below we will demonstrate that the lower peak in Fig.~\ref{sw_ed1} develops with increasing system size into a $\delta$ function at energy $\omega \propto 1/N$,
reflecting the lowest excitation in what is known as the tower of Anderson quantum rotor states associated with the breaking of the O($3$) symmetry of the N\'eel
order parameter \cite{Anderson52,Manousakis91,Sandvik10}. This peak is very sharp and close to $\omega=0$ in Fig.~\ref{schematic}, which is based on results
for $L=64$. In a translationally invariant system, the $S=1$ rotor excitation of
the ground state is present in $S({\bf q},\omega)$ only at $(\pi,\pi)$. However, in a diluted system, where the momentum is not conserved, it spreads
out over the entire BZ in a way reflecting inhomogeneity in the $\omega=0$ long-range order.
The energy of the rotor excitation must then be independent of ${\bf q}$,
as indeed can be observed in Fig.~\ref{sw_ed1} (note the different scales on the $\omega$ axis for different ${\bf q}$), where at ${\bf q}=(\pi,\pi)$ the
rotor peak contains the majority of the total spectral weight. We will show that the distribution of the spectral weight of the rotor mode, in both
momentum space and real space, provides valuable information on the nature of the disordered N\'eel state.

For a finite system with rotationally averaged N\'eel ordered ground state (i.e., with the amplitude of the order parameter formed but the symmetry
not broken), the lowest $\delta$ function in the exact spectrum represents the rotor excitation. We can specifically confirm that the Gaussian like
broadening of the low-energy peaks in Fig.~\ref{sw_ed1} reflects the sample-to-sample fluctuations of the lowest peak in the exact spectra.
Fig.~\ref{first_peak}(a) and Fig.~\ref{first_peak}(b) show the exact form Eq.~(\ref{sqwexact}) of $S({\bf q},\omega)$ for two vacancy samples. Here the
amplitudes of the $\delta$ functions correspond to the heights of the bars at the locations $\omega_n=E_n-E_0$. Separating the contributions from the
lowest $\delta$ function and all other spectral weight, we see in Fig.~\ref{first_peak} that the low-energy peak indeed arises exclusively from the
lowest excitation of each sample. In Sec.~\ref{sec:sqw} we will demonstrate that the width of the peak shrinks as the system size is increased while
its location scales toward $\omega=0$ in the way expected for a rotor excitation.

The ground state of  a cluster with equal number of spins on both sublattices has total spin $S=0$ and the gap to the rotor states are of the
form $\Delta_S \propto S(S+1)/N$. Only the $S=1$ state is visible in the dynamic spin structure factor because of the selection rules for the spin operators.
The clusters formed in a diluted system typically have $S>0$ ground states, however, because of sublattice imbalance. In the example in Fig.~\ref{gtau1},
the ground states of the large connected cluster and the isolated spin both have spin $S=1/2$, which implies that the ground state of the entire system
is degenerate, with $S=0$ and $S=1$. For larger systems with many isolated spins, where the main connected clusters can have ground state spin $S>0$,
there can be rotor states with both $S+1$ and $S-1$. They will typically be very close to each other in energy, and we therefore still expect a single narrow
rotor peak in the disorder averaged $S({\bf q},\omega)$ at $\omega \propto 1/N$. The trivial $\delta$-peak resulting from ground state degeneracy are
always removed from the imaginary-time correlation functions, as we explained above (Fig.~\ref{gtau1}).

Another potential complication for large systems is that there can also be relatively large isolated clusters of many spins in addition to the system spanning
majority cluster. The emergent rotor states of these smaller clusters will have relatively high energy, and their contributions to $S({\bf q},\omega)$
can then mix in with the other spectral features. However, at the low dilution fractions we consider here, isolated clusters of size larger than $1$
are rare, and $S({\bf q},\omega)$ is completely dominated by the largest cluster. We will see in Sec.~\ref{sec:sqw} that a very clear separation
forms with increasing system size between the low-energy rotor peak and the features at higher energy, even when those other features also move down in energy (in
particular the localization peak), because the rotor peak narrows at the same time.

\subsection{Dimer-monomer model}
\label{sec:dimer}

Even in a cluster with equal number of vacancies on the two sublattices, effectively localized spin degrees of freedom can emerge due to local sublattice
imbalance, where bipartite antiferromagnetic order in combination with local excess of sites on one of the sublattices leads to an effective nonzero total
spin of an isolated region of the lattice. These effective moments are often referred to as "dangling spins", though they can form at arbitrary locations
of the lattice and are not necessarily (though often) obviously related to spins with very low coordination numbers; they are in many cases effectively
spread out over large regions of the lattice, with different regions essentially isolated from each other. Regions with sublattice imbalance
exhibit elevated magnetic response.

The presence of regions with net moments is a key aspect of the diluted Heisenberg model. The formation of local sublattice imbalance and the so induced
magnetic properties can often be surprisingly well modeled by a classical dimer-monomer model \cite{Wang10,Changlani13a,Changlani13b,Ghosh15,Bhola22}. In
this simplified heuristic description, each monomer represents an ``unpaired'' spin and the dimers represent singlets formed by two nearest-neighbor spins
that reduce the background N\'eel order in which they reside. The unpaired spins associated with the monomers are regarded as fully aligned with the overall
N\'eel order, with, say, up and down orientation on sublattice A and B respectively. Regions with nonzero density of monomers therefore correspond to increased
N\'eel order and magnetic response. The basic dimer-monomer model is purely entropic, i.e., all configurations of dimers and monomers have equal weight in the
partition function and the configurations can be easily generated using Monte Carlo simulations. We will take this approach here with a diluted dimer-monomer
model to explain some of the qualitative features of the distribution in the BZ of the weight of the rotor state in $S({\bf q},\omega)$, and also to discuss
the potential connection between regions of local sublattice imbalance and localized excitations. For a given set of static vacancies, we are interested in
configurations with the minimum number of monomers; those monomers will then represent both local and global sublattice imbalance.

\begin{figure}[t]
\includegraphics[width=60mm]{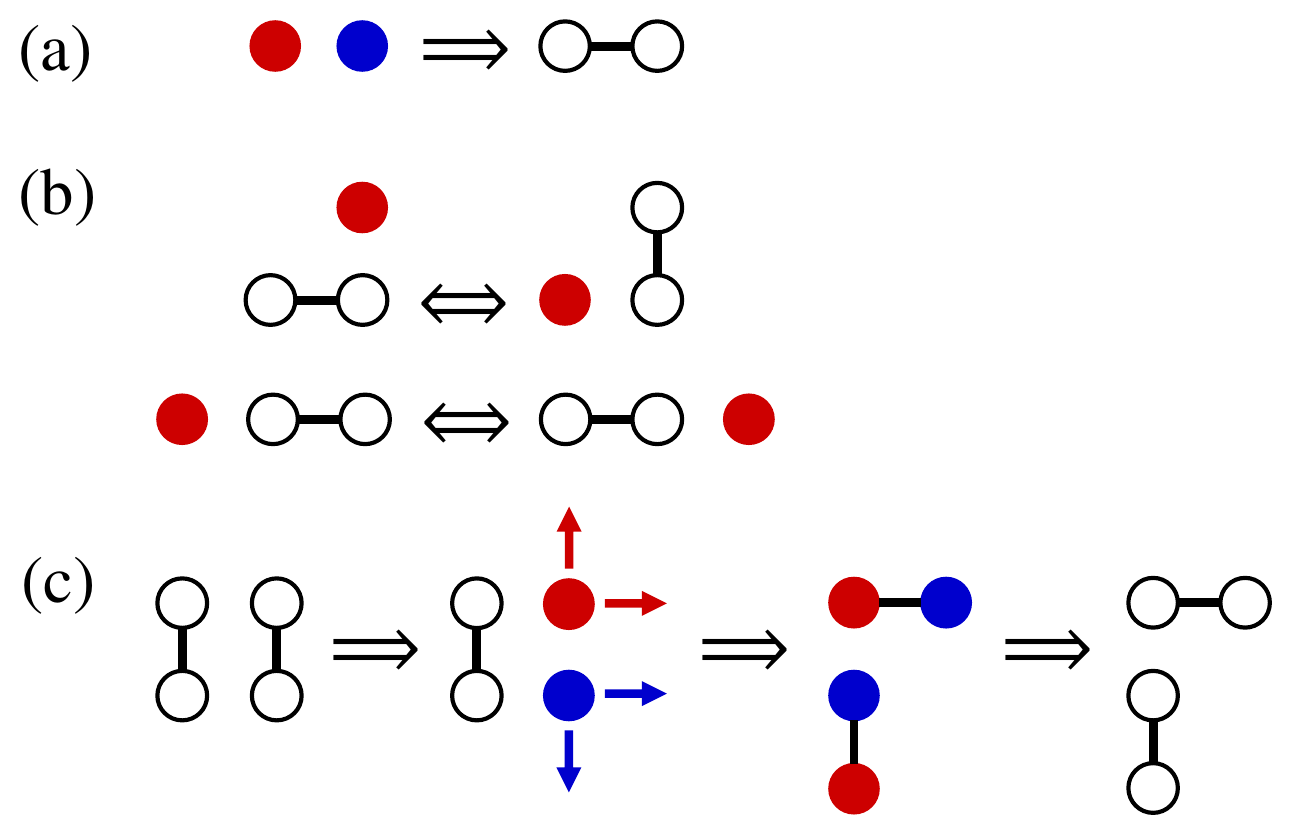}
\caption{Updating processes of the dimer-monomer model, with red (blue) circles representing sites of sublattice A (B). (a) Two nearest-neighbor monomers
are annihilated, leading to a dimer. (b) Two different joint moves of a monomer and a dimer. Note that a given monomer can only move within one sublattice.
(c) Breaking a dimer, then moving the two monomers until they can be annihilated with other monomers as in (a). Starting from a configuration of random
dimers and monomers (where we place dimers first until no longer possible, whence the remaining sites are assigned as monomers), a series of updates are
performed and the number of monomers gradually decreases by the process in (a). After a sufficiently long time the minimum number of monomers has been
reached and the process continues for the purpose of collecting the spatial distribution of the remaining monomers. The update in (c) is strictly not needed
for ergodicity, but it significantly speeds up the sampling after the minimum number of monomers has been reached.}
\label{dimer_update}
\end{figure}

\begin{figure}[t]
\includegraphics[width=50mm]{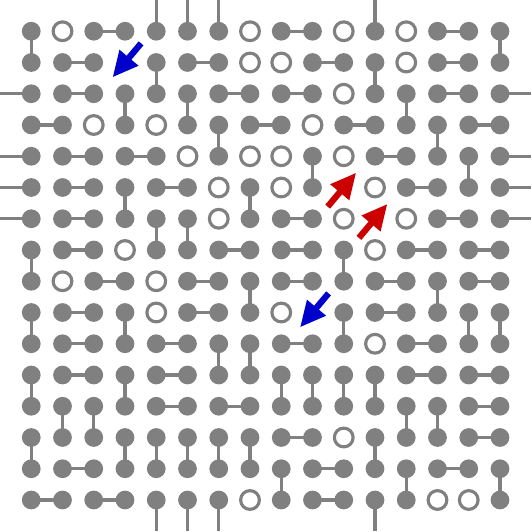}
\caption{A dimer-monomer configuration on a $16 \times 16$ lattice, with vacancies indicated by open circles. Dimers connect nearest-neighbor sites
and monomers are indicated by arrows, with red (up) and blue (down) corresponding to the A and B sublattices, respectively.}
\label{sketchmap_dm}
\end{figure}

The monomers undergo stochastic (Monte Carlo) dynamics along with the dimers, with the required ergodic updates of the dimer-monomer configurations 
illustrated and further explained in Fig.~\ref{dimer_update}. During the equilibration stage of this process, any pair of monomers occupying nearest-neighbor
sites is eliminated; thus replaced by a dimer. Once a steady state with the minimal number of monomers remaining has been achieved, the spatial monomer
distribution is accumulated; for details see Ref.~\onlinecite{Wang10}. In Fig.~\ref{sketchmap_dm} we show one configuration from such a simulation in which there
are two monomers on each sublattice, with the sublattices associated with different spin-$z$ excess. Upon running the simulation, it is found that the
``up'' monomers are restricted to a small region of the lattice (one of them being completely isolated by surrounding vacancies), while the ``down'' ones
can occupy most of the lattice but cannot reach the sites neighboring the monomers on the opposite sublattice, in which case they would be annihilated.

As already mentioned, in a quasi-classical heuristic picture of the N\'eel state, the monomers are associated with spin up or down according to the sublattice within
which they reside, to account for the spins in a perfect checkerboard pattern that are not paired up (sublattice-balanced) with a nearest neighbor. The dynamic (sampled)
monomer distribution typically forms isolated regions where the monomers can move according to the updates in Fig.~\ref{dimer_update}. These regions define local
sublattice imbalance and associated net magnetic moments. The dimer-monomer model only concerns the spatial density distribution $\rho_{\bf r}$ of these moments,
while the background N\'eel order  defines the assignment of spin up or down to the monomers, i.e., net moments $\pm \rho_{\bf r}$. More quantitatively,
the spatially varying N\'eel order parameter $m_s({\vec r})$ can be written as
\begin{equation}\label{msdimermodel}
m_s({\vec r})=(-1)^{r_x+r_y}[1/2 + d(\rho_{\bf r}-1)],
\end{equation}  
so that $m_s({\bf r})=1/2$ for a site with the maximum monomer density, $\rho_{\bf r}=1$, and $m_s({\bf r})=1/2-d$ for sites with no monomers. Here $d \in [0,1/2]$
represents the suppression of the order by quantum fluctuations, which should increase with the dilution fraction $p$ on account of the singlet formation associated
with the dimers.

In a more sophisticated treatment of a quantum state expressed with dimer-monomer configurations, each dimer connecting sites $a$ and $b$ could in principle be
regarded as a ``N\'eel biased singlet'';
\begin{equation}\label{dimersinglet}
|\Psi_{ab}\rangle = |\uparrow_a\downarrow_b\rangle + c_{ab}(|\uparrow_a\downarrow_b\rangle - |\downarrow_a\uparrow_b\rangle),
\end{equation}
and each dimer-monomer configuration would then be a normalized product state of $|\Psi_{ab}\rangle$ on all the dimers together with the $|\uparrow_a\rangle$ and
$|\downarrow_b\rangle$ states associated with the monomers. The sampled dimer-monomer configurations then correspond to the different terms of a superposition. In
practice, however, the basis is overcomplete, like the conventional valence-bond basis, and state overlaps would have to be taken into account in the sampling
weights and matrix elements of observables. The constant $c_{ab}$ in Eq.~(\ref{dimersinglet}) can be variationally optimized, and the flexibility of the state
could be further extended by introducing longer-range dimers. Here we only compute $\rho_{\bf r}$ using classical sampling within the standard dimer-monomer model,
which corresponds to neglecting the non-orthogonality of the basis. Instead of discussing the sublattice magnetization in the form (\ref{msdimermodel}), for simplicity
we will just show results for $\rho_{\bf r}$.

\begin{figure}[t]
\centering
\includegraphics[width=60mm]{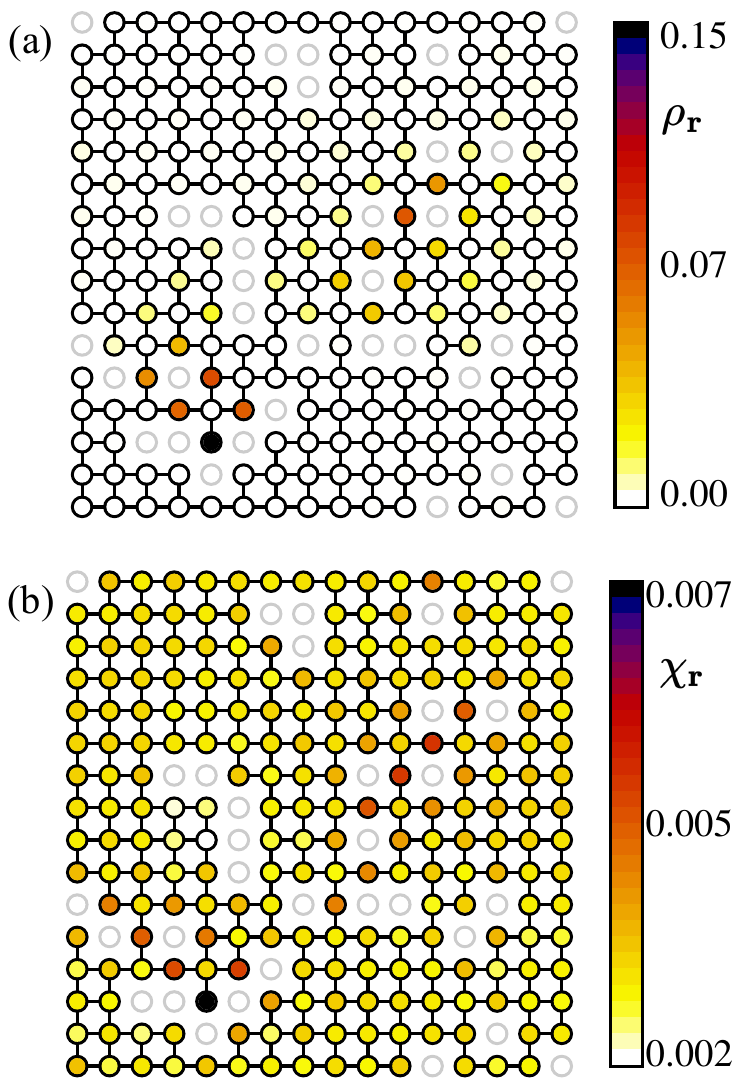}
\caption{Visualization of effective magnetic moments formed due to local sublattice imbalance in an $L=16$ system at dilution fraction $p=1/8$.
  The dim circles not linked to other circles represent vacancies. (a) The monomer density obtained by sampling dimers and monomers according to the
  rules in Fig.~\ref{dimer_update}. (b) The local spin susceptibility, Eq.~(\ref{locsus}), of the spin-1/2 Heisenberg model, Eq.~(\ref{ham}).}
\label{monomer_locx}
\end{figure}

An important measure of nonuniformity in the diluted system is provided by the local spin susceptibility $\chi_{\bf r}$,
\begin{eqnarray}
	\chi_{\bf r}=\int_{0}^{\beta}d\tau\langle S_{\bf r}^z(\tau)S_{\bf r}^z(0) \rangle,
\label{locsus}
\end{eqnarray}
where the integral can be performed analytically without approximations within the SSE method \cite{Sandvik92}.
Previous studies have demonstrated a strong correlation between the spatial nonuniformity of the low-energy excitations and the monomer density at
the 2D Heisenberg model at percolation point and other fractal clusters \cite{Wang10,Changlani13a,Changlani13b,Ghosh15}. We provide a similar example in the
context of the 2D Heisenberg model at low dilution in Fig.~\ref{monomer_locx}. Here we clearly observe enhanced local magnetic response $\chi_{\bf r}$ at the
sites with high monomer density, though there are also sites with vanishing monomer density but nevertheless large response. Note again that the monomer
density represents enhanced response relative to a background value of the frozen N\'eel order in the regions without monomers. An enhanced response due
to locally isolated moments is intuitively clear and demonstrated by our results here and in the previous works cited above. Given the simplicity of the
dimer-monomer and the lack of rigorous connections to the quantum magnet, it is not possible to write down a precise relationship between $\rho_{\bf r}$ and
$\chi_{\bf r}$.

In Sec.~\ref{sec:rotor} the purpose of studying the dimer-monomer model will be in the context of the weight distribution of the Anderson rotor
excitation in the BZ, where we will similarly find that some aspects of the nonuniform symmetry breaking can indeed be explained by local sublattice
imbalance, modeled by monomers, while there is also structure not captured by the clearly very simplistic classical description.
The monomer regions of the dimer-monomer model also provide insights into the localized modes of the quantum system at large length scale
for low $p$. In Sec.~\ref{sec:tomo} we will indeed find that the spectral weight in real space of the localized excitations is concentrated
primarily on some of the corresponding monomer-rich sites. However, the quantitative match between the classical and quantum models is not obvious,
as we will discuss further also in Sec.~\ref{sec:conc}.

\section{Dynamic Structure Factor}
\label{sec:sqw}

In Sec.~\ref{sec:qdep} we present our results for $S(\boldsymbol{q},\omega)$ computed on system sizes up to $L=64$, averaged over 1000 disorder
realizations typically. In Sec.~\ref{sec:local} we discuss the local dynamic structure factor $S_0(\omega)$, i.e., the ${\bf q}$ averaged
$S(\boldsymbol{q},\omega)$. In addition to results for $p=1/16$ and $p=1/8$ that we mainly focus on, in Sec.~\ref{sec:single} we consider the case
of a single vacancy and two far separated vacancies, comparing these with results for the clean system. In Sec.~\ref{sec:tomo} we show results for the
real-space resolved spectral function $S(\boldsymbol{r},\omega)$ of the diluted systems in different frequency windows (``tomography''), which allows
us to formulate a more complete picture of the nature of the different types of excitations.

\subsection{Multiscale excitation structure \\ and momentum dependence}
\label{sec:qdep}

\begin{figure}[t]
\includegraphics[width=55mm]{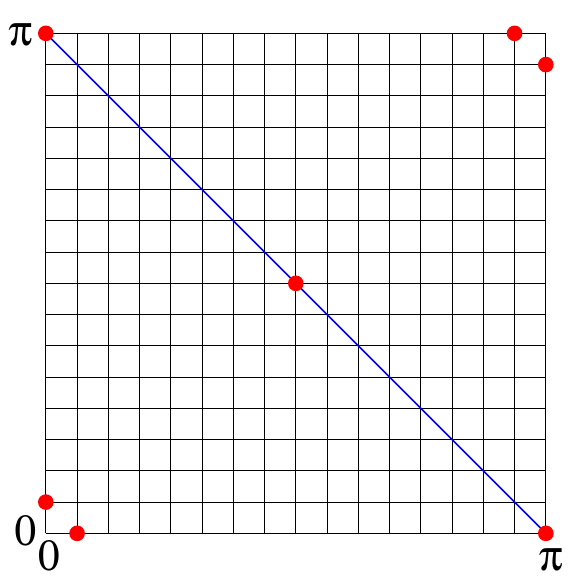}
\caption{One quadrant of the BZ for a lattice of size $L=32$. In the uniform Heisenberg model there are gapless excitations at $(0,0)$ and $(\pi,\pi)$,
with linearly dispersing spinwave around these points. Within linear spin-wave theory \cite{Anderson52,Manousakis91} the highest-energy excitations are on
the magnetic zone boundary (in the folded BZ of the N\'eel state), indicated by the blue line, with the degeneracy on this line broken in higher-order
spinwave theory \cite{Igarashi92,Canali93,Igarashi92} and in nonperturbative treatments \cite{Singh95,Sandvik01,Shao17}, with the highest energy being
at $(\pi/2,\pi/2$). The red circles mark points (some of them symmetrically equivalent)
neighboring the gapless points and on the zone boundary, for which results for $S({\bf q},\omega)$ are shown in Fig.~\ref{sw_p}. Other results will
be presented on lines connecting the gapless points and the zone boundary.}
\label{bz}
\end{figure}

\begin{figure}[t]
\includegraphics[width=26em]{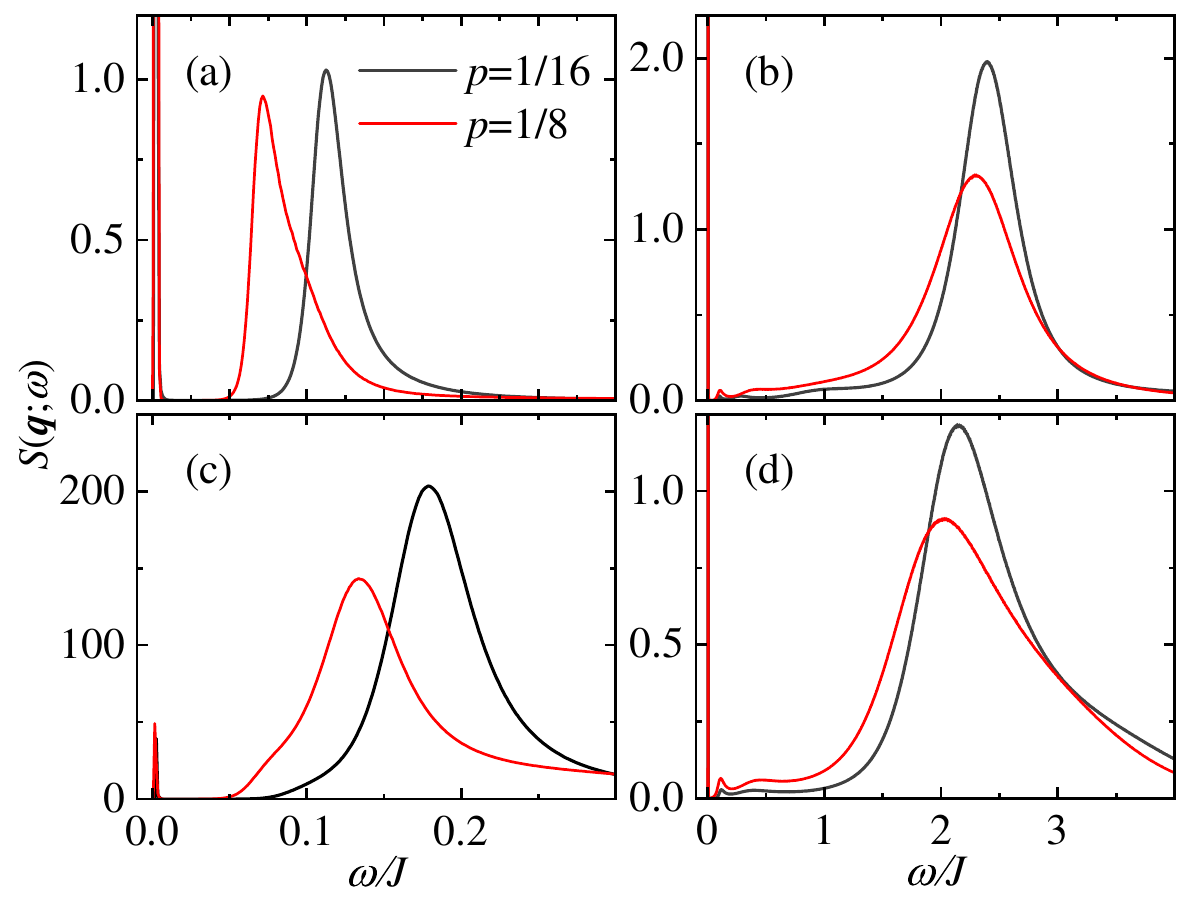}
\caption{$S(\boldsymbol{q},\omega)$ for $L=64$ at the $\boldsymbol{q}$ points marked in Fig.~\ref{bz}. Results are shown in each panel for both $p=1/16$
(black curves) and $p=1/8$ (red curves).
The ${\bf q}$ points are: (a) $\boldsymbol{q}=(2\pi/L,0)$ (i.e., long wavelength), (b) $\boldsymbol{q}=(\pi/2,\pi/2)$ (on the zone boundary),
(c) $\boldsymbol{q}=(\pi-2\pi/L,\pi-2\pi/L)$ (close to the antiferromagnetic wave number), (d) $\boldsymbol{q}=(\pi,0)$ (also on the zone boundary). 
Note that the $\omega$ scale is the same for (a), (c), where there is no high-energy spectral weight, and also for (b), (d), where weight extends
to high energy. Note that the small peaks (identified as the localized mode) and flat continua above them in (b), (d) should in principle be present
also in (a), (c) but are masked by the damped spin waves that extends down to the same energy scale.}
\label{sw_p}
\end{figure}

We will present results in the BZ of the uniform system, depicted in Fig.~\ref{bz} with special points marked and with some basic facts on the spin waves
of the uniform system reviewed in the caption. Though the momentum is not conserved for individual disorder realizations of the diluted system, the translational
symmetry is restored in correlation functions upon disorder averaging, and $S({\bf q},\omega)$ therefore is defined on the same BZ. For some initial observations
at high- and low-energy points in the BZ, in Fig.~\ref{sw_p} we show $S(\boldsymbol{q},\omega)$ for $L=64$ at four representative momenta (those marked in
Fig.~\ref{bz}), comparing results for the two different vacancy fractions. In all cases, the rotor mode at very low frequency, discussed in the case of a small
system in Sec.~\ref{sec:test}, is apparent, and there is a clear separation between this very narrow peak and the other features of the spectrum. The rotor peak
is clearly much narrower than in Fig.~\ref{sw_ed1}, and we will show below that it systematically narrows with increasing system size while moving down
toward $\omega=0$.

At the momenta closest to ${\bf q}=(0,0)$ and $(\pi,\pi)$, shown in Fig.~\ref{sw_p}(a) and Fig.~\ref{sw_p}(c), respectively, there is an asymmetric main
peak that moves down in frequency upon increasing $p$ while also broadening somewhat. On the magnetic zone boundary, Fig.~\ref{sw_p}(b) and \ref{sw_p}(d),
the dominant peak is also slightly broader for the larger $p$ value, though less so in relation to the peak energy. In addition to what is most
naturally interpreted as a highly damped magnon peak with a substantial multimagnon tail (extending to $\omega \approx 5J$ at the zone boundary),
there is also a substantial almost flat portion at low energies, ending at a small peak that is clearly separated from the rotor peak. As we will show
below, the peak corresponds to the localized mode discussed by Chernyshev et al.~\cite{Chernyshev02}.

The dispersionless localization peak should also be present in Figs.~\ref{sw_p}(a) and \ref{sw_p}(c), but its relative weight is small and its
energy scale falls within the much larger main peak. At these momenta, which correspond to long-wavelength uniform and staggered fluctuations, separating the
expected different features of the spectrum above the low-energy rotor peak is not possible within the resolution of our method. This state of affairs also
likely reflects the expectation that all the features corresponding to localized excitations and propagating modes should merge close to ${\bf q}=(0,0)$
and $(\pi,\pi)$, reflecting fully incoherent quantum dynamics \cite{Chernyshev02}. All the different features of the multiscale spectral functions
can be most clearly observed at momenta away from the extreme points considered in Fig.~\ref{sw_p}, as already exemplified in Fig.~\ref{schematic}.

\begin{figure}[t]
\includegraphics[width=65mm]{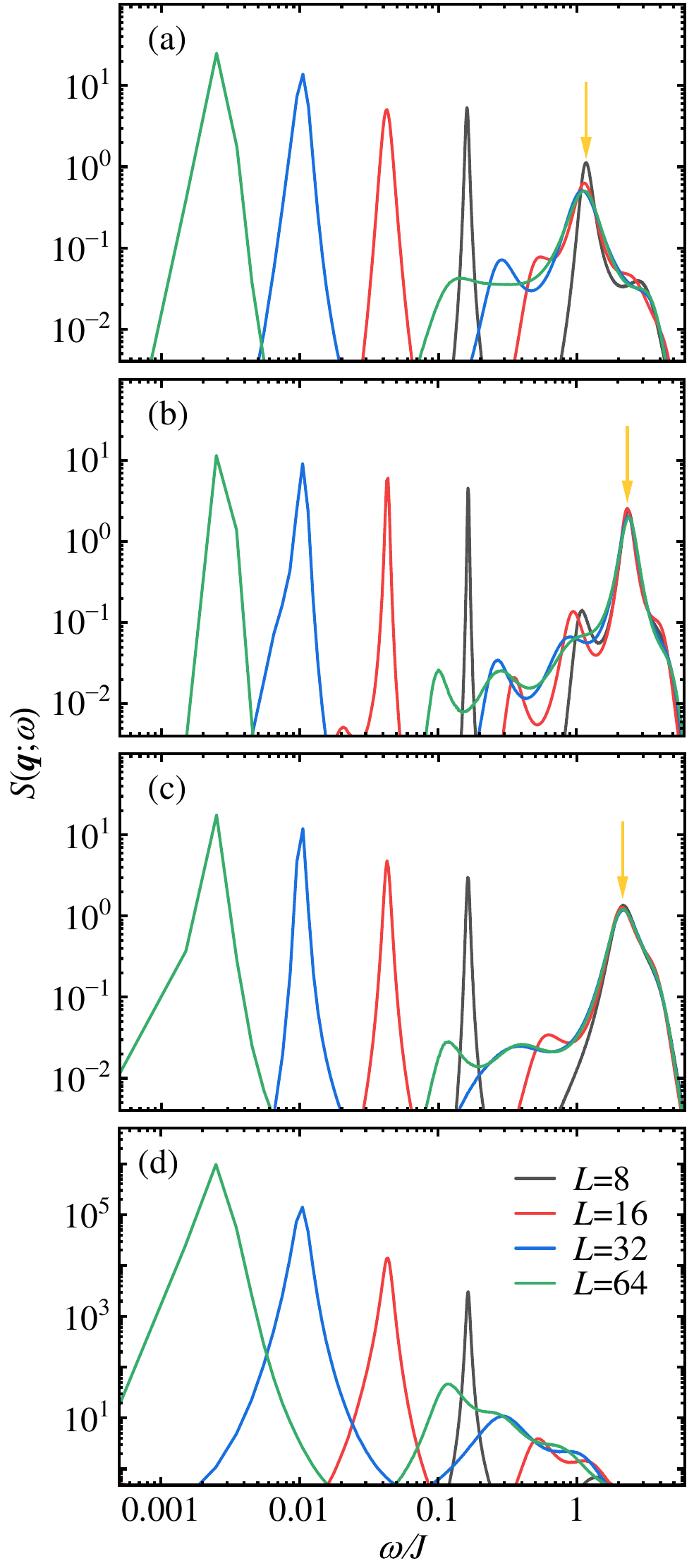}
\caption{Log-log plots of $S(\boldsymbol{q},\omega)$ for $p=1/16$ at (a) $\boldsymbol{q}=(\pi/4,0)$, (b) $\boldsymbol{q}=(\pi/2,\pi/2)$,
(c) $\boldsymbol{q}=(\pi,0)$, (d) $\boldsymbol{q}=(\pi,\pi)$. The dominant broadened magnon peaks in (a)-(c), which converge quickly with increasing
$L$, are marked with arrows. In (d), at the gapless (for $L \to \infty$) ordering wave-vector, the rotor weight is very large and there is no sign
of a well defined single-magnon peak.}
\label{sw_p16}
\end{figure}

The very substantial broadening of the dominant $\omega \approx 2J$ peaks in Figs.~\ref{sw_p}(b) and \ref{sw_p}(d), for ${\bf q}$ close to the
magnetic zone boundary, is absent in the $T$-matrix studies \cite{Brenig91,Chernyshev02}, where the interactions between the magnons and the local
impurity states were not accounted for. Some broadening was observed in a numerical diagonalization study of the linear spin wave Hamiltonian on
finite lattices with vacancies \cite{Mucciolo04}, where, for each disorder realization,  the interactions between the spin waves and the impurities
are fully accounted for at the level of the linear spin-wave Hamiltonian, in contrast to the further approximations made within the $T$-matrix approach
(which is formulated for the infinite system with a different form of disorder averaging). The results of numerical spin wave theory still do not agree
well with our SAC results. In particular, the long tail up to $\omega/J \approx 5$ [beyond the $\omega$ scale used in Figs.~\ref{sw_p}(b) and \ref{sw_p}(d)]
is missing, which indicates that it arises from multimagnon excitations (which are naturally not present within linear spin-wave theory) in the fully
interacting system.

To study the size dependence, in Fig.~\ref{sw_p16} we graph results for $p=1/16$ in log-log plots so that the low energy features can
be examined more clearly. Since the spectral functions are accumulated in histograms with uniform bin width $\Delta_\omega$, at the very low energy
scales, where $\omega$ approaches $\Delta_\omega$, the low-energy peak occupies only a small number of bins, causing the peaks to appear less smooth.
It is nevertheless apparent that the lowest peak moves down rapidly as $L$ increases, exactly as would be expected for an Anderson quantum rotor
mode of a system with long-range N\'eel order.

\begin{figure}[t]
\includegraphics[width=20em]{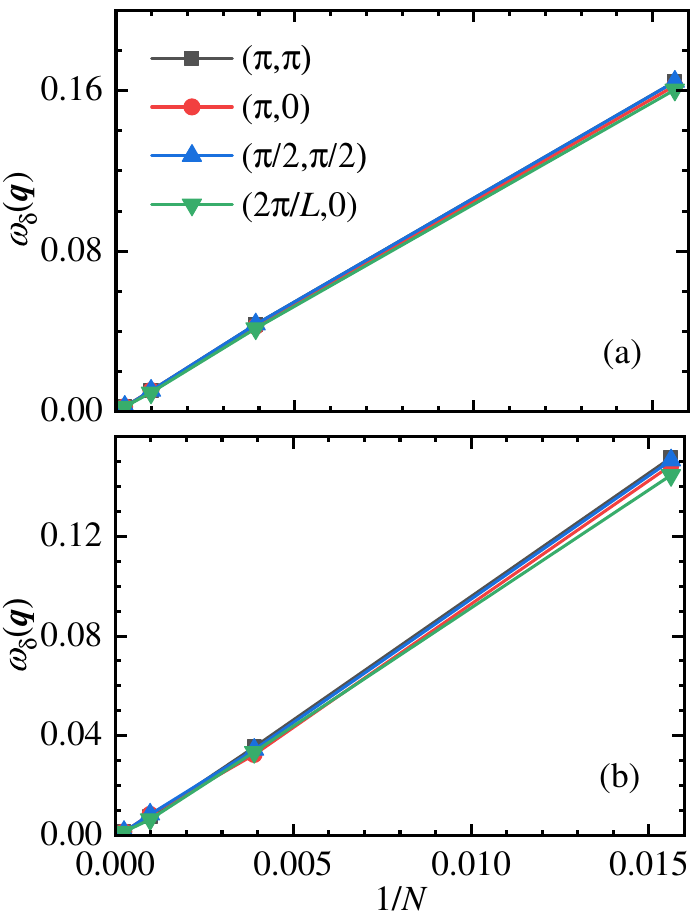}
\caption{Energy of the lowest-energy peak in $S({\bf q},\omega)$ at four different momenta (from results such as those in Fig.~\ref{sw_p16})
vs the inverse lattice volume $1/N=1/L^2$. The dilution fraction is $p=1/16$ in (a) and $p=1/8$ in (b). The energy being the same at all
momenta at fixed $N$ and the asymptotic $1/N$ scaling confirm the identification of the peak as the lowest Anderson quantum rotor excitation.}
\label{rotor_ens}
\end{figure}

Considering the log scale for $\omega$ in Fig.~\ref{sw_p16}, the rotor peak also narrows with increasing $L$, reflecting less fluctuations
in the lowest excitation (recall Fig.~\ref{sw_ed1}) as the magnitude of the N\'eel order stabilizes. For fixed system size the peak is located
at the same energy at all momenta. As shown in Fig.~\ref{rotor_ens}, this energy scales with the system size in the expected way as
$1/N=1/L^2$ for an Anderson quantum rotor state \cite{Anderson52,Neuberger89,Fisher89,Bernu94}.

\begin{figure}[t]
\includegraphics[width=75mm]{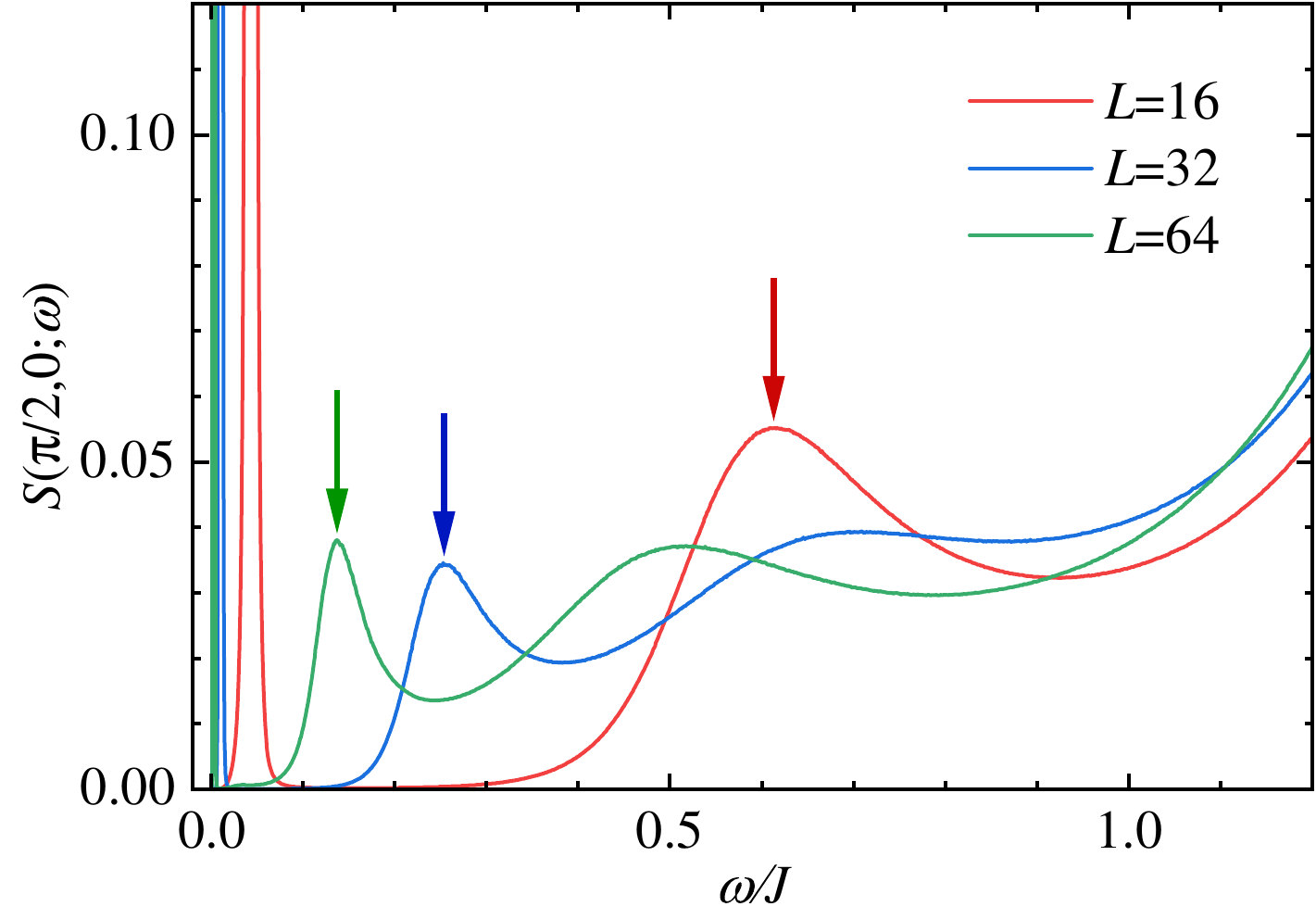}
\caption{Low-energy part of the dynamic structure factor at $\boldsymbol{q}=(\pi/2,0)$ and dilution fraction $p=1/16$, shown for system sizes
$L=16$, $32$ and $64$. The position of the size dependent localization peak is marked by arrows.}
\label{locpeak_mark}
\end{figure}

Returning to Fig.~\ref{sw_p16}, the next low-energy feature above the rotor peak, at $\omega \approx 0.1$ for $L=64$, also moves down in frequency
with increasing system size while becoming more well defined as a peak. Results at $p=1/16$ for a different momentum, ${\bf q}=(\pi/2,\pi/2)$
is shown in more detail on a linear scale in Fig.~\ref{locpeak_mark}. Here as well the size shift is roughly $\propto 1/N$, but we are not able to
carry out a reliable extrapolation to infinite size. Within estimated uncertainties, the peak location extrapolates to zero, but most likely it
tends to a very small non-zero $\omega$ value. The predicted energy scale $\omega_{\rm loc} \propto {\rm exp(-\pi/4p})$ of with the localized mode 
\cite{Chernyshev02} would at $p=1/16$ indeed be too small to resolve using our $S({\bf q},\omega)$ results for sizes up to $L=64$. The clear
separation between the rotor peak and the emergence of a second low-energy peak above it nevertheless lend support to the second feature reflecting a
specific type of excitation different from the quantum rotor below and magnon above. Note that the energy of the damped magnon peak is at
$\omega/J \approx 2$ for ${\bf q}$ on the the zone-boundary, as in Figs.~\ref{sw_p16}(b) and (c), and also at ${\bf q}=(\pi/2,0)$ in Fig.~\ref{locpeak_mark},
it falls above the highest $\omega$ graphed---we will analyze the spin wave energies in detail further below. The emergence of a peak at very low energy
(but clearly above the rotor peak) throughout the BZ with no detectable dispersion suggest that it indeed reflects the predicted localized excitations of
large spatial extent. In further support of a dispersionless mode, in Sec.~\ref{sec:local} we will discuss a much sharper peak in the local (${\bf q}$
integrated) structure factor, allowing for $L\to \infty$ extrapolation of the localization energy at $p=1/8$.

Yet another important aspect of the results in Fig.~\ref{sw_p16} is that the broad high-energy peak, which we will below identify as the predicted
\cite{Chernyshev02} damped magnon with anomalous dispersion, converges rapidly with increasing system size, while the tail below it, with an edge
eventually forming the localization peak, builds up gradually. The consistent evolution of the spectral weight at low energy is natural in light of
the exact form of the spectral function Eq.~(\ref{sqwexact}), where a higher density of $\delta$-functions in regions of low but ultimately nonzero
spectral weight is expected with increasing $L$. An excitation associated with a large length scale will also be fundamentally limited by the inverse
system length, leading to higher energy for smaller $L$ (as we have seen quantitatively in the case of the quantum rotor state). In contrast, the fast
convergence of the broad magnon peak has a high density of discrete contributions already for relatively small system sizes and a further
increase in the the number of these $\delta$-functions with $L$ does not affect the overall shape of the spectrum. Thus, the observed convergent
high-energy features and evolving low-energy tail are consistent with expectations and unlikely artifacts of the analytic continuation procedures.

Note that there is no apparent magnon peak at the ${\bf q}=(\pi,\pi)$ in Fig.~\ref{sw_p16}(d). This is expected in light of the uniform system having
true spin wave excitations only at the momenta away from $(0,0)$ and $(\pi,\pi)$, with the ground state being in the former sector and the latter hosting
the $S=1$ rotor state that becomes degenerate with the ground state when $L \to \infty$. The lowest spin wave is at ${\bf q}=(\pi,\pi)$. In the diluted
system there can also be no propagating magnon at $(\pi,\pi)$ and the large continuum (relative to its size at other momenta) that builds up with
increasing $L$ should reflect large matrix elements between the ground state and the localized modes and other incoherent excitations. Note again that
the left edge of the continuum, at $\omega \approx 0.1$ for $L=64$, is very close in energy for all momenta and system sizes in Fig.~\ref{sw_p16}, thus
suggesting its origin from dispersionless excitations.

\begin{figure*}[t]
\includegraphics[width=150mm]{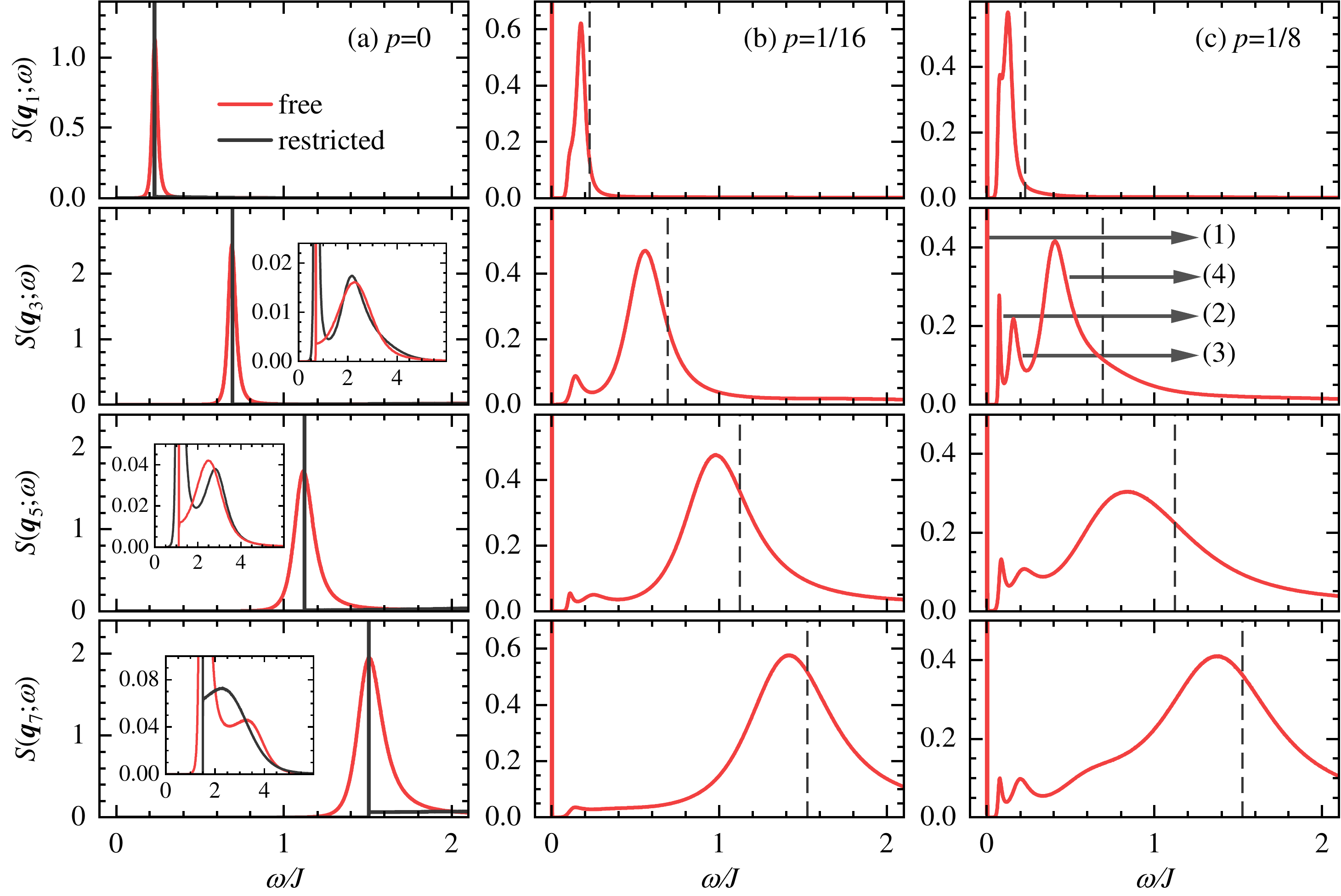}
\caption{$L=64$ results for $S(\boldsymbol{q},\omega)$ with ${\bf q}$ on a diagonal of the BZ,
$\boldsymbol{q}_x =2\pi(x,x)/L$, $x=1,3,5,7$. Results for a pristine system ($p=0$) are shown in (a), while
(b) and (c) are for $p=1/16$ and $p=1/8$, respectively. The red curves show results of unconstrained SAC sampling, while the black curves in
(a) include a macroscopic $\delta$ function (modeling a sharp quasi-particle peak) with optimized weight and continuum only above the peak. The resulting
multimagnon continua are shown in more detail in the insets. The location of the peak in (a) is indicated by the dashed lines in (b) and (c). All the
different peaks are marked in (c),  the second panel from top, as follows: (1) is the quantum rotor peak, (2) the localization peak, (3) the peak
likely resulting from very weakly dispersing, almost diffusive modes, and (4) the disorder-broadened magnon peak. Features (1) and (4) are present in
all panels in (b) and (c), while features (2) and (3) are visible to varying degree. Note that the rotor peak is very close
to $\omega=0$ (as is clearer from the $L=64$ results in Fig.~\ref{sw_p16}); a small range of negative values are included on the $\omega$ axis
to make this sharp peak better visible.}
\label{normalize_sw}
\end{figure*}

In Fig.~\ref{normalize_sw} we compare $p=0$, $1/16$, and $1/8$ results for $S(\boldsymbol{q},\omega)$ in our largest systems, $L=64$. As in
many plots in Ref.~\onlinecite{Chernyshev02}, we here focus on ${\bf q}$ points on the diagonal $(x,x)$ line of the BZ. In the case of $p=0$, we also compare
results of unrestricted SAC, discussed in Sec.~\ref{sub:sac}, and a constrained parametrization with a macroscopic $\delta$ peak imposed at the lower edge in
addition to the large number of ``micro'' $\delta$-functions representing the continuum above it (i.e., with the constraint that the continuum does not extend
below the $\delta$-edge). The relative weight $A_0$ of the macroscopic $\delta$ function is fixed in the SAC process while its location is also sampled (though
it fluctuates very little after the sampling has equilibrated). A scan over $A_0$ reveals its optimal amplitude as quantified by $\langle \chi^2\rangle$.
For technical details of SAC with the $\delta$-edge parametrization we refer to Refs.~\onlinecite{Shao17,Shao23}.

At $p=0$, the $\delta$-edge parametrization separates the expected sharp magnon peak and the multi-magnon continuum, which is not possible with the broadening
effects of the unrestricted SAC, though the peak locations are nevertheless almost the same. The broadening of the sharp peak with unrestricted sampling reflects
the $\omega$ dependent resolution (better for smaller $\omega$) imposed by the statistical errors of the QMC data. The reason for the better resolution at lower
energy is simply because the contributions to $G({\bf q},\tau)$ from small $\omega$ remain statistically significant up to larger $\tau$, i.e., there is more
information on low-energy features in the available QMC data.

The broadened peaks in Fig.~\ref{normalize_sw}(a) are also somewhat asymmetric, reflecting the continuum above the magnon peak that cannot always be well
resolved as a separate part of the spectrum unless the $\delta$-edge is imposed. A small secondary maximum is still also present and often close in
shape to what is obtained with the $\delta$-edge constraint, as seen in the insets of Fig.~\ref{normalize_sw}(a). The spectral weight above the sharp magnon
peak is most naturally interpreted as a multimagnon continuum \cite{Powalski18}. However, for ${\bf q}$ in the neighborhood of $(\pi,0)$ (which is not shown here),
the main magnon peak is greatly diminished and the continuum is more prominent, which is interpreted as a precursor to spinon deconfinement. We refer
to previous works \cite{Singh95,Sandvik01,Piazza15,Shao17} for detailed discussions of the excitations of the uniform system; we will still touch on
its dispersion relation further below when comparing with the diluted system. The deconfinement process will be discussed in Sec.~\ref{sec:jq},
where other interactions beyond the Heisenberg exchange is are considered.

In Figs.~\ref{normalize_sw}(b) and \ref{normalize_sw}(c) we observe the extremely sharp quantum rotor excitation at the lowest energies. In
Fig.~\ref{normalize_sw} its $\propto 1/N$ energy is practically zero on the scale used ($\approx 0.002$ according to Fig.~\ref{sw_p16}). As already
discussed, spectral weight below the large broad peak extends to very low energy, with a gap remaining to the rotor peak. We can identify the
localization peak using the interpretation of Ref.~\onlinecite{Chernyshev02}. We now discuss the features above this peak, starting from high energy.
The clearest illustration of the generic shape of the spectrum is provided by Fig.~\ref{schematic}, which is based in the data in the second panel
from the top in Fig~\ref{normalize_sw}(c).

The dominant peaks are much broader than the resolution limited magnons in Fig.~\ref{normalize_sw}(a) obtained with unrestricted SAC. Thus, at least
the high-energy features should not be affected by the resolution of the method, given also that the QMC data quality is similar for all $p$ values.
The ability of our method to accurately resolve both high-energy and low-energy features is also supported by the tests in Fig.~\ref{sw_ed1}, though
in that small system there is no structure between the rotor peak and the rest of the spectrum---there are also no spurious features in the SAC
spectra---a fact which also can be taken as support for the emergent features for the larger system being correct.

\begin{figure}[t]
\includegraphics[width=78mm]{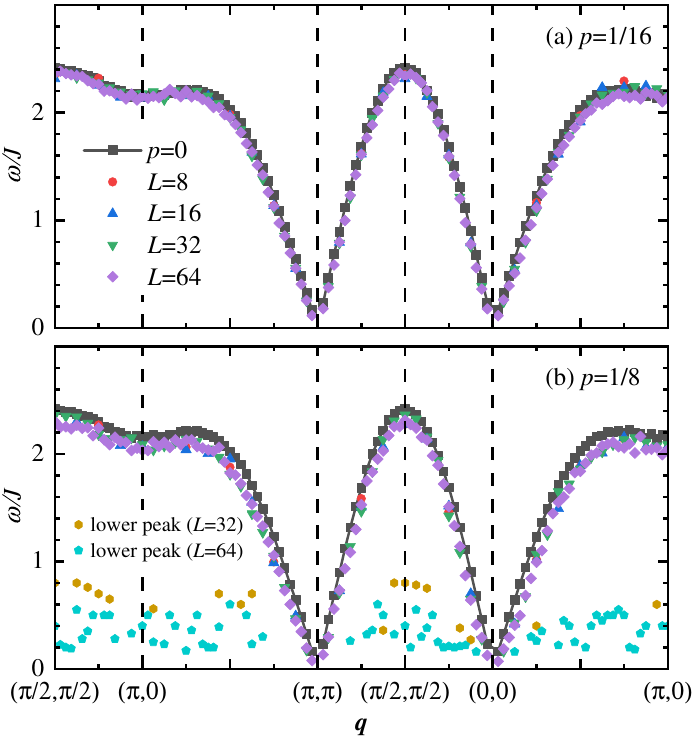}
\caption{Dispersion relations along a cut through the BZ. The magnon peak location for system sizes $L=8$, $16$, $32$, and $64$ is shown for $p=1/16$ in (a)
and $p=1/8$ in (b). Results for the uniform $L=64$ Heisenberg model are shown as a reference in both cases. In (b), the location of the peak below the magnon,
i.e., the almost dispersionless diffusive-like excitation marked as (2) in Fig.~\ref{normalize_sw}(c), is shown for $L=32$ (golden symbols) and $64$ (teal
symbols) at wavevectors where the feature is clearly discernible. At $p=1/16$, this peak is much less pronounced and it is not possible to reliably extract its
location.}
\label{dispersion}
\end{figure}

For all momenta in Figs.~\ref{normalize_sw}(b) and \ref{normalize_sw}(c), the dominant peak is shifted to lower frequency relative to the $p=0$ magnon,
with the relative shift larger for smaller $q$. The dispersion relations for $p=1/16$ and $p=1/8$ along a cut in the BZ are graphed in Fig.~\ref{dispersion}.
Here we define the excitation energy at $p>0$ based on the peak maximum, while at $p=0$ it is the location of the $\delta$ function edge shown in
Fig.~\ref{normalize_sw}(a). Similar $p=0$ results were obtained in Ref.~\onlinecite{Shao17}, but only for system sizes up to $L=48$. The size dependence 
at the ${\bf q}$ points available for given $L$ is rather weak except at $(\pi,\pi)$, where there is no magnon and data are not shown in Fig.~\ref{dispersion}.

Despite the substantial broadening of the peaks for $p>0$, the energy at and around the zone boundary is very close to that of the magnon at $p=0$.
It is particularly noteworthy that even the characteristic dip in energy at ${\bf q}=(\pi,0)$ of the clean system \cite{Singh95,Sandvik01} remains at vacancy
fraction as high as $p=1/8$. This dip can be related to fluctuations of local singlets and was argued to be a precursor to the deconfined quantum phase
transition into a non-magnetic valence-bond solid state in an extended parameter space \cite{Shao17}. The dip can be enhanced by adding an interaction that
brings the system closer to the phase transition, as we will also find here for the diluted systems in Sec.~\ref{sec:jq}. The presence of the ``spinon precursor''
also for the diluted system is another indication that the high-energy excitations are similar to those in the uniform system, though with significant damping and
more multimagnon contributions to $S({\bf q},\omega)$.

Naively, the robust high-energy excitation reflects uniformity of the system on length scales less than the typical separation
between vacancies. The broadening reflects the substantial damping effects of the disordered environment surrounding uniform patches.
When the energy drops well below the $p=0$ magnon, the peak also narrows significantly, as seen in in Figs.~\ref{normalize_sw}(b) and \ref{normalize_sw}(c),
while a long thin high-energy tail still remains. A broadening of order $pk$ (with $k$ the momentum relative to either of the two low-energy points) was
predicted \cite{Chernyshev02}, and in Figs.~\ref{normalize_sw}(b) and \ref{normalize_sw}(c) we indeed also observe systematically reduced broadening as $q$
approaches $0$, though the proportionality to $p$ is not apparent in all cases.

\begin{figure}[t]
\includegraphics[width=82mm]{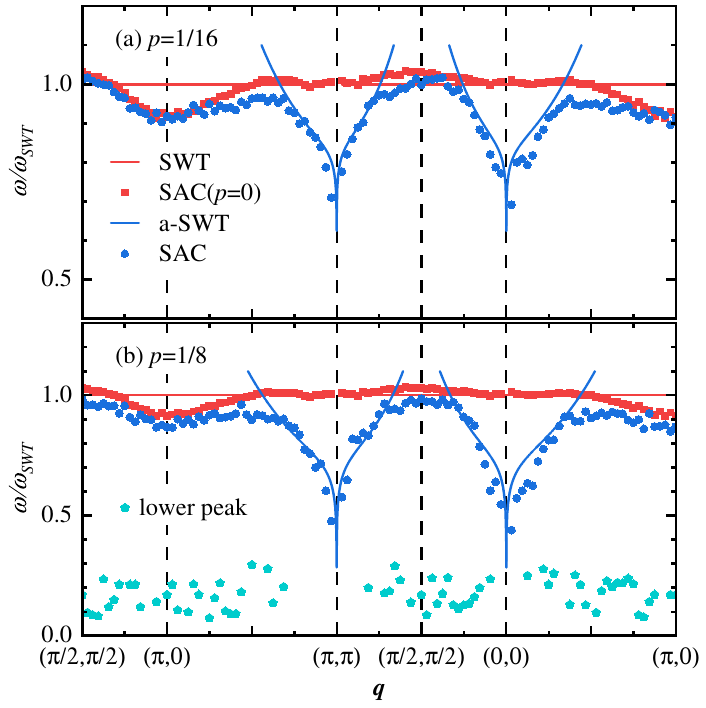}
\caption{The $L=64$ peak energies from Fig.~\ref{dispersion} shown relative to the dispersion relation $\omega_{0}({\bf q})$ in linear spin wave theory
(including the estimated spin-wave velocity renormalization factor $1.16$ \cite{Singh89,Shao17}). The magnon energies at $p=0$ and $p > 0$ are shown
with red squares and blue circles, respectively, and the energy of the diffusive-like random network mode is shown with the teal hexagons. The blue
curves show the predicted anomalous magnon dispersion relation Eq.~(\ref{wklog2}), where in (a) and (b) we have set the spin wave velocity to $c=1.58$ and
$c=1.54$, respectively, i.e., the renormalization factor $c/c_0=c/\sqrt{2} \approx 1.12$ in (a) and $1.09$ in (b), smaller than $c/c_0 \approx 1.16$ at $p=0$.}
\label{relativedisp}
\end{figure}

The significant drop in magnon energy close to ${\bf q}=(0,0)$ and $(\pi,\pi)$ is further highlighted by Fig.~\ref{relativedisp}, where the dispersion
relations are shown relative to $\omega_0({\bf q})$ of the clean system in linear spin wave theory. Here we also show the predicted anomalous dispersion relation
\begin{equation}
\omega_{\rm ano}(k) = c k[1-p(\pi/2-1)+\ln(ck/4)/\pi],
\label{wklog2}  
\end{equation}  
which is Eq.~(51) from Ref.~\onlinecite{Chernyshev02} with the limit $k \to 0$ taken but the full $p$ dependence kept---in Eq.~(\ref{wklog}) the limit
$p \to 0$ was also taken. However, the expression is still only expected to be valid for small $p$ and then the further simplified form in Eq.~(\ref{wklog})
is also very similar.

Our results for $p=1/16$ in Fig.~\ref{relativedisp}(a) indeed match the predicted form slightly better than those for $p=1/8$ in Fig.~\ref{relativedisp}(b).
To optimize the fits, we adapted the spin wave velocity for both $p=1/16$ and $p=1/8$. The optimal values, listed in the caption of Fig.~\ref{relativedisp},
are marginally larger than the linear spin-wave value $c_0=\sqrt{2}J$ but smaller than the estimate $c \approx 1.63$ for $p=0$ (renormalization factor $1.16$)
\cite{Singh89,Shao17}. The smaller velocities for $p>0$ are natural in light of the smaller mean couplings, though it is difficult to judge exactly what
the optimal $c$ values are, given that the agreement with the theory is never perfect. The agreement is in general slightly better near $(\pi,\pi)$
than near $(0,0)$.

The predicted anomalous form of the dispersion should be valid only for ${\bf q}$ in the neighborhood of $(0,0)$ and $(\pi,\pi)$ and must also eventually
break down very close to these points, reflecting the incoherent dynamics at the lowest energies, where the magnon should merge with the localized
excitations. We have also tried to fit our results to the full form Eq.~(51) of Ref.~\onlinecite{Chernyshev02}, which avoids that continued increase above
the baseline spin wave dispersion seen in Fig.~\ref{relativedisp}. However, the agreement with our data around ${\bf q}=(0,0)$ and $(\pi,\pi)$ is not
further improved and the zone boundary energy cannot be matched well (it is too low) when $c$ is adapted to the anomalous dispersion;
we do not show graphs here.

The reasonable good match between the anomalous magnon dispersion and our numerics over a substantial part of the BZ provide strong support for the
validity of both approaches. However, the peak present in between the localization and magnon peaks, most clearly visible in \ref{normalize_sw}(c), was not
predicted by the $T$-matrix theory. As already mentioned in Sec.~\ref{sec:summary}, we believe that this peak reflects diffusive-like excitations
arising from randomly located effective moments adjacent to vacancies. The peak and the associated weight below it seem to connect smoothly to the
localization peak, suggesting that they both could be features of the random spin network. Diffusive behavior was also mentioned as an interpretation
of the continuum below the magnon energy produced in the $T$-matrix calculations \cite{Chernyshev02}. However, in that approximation the continuum is
featureless before merging with the magnon peak, in contrast to the intermediate peak found here. It is plausible that the peak reflects very weakly
dispersing excitations of the random network, with the modes becoming more diffusive as the energy is lowered and finally localize fully
at $\omega_{\rm loc}$. 

In Figs.~\ref{dispersion}(b) and Figs.~\ref{relativedisp}(b) we have also included the location of the peak between the localization peak and magnon at $p=1/8$,
where the peak is larger and appears more consistently in our results than at $p=1/16$ (as is apparent in Fig.~\ref{normalize_sw}). Because of the very significant
scatter, it is not possible to make any strong statements about the dispersion, however, other than it being weak or absent. There is also some size dependence,
as is apparent for $L=32$ and $64$ in Fig.~\ref{dispersion}(b). As seen more clearly in Fig.~\ref{relativedisp}(b), the results are less noisy close to ${\bf q}=(0,0)$
and appear to show a trend of increasing energy for small $q$ relative to the linearly dispersing spin waves. We lack reliable data close to ${\bf q}=(\pi,\pi)$, but
the existing points are at least also consistent with a maximum there as well. Given that these are results relative to the linear spin wave dispersion, the upturn may
possibly indicate that the weak dispersion of these low-energy excitations is effectively sublinear before they, along with the main magnon peak, merge with the localized
excitations at $\omega_{\rm loc}$. A small subset of essentially randomly located spins adjacent to vacancies that participate in the excitations in the energy window
from the localization peak to the magnon can be clearly observed in the real-space dynamic structure factor $S({\bf r},\omega)$, as will be shown in Sec.~\ref{sec:tomo}.
This finding motivates the proposal of diffusive-like modes of a random spin network, leading to a weakly dispersive peak in addition to the dispersionless localization
peak. To construct a meaningful effective random network model of the subsystem would require knowledge of the interactions mediated by the bulk medium.

It should be noted that, the number of disorder realizations used here (1000) is not sufficient to make the vacancy distribution at $p=1/16$ very uniform
on average, and the density fluctuations also translate to momentum space. While the localization peak and the diffusive-like peak are consistently observed
at $p=1/8$ in Fig.~\ref{normalize_sw}(c), as indicated by (2) and (3) in the second panel from top, at $p=1/16$ there is often only a flat portion with
a small maximum at its lower edge, as exemplified by the results in Fig.~\ref{normalize_sw}(b). This apparent problem in resolving the full structure of
the random-network mode is not necessarily just a reflection of limited frequency resolution of the SAC method, but may very well be the true behavior of
$S({\bf q},\omega)$ averaged over the finite set of vacancy realizations used. For $p=1/8$ the mean vacancy distribution is naturally more uniform, and
we can extract the energy of the random network mode more reliably, though still with significant statistical fluctuations. The features also tend to become
clearer with increasing system size, as we will observe in detail further below in the local spectral function, and $L=64$ may not be quite large enough
to produce clearly emergent momentum resolved modes at $p=1/16$.

To contrast these results with those the $T$-matrix theory \cite{Chernyshev02}, we note that disorder averaging is there carried out at the level of the Green's
function before it is used to generate the spectral function. No structure in $S({\bf q},\omega)$ was found in this calculation between the localization and
magnon peaks, in contrast to the intermediate peak observed here. The localized excitations and continuum above it (including the second peak) are associated with
a very small number of spins at vacancies (Sec.~\ref{sec:tomo}). Such specific sparse local degrees of freedom may not be reflected when using the disorder averaged
Green's function, and it is then possible that the localized mode obtained with the $T$-matrix approach is not exactly of the same nature as in the fully
interacting spin model considered here. The extrapolated localization energy at $p=1/8$, which we obtain below from the local dynamic structure factor, is in
fact significantly higher than the predicted value.

Another discrepancy between our results and the theory is that a double-peaked $S({\bf q},\omega)$ forms in the latter  when the magnon resonates with
local impurity states at energy of order $J$ \cite{Brenig91,Chernyshev02}. The interactions are not fully taken into account by the $T$-matrix approach,
and the zone boundary magnons therefore appear essentially as a band of excitations different from those with the anomalous dispersion at lower energy.
A double peak is also consistently seen in the numerical spin wave calculations in Ref.~\onlinecite{Mucciolo04}, even beyond the ${\bf q}$ range where it appears
in the $T$-matrix calculations. We always find just a single peak, which suggests that, indeed, the neglected interactions cause an artificial feature in
the spin wave calculation of the spectral function. Regardless of the interpretation, we now also have an unbiased quantitative characterization of
the high-energy part of $S({\bf q},\omega)$, which was stated as an open issue previously \cite{Chernyshev02}.

\subsection{Local dynamic structure factor}
\label{sec:local}

Here we examine the local dynamic structure factor $S_0(\omega)$, where the operator in Eq.~(\ref{sqwexact}) is $O=S^z_{\bf r}$ and we take the average over
all lattice points ${\bf r}$. This local quantity also is equivalent to $S({\bf q},\omega)$ averaged over all ${\bf q}$, a fact which represents another opportunity
to test the SAC approach: We can either run the SAC for all momenta separately and then average the resulting spectra over ${\bf q}$, or we can average
$G({\bf q},\tau)$ over ${\bf q}$ first, resulting in $G_0(\tau)$ (where now the subscript $0$ refers to the distance-$0$ case, i.e., the on-site correlation
function), which can be computed directly in real space (averaged over locations) and analytically continued for $S_0(\omega)$. The two different ways of
computing $S_0(\omega)$ should in principle be equivalent, but in practice, because of the noisy imaginary-time data and imperfections of the analytic
continuation, the results will differ to some extent, thus providing some measure of the reliability of the SAC approach. This kind of test was previously
applied with the MEM to study the local dynamic response of the Heisenberg chain ~\cite{Starykh97}, and, more recently, also with the SAC method in the case
of the random Heisenberg chain \cite{Shu18}.

\begin{figure}[t]
\includegraphics[width=84mm]{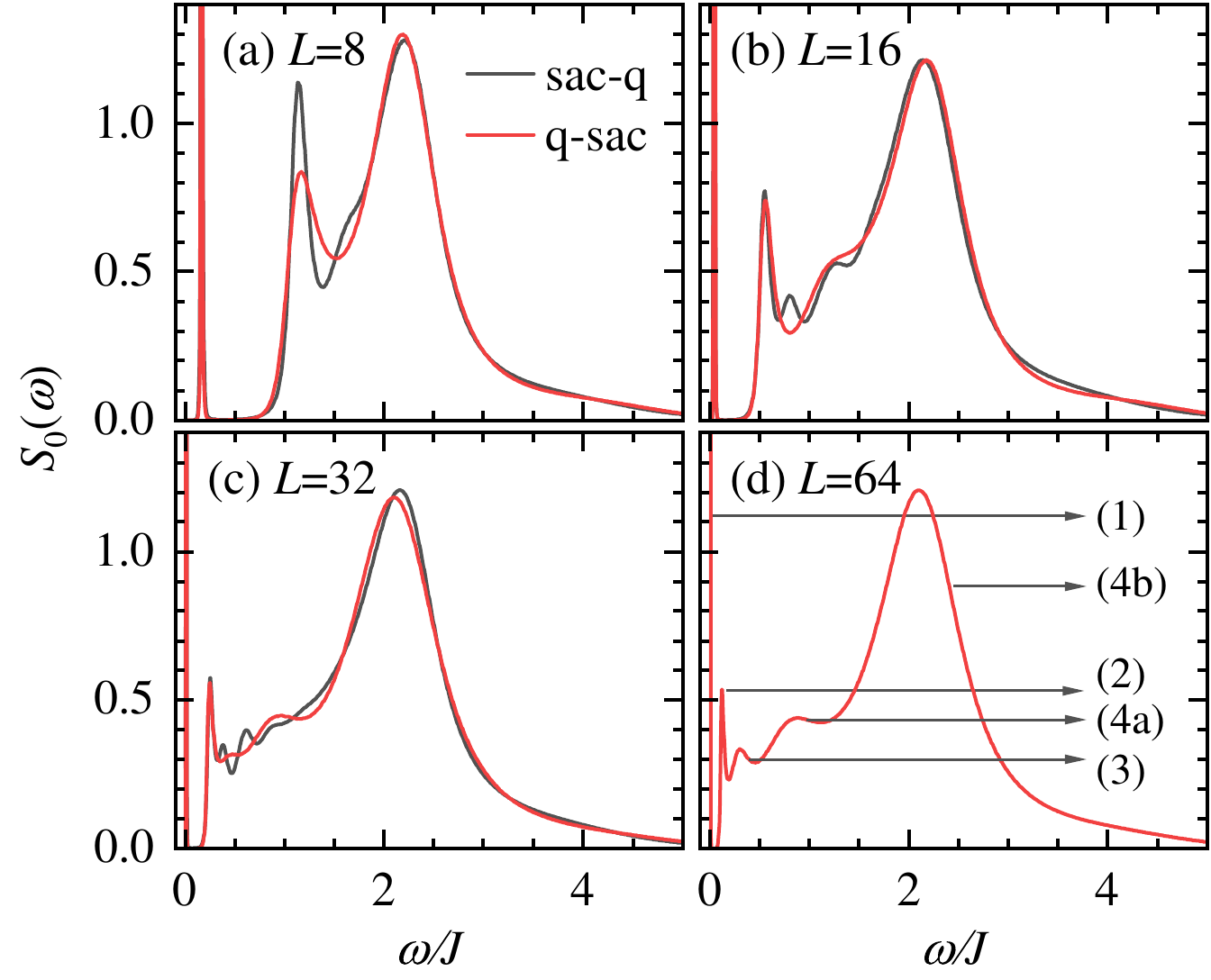}
\caption{Local spectral function averaged over sites and disorder realizations at $p=1/16$ for system sizes (a) $L=8$, (b) $L=16$, (c) $L=32$, and (d) $L=64$.
The red curves correspond to momentum summation before analytic continuation, i.e., applying SAC to the local correlation function $G_0(\tau)$ averaged over
all spins, while the black curves were obtained by applying SAC to all $G({\bf q},\tau)$ and averaging after. In (d), the different peaks are marked for
correspondence with Fig.~\ref{schematic} as follows: (1) is the quantum rotor peak, (2) the localization peak, (3) the peak arising from the momentum
averaged weakly dispersing, diffusive-like modes, (4a) and (4b) from the magnons, with (4a) likely a consequence of relatively large spectral weight for
the magnons close to $(\pi,\pi)$ and (4b) from the high-energy magnons.}
\label{sw0_16}
\end{figure}

Figure \ref{sw0_16} shows $p=1/16$ results for system sizes $L=8$, $16$, $32$, and $64$. For the largest system we did not compute $G({\bf q},\tau)$ for all
${\bf q}$ (only on high-symmetry lines for which results were presented in Sec.~\ref{sec:qdep}). For the other systems, the results of the two calculations agree
very well on the location and width of the quantum rotor peak, the low-energy localization edge and peak above it, and also in the entire frequency range of the
main peak. For $L=16$ and $32$, the results from averaging $S({\bf q},\omega)$ exhibit wiggles that are likely finite-size effects resulting from the peaks in the
individual ${\bf q}$ spectra, which are not sufficiently dense on the ${\bf q}$ grid for small $L$ (as was also found for the random Heisenberg chain in
Ref.~\onlinecite{Shu18}). It is not realistic to expect SAC applied to $G_0(\tau)$ to resolve these fine structures. Since such oscillations are just a
consequence of the finite momentum grid and not expected in the thermodynamic limit, not resolving them may in a sense correspond to faster convergence with $L$.

From now on we discuss only the method of applying SAC to $G_0(\tau)$; the red curves in Fig.~\ref{sw0_16}. It is interesting to see that more structure
gradually develops above the quantum rotor peak as $L$ increases. For $L=8$, there are just two broadened peaks above the already well isolated and narrow
rotor peak, while for $L=16$ there is a hint of an intermediate peak forming, which is more completely developed for $L=32$ and further grows in size
for $L=64$. For $L=64$ there is another small peak above the edge.

\begin{figure}[t]
\includegraphics[width=84mm]{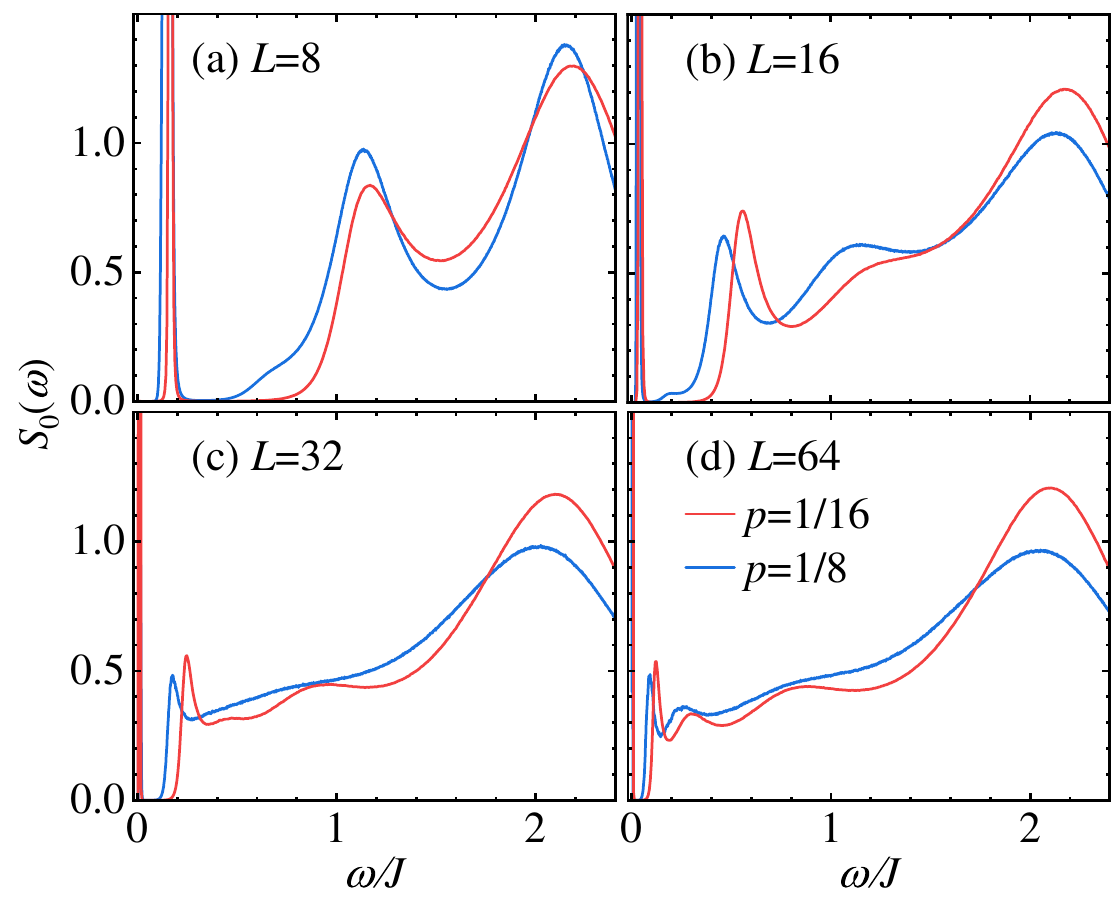}
\caption{Local spectral function obtained from the site averaged $G_0(\omega)$ at $p=1/8$ (blue) and $p=1/16$ (red) for system sizes
(a) $L=8$, (b) $L=16$, (c) $L=32$, and (d) $L=64$.}
\label{sw0_8}
\end{figure}

The four peaks that are clearly formed for $L=64$ above the obvious rotor peak are marked with arrows and labels in Fig.~\ref{sw0_16} in a way corresponding
to $S({\bf q},\omega)$ in Fig.~\ref{schematic} for an individual momentum. These features can all be understood based on the picture of the excitations developed in
Sec.~\ref{sec:qdep}. The peak at the low-energy localization edge clearly becomes sharper and diminishes in relative weight as $L$ increases---a rough
extrapolation shows that the peak height remains finite as $L \to \infty$. The energy scale of the peak marked (3) is consistent with the ${\bf q}$ averaged
weakly dispersing diffusive-like peak, as is apparent by comparing with Fig.~\ref{dispersion}(b), despite the large scatter in the location of the
lower peak and the results being for $p=1/8$ in that figure---we will see below that the location of the peak is not much different for $p=1/8$ for
these system sizes. The remaining features, marked (4a) and (4b) in Fig.~\ref{sw0_16}, must then be due to the magnons, with the small bump on the
left side of the main broad peak roughly consistent in its energy scale with contributions from ${\bf q}$ close to $(\pi,\pi)$, where the spectral
weight of the anomalously dispersing magnon peak (Fig.~\ref{relativedisp}) is large. The dominant peak marked (4b) contains mainly the contributions
from the high-energy magnons.

Figure \ref{sw0_8} shows similar results at vacancy fraction $p=1/8$, only for the case of analytical continuation of the local function $G_0(\tau)$.
We also include the $p=1/16$ results from Fig.~\ref{sw0_16} in order to compare the two cases. At $p=1/8$ the spectral features between the large magnon
peak and the localization peak become less clear with increasing $L$, which can be understood as due to broadening of peaks with increasing $p$. Nevertheless,
for $L=64$ we observe all the same features qualitatively at both $p$ values. In particular, the first sharp peak above the rotor peak is consistent with the
localization peak or edge seen at very similar energy in Figs.~\ref{normalize_sw}(b) and \ref{normalize_sw}(c), and the maximum above the localization peak
is consistent with the overall energy scale of the random network mode according to the results in Fig.~\ref{dispersion}(b).

\begin{figure}[t]
\includegraphics[width=75mm]{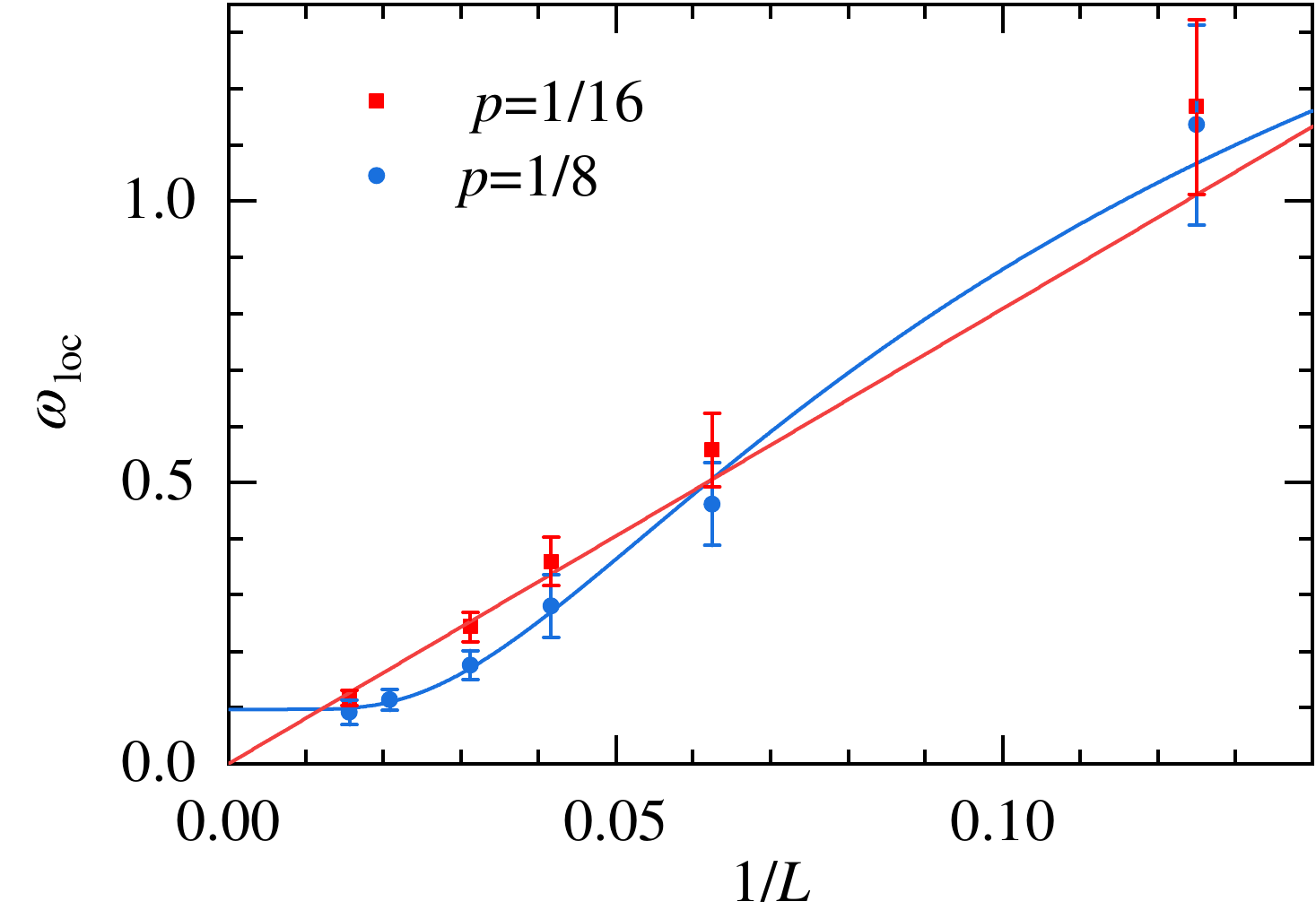}
\caption{Inverse system size dependence of localization energies determined using $S_0(\omega)$ data such as those in Fig.~\ref{sw0_8} at $p=1/16$
(red squares) and $p=1/8$ (blue circles). The error bars correspond to half of the width at the half maximum on the left side of the $S_0(\omega)$ peaks.
The fitted curve for $p=1/8$ is of the form $\omega_{\rm loc} = a+b{\rm e}^{-cL}$ with free parameters $a,b,c$ and including all system sizes. For $p=1/16$
a linear fit with $\omega_{\rm loc} \sim L^{-1}$ is shown.}
\label{locapeakw0}
\end{figure}

The localization peak in $S_0(\omega)$ is much sharper than the corresponding momentum resolved peaks, e.g., in Figs.~\ref{normalize_sw}(b) and
\ref{normalize_sw}(c). We have extracted the peak locations for the system sizes used above, and also for $L=24$ and $48$. It is difficult to establish an
absolute statistical error on the peak location, and instead we use the half width at the half maximum on the left side of the $S_0(\omega)$ peaks as an
estimate of the uncertainty of the finite-size localization energy. The results for both dilution fractions are graphed versus $1/L$ in Fig.~\ref{locapeakw0}.

We should expect the localization energy to increase with $p$ \cite{Chernyshev02}, but for each individual size the $p=1/16$ energy is actually above that
for $p=1/8$, as also seen in Fig.~\ref{sw0_8}.
However, the results for $p=1/8$ indicate a flattening-out for the larger sizes, suggesting an eventual swapping of the order of the peaks for
larger $L$. Given that the localized excitations should have a typical size, an exponential convergence of the energy $\omega_{\rm loc}(L)$ versus the system
size should be expected asymptotically for sufficiently large $L$. For $p=1/8$, the observed flattening of the data for the largest system sizes in
Fig.~\ref{locapeakw0} clearly demonstrates a non-zero energy in the thermodynamic limit, and an exponential form fits the data well from the smallest
to the largest system size ($8 \le L \le 64$). This fit extrapolates to $\omega_{\rm loc}= 0.10 \pm 0.01$, which is about 50 times larger than the predicted
\cite{Chernyshev02} order of magnitude of the localization energy, $\omega_{\rm loc} \propto {\rm e}^{-\pi/4p} \approx 0.0019$. However, the prefactor of the
exponential was not determined, and the form may strictly apply only at much smaller dilution fractions.

For $p=1/16$, the results in Fig.~\ref{locapeakw0} do not show any tendency to an exponential flattening out, suggesting that the system sizes $L \le 64$
are still much smaller than the typical localization length. Given the prediction that $\omega_{\rm loc} \propto l^{-1}$, where $l$ is the localization length
\cite{Chernyshev02}, when the latter is limited by the system size $L$ we would expect the localization peak to scale as $\omega_{\rm loc}(L) \propto L^{-1}$.
Indeed, this relationship is apparent in Fig.~\ref{locapeakw0}, where we include a linear fit in $L^{-1}$.
The linear form should eventually cross over to a small constant $\omega_{\rm loc}$ as $L \to \infty$, as it
does for $=1/8$. For $p=1/16$ the asymptotic value simply too small to determine based on the available system sizes, which would be a natural conclusion also
under the assumption that $\omega_{\rm loc} = A{\rm e}^{-\pi/4p}$ with $A \approx 50$ from the result for $p=1/8$. This form with constant $A$ would imply
$\omega_{\rm loc} \approx 0.00017$ for $p=1/16$, far below the resolution of the data in Fig.~\ref{locapeakw0}.

It should be noted that the sharp localization peak in $S_0(\omega)$ does not necessarily imply that the system is gapped. In fact, the low-energy tail
of the peak should be expected to extend down to $\omega=0$ in the thermodynamic limit, reflecting the existence of arbitrarily large local excitations
despite there being a typical size (predicted to be of the order $l \propto {\rm e}^{\pi/4p}$ \cite{Chernyshev02}). The probability of excitations larger
than the typical scale will be exponentially small, and the spectral tail should then be exponentially thin for $\omega \to 0$. Many
thermodynamic quantities at $T < \omega_{\rm loc}$ will in practice appear to exhibit gapped behaviors.

\subsection{Single vacancy}
\label{sec:single}

\begin{figure}[t]
\includegraphics[width=75mm]{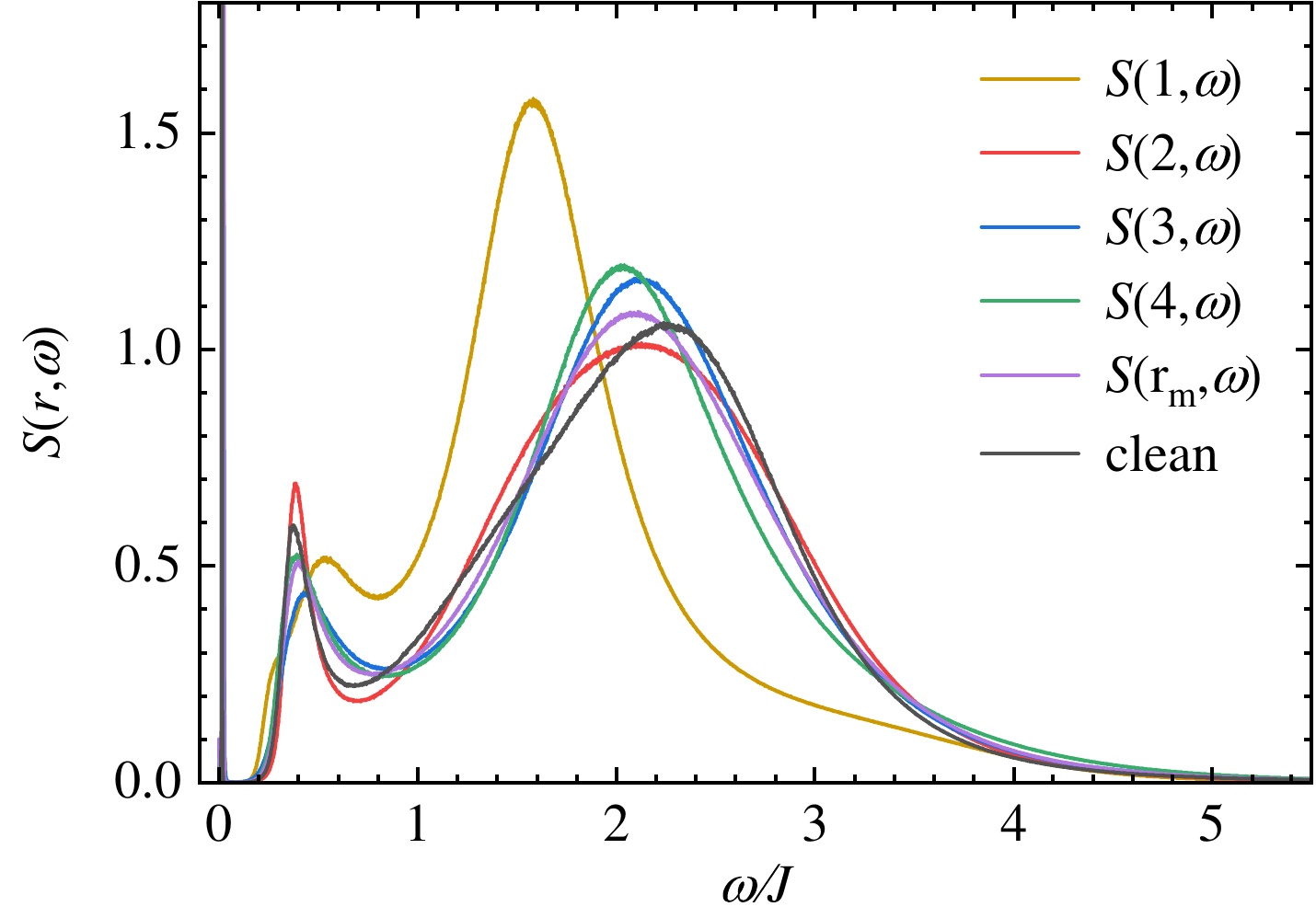}
\caption{Local spectral function at several short distances away from a single vacancy in a system of size $L=32$. The numbers $n=1,2,3,4$  in the legends
indicate the $n$th closest neighbor of the vacancy, with $r_{\rm m}$ the farthest separation on the $L=32$ lattice. The result for the clean system
is also shown for comparison.}
\label{sw0_0}
\end{figure}

It is also useful to examine the local spectral function in the case of a single vacancy, which we show in Fig.~\ref{sw0_0} for several locations close to
the vacancy along with results for the clean system. The significant quantum rotor peak is seen clearly in all cases. All the spectra look rather similar,
expect at the sites closest to the vacancy, $r=1$, where we observe a large shift of the dominant peak to lower $\omega$ and also a less sharp peak
at lower energy. In all other cases, we find a sharp peak at $\omega \approx 0.5$---it is the tallest at $r=2$, significantly suppressed at $r=3$,
and gradually approaches the height in the clean system.

In linear spin wave theory, the density of states for the clean system exhibits a van Hove like singularity at the band edge, due to the degeneracy of all
states on the magnetic zone boundary. This degeneracy is broken for the true magnon, as seen here in Fig.~\ref{relativedisp}. Moreover, there is a substantial
continuum above the main magnon peak. Thus, the singularity in the density of states  should be replaced by a broad peak, resulting in the high-energy profile
seen in Fig.~\ref{sw0_0} for both the clean system and the single vacancy. The sharp peaks at $\omega \approx 0.5$ are caused by the large matrix elements at
momenta close to $(\pi,\pi)$, which are not included in the density of states calculated for both the clean system and the single vacancy, e.g., in
Ref.~\onlinecite{Bulut89}. The impurity feature found there in the density of states at distance $r=1$ can still be roughly identified with the shifted
main peak in Fig.~\ref{sw0_0}.

Overall, comparing $S_0(\omega)$ for a single vacancy with the profiles at $p=1/16$ and $1/8$ in Fig.~\ref{sw0_8}, the low energy features bear little
resemblance. The high-energy features are similar, but with more weight in the tail above $\omega \approx 4$ at $p>0$. The static spin structure around
a vacancy was previously studied with QMC simulations \cite{Anfuso06}. It corresponds to the  frequency integrated local structure factor
in Fig.~\ref{sw0_0} and does not provide detailed information about the excitations.

\subsection{Real-space tomographic imaging}
\label{sec:tomo}

To gain insights into the spatial distribution of the different types of excitations at $p>0$, we here study the real-space resolved spectral function
$S({\bf r},\omega)$ for individual vacancy realizations. In a tomographic procedure, we compute integrals over energy windows
\begin{equation}
D_k({\bf r}) = \frac{N(1-p)\int_{a_k}^{b_k}S({\bf r},\omega)d\omega}{\sum_{\bf r}\int_{a_k}^{b_k}S({\bf r},\omega)d\omega}.
\label{stomo}
\end{equation}
This function represent the relative participation of each individual spin in all the excitations with energy within the chosen windows $[a_k,b_k]$.
We have normalized the weight distribution in Eq.~(\ref{stomo}) so that $D_k({\bf r})$ on average equals $1$ on the sites with spins.

From a real-space map we can also construct a participation ratio, defined as
\begin{equation}
R_k = \frac{1}{N(1-p)}\frac{\left ( \sum_{{\bf r}} D_k({\bf r}) \right )^2}{\sum_{{\bf r}} D^2_k({\bf r})},
\label{rkdef}
\end{equation}  
where we have also normalized by the number of spins. Then $R_k=1$ if the spectral weight is evenly distributed on all spins, and, in the opposite
extreme, $R_k={1}/{N(1-p)}$ if all the weight is concentrated in on a single spin. A similar inverse participation ratio $1/R$ was computed versus $\omega$
in the real-space spin wave diagonalization study in Ref.~\onlinecite{Mucciolo04} and also in the 2D Heisenberg model at the percolation point \cite{Wang10},
in the latter case not using an energy window but the local susceptibility defined in Eq.~(\ref{locsus}).

\begin{figure}[t]
\includegraphics[width=80mm]{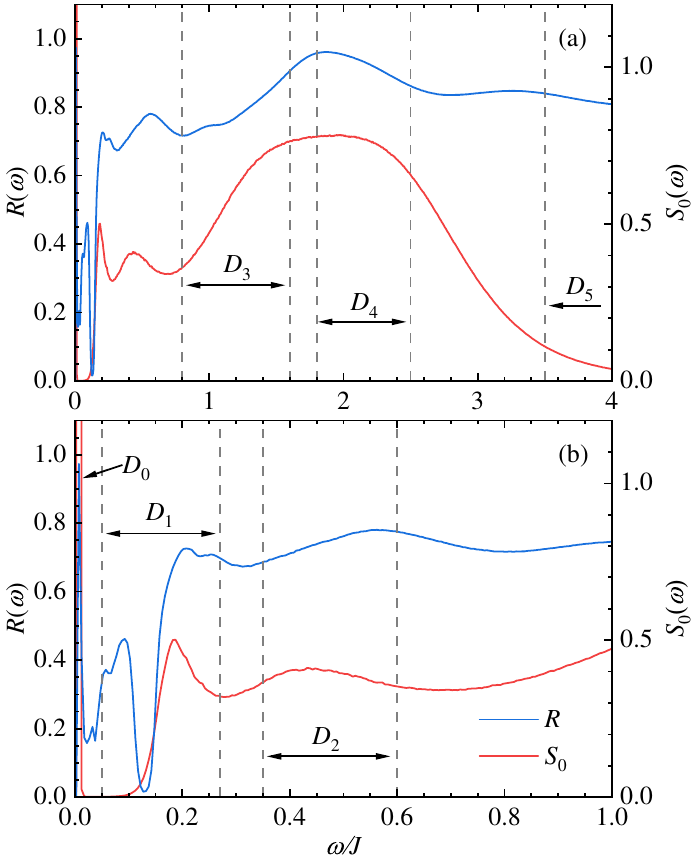}
\caption{Local spectral function (red curves) and participation ratio (blue curves) for an individual $L=32$, $p=1/8$ vacancy realization.
The results are shown on different energy scales in (a) and (b). The frequency windows used to compute the integrals $D_k$ (shown in Fig.~\ref{tomo1234},
where the vacancy locations can also be seen) are also indicated.}
\label{tomography}
\end{figure}

Here we will use a large number of thin energy window for a function $R(\omega)$ of an essentially continuous $\omega$, while maps $D_k({\bf r})$ will be
shown for a few larger windows with $\omega \in [a_k,b_k]$, which for given ``slice'' $k$ is selected so that it is dominated by one of the excitation
types discussed previously. The selection of appropriate windows is aided by the ${\bf r}$ average $S_0(\omega)$ of $S({\bf r},\omega)$, which, for a
sufficiently large system, we expect to contain similar structure as the disorder averaged local dynamic structure factor graphed in Fig.~\ref{sw0_8}. The
participation ratio also will be useful in this regard, as we will see. We here consider a single $L=32$ vacancy realization at $p=1/8$ and present similar
results for $p=1/16$ in Appendix \ref{sec:l32maps}. 

\subsubsection{Participation ratio}

Figure \ref{tomography} shows both the average dynamic structure factor $S_0(\omega)$ and the participation ratio $R(\omega)$, with \ref{tomography}(a)
showing almost the full energy range $\omega \in [0,4]$ and \ref{tomography}(b) focusing on details in the low-energy part. The spectrum for this single
sample has a peak structure similar to the disorder average in Fig.~\ref{sw0_8}, though with differences in details as expected.

The participation ratio shows an interesting structure, being close to $1$ in the very narrow energy range corresponding to the rotor peak, reflecting the
fact that most of the spins participate in this excitation. On the left side of the localization peak in $S_0(\omega)$, there is a very low minimum in
$R(\omega)$, showing that there is a narrow energy range were only a few spins are excited. The peak structure in $R(\omega)$ between the rotor peak and
the global minimum may not have much significance because the spectral weight there is very small. Moving further into the localization peak, $R(\omega)$ grows
sharply, then decreases somewhat before increasing slowly in the $\omega$ region where the spectral weight of the diffusive-like mode is concentrated.
At higher energy still, a maximum forms within the energy range corresponding to the magnons, after which the behavior is rather flat all the way to the far
end of the tail of the spectrum.

Comparing the participation ratio $R$ with $1/R$ computed in the numerical spin wave calculation, the closest case to our $p=1/8$ is $p=p_c/2 \approx 0.2$
in Fig.~4(b) of Ref.~\onlinecite{Mucciolo04}, where $L=36$. Here also there is a tendency to a sharp drop in $R$ at a low energy scale compatible with the
localization scale that we find here. However, $S({\bf q},\omega)$ at ${\bf q}=(0.4\pi,0)$, graphed for an $L=32$ system in Fig.~8 of Ref.~\onlinecite{Mucciolo04},
does not show any signs of a localization peak at any $p$. There is also no peak corresponding to the diffusive-like mode.

\subsubsection{Tomographic maps}

\begin{figure*}[t]
\includegraphics[width=165mm]{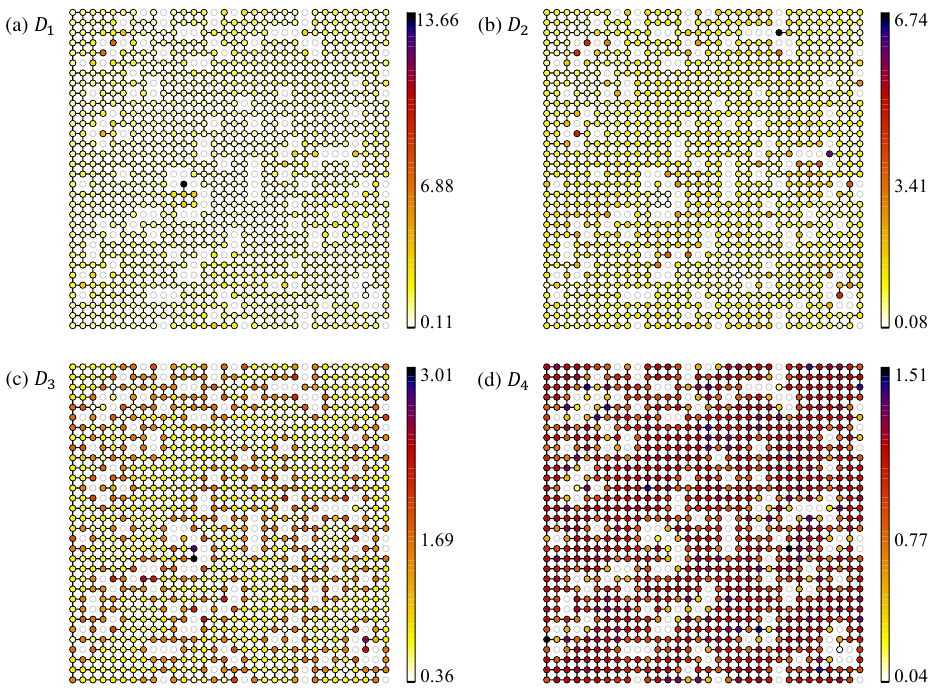}
\caption{Tomographic maps $D_1$-$D_4$ defined in Eq.~(\ref{stomo}) for an $L=32$ system at $p=1/8$. The frequency windows are shown in Fig.~\ref{tomography}
and the color maps correspond to the highest to lowest spectral weight for each case, as indicated. The dim circles represent vacancies. The mean spectral
weight on the spins in all maps is unity.} 
\label{tomo1234}
\end{figure*}

Based on the results for both $S_0(\omega)$ and $R(\omega)$, we have selected energy windows $k=0,\ldots,5$, for constructing color coded real space maps
$D_k({\bf r})$. These windows are also indicated in Fig.~\ref{tomography}, and the resulting maps $D_1,\ldots, D_4$ are shown in Fig.~\ref{tomo1234}. We will
discuss the map $D_0$ of the rotor state in Sec.~\ref{sec:rotor}, and the high-energy window $D_5$, which we believe contains mainly multimagnon excitations,
is discussed in Appendix \ref{sec:l32maps}.

It should be kept in mind that the tomographic maps include all the excitations within the chosen window, and so informs us about the spins involved
in forming an entire band of excitations, not individual excitations. The maps are not very sensitive to exactly how the energy windows are selected,
as long as they each correspond to a region dominated by one of the excitations. Clearly these regions cannot be completely separated from each other,
except that of the rotor map $D_0$, because all excitation modes are subject to broadening and we do not resolve the momentum structure here.
Nevertheless, we will see that reasonably clear pictures of the different excitations emerge from the tomographic approach.

The map $D_1$ should correspond to the localized mode, while $D_2$ should roughly capture the low-energy diffusive-like mode. We will argue here that
both of these features arise from a random spin network. The maps $D_3$ and $D_4$ both cover the magnons, with the subdivision chosen so as to separate
the contributions from the part of the BZ where the energy is close to $\omega({\bf q})$ of the clean system (Fig.~\ref{relativedisp}), in $D_4$, and when
the dispersion is anomalous, in $D_3$. It should be noted that the low-energy magnons will also to some extent reach down to the random network excitations,
since they are both damped modes and should merge together somewhere close to the localization energy.

The weight $D_1({\bf r})$ of the localized excitations, shown in Fig.~\ref{tomo1234}(a), is concentrated on a rather small number of sites
adjacent to the vacancies (i.e., those with less than four nearest neighbors), which explains the small participation ratio in most of this window.
Almost all of the sites with large weight belong to monomer rich regions---the results of the dimer-monomer simulation of this system are presented
in Appendix \ref{sec:l32maps}. Thus, it appears that the localized states are formed primarily by spins in regions of local sublattice imbalance.

Moving to $D_2$ in Fig.~\ref{tomo1234}(b), a large fraction of the spectral weight here is distributed on a subset of the spins next to vacancies---but
many more spins than in $D_1$ and most of them not belonging to monomer rich regions. Note also that some spins that have very little spectral weight in
$D_2$ have much larger weight in the localization map $D_1$, and vice versa. Another interesting aspect of the spins with large spectral weight in $D_1$
is that most of them have only one neighbor, as seen in Fig.~\ref{tomo1234}(a). We have also computed separately the relative spectral weight on the
spins with a single neighbor, finding a very sharp peak at the energy where the participation ratio is very small in Fig.~\ref{tomography}. This behavior
is another indication of the localized states being qualitatively different from the states immediately above them. However, as already mentioned, the
$D_2$ window should also be to some extent influenced by the low-energy magnons, and it is possible that the weak, almost uniform background spectral
weight in this map arises primarily from the magnons. Thus, the excitations may be even more concentrated on the subset of spins with large $D_2$ spectral
weight. Shifting the $D_2$ window to lower energy, to reduce the influence of the magnons, indeed leads to less weight on the patches between vacancies.
This behavior is also clear from the behavior of $R(\omega)$ in Fig.~\ref{tomography}(b) for $\omega \in [0.3,0.6]$, where $R$ decreases (less spins
participate) as the energy is lowered. While the fully localized excitations corresponding to the $D_1$ window does involve fewer spins than those seen
in the $D_2$ maps, the evolution clearly must be gradual, considering that the localization peak is not isolated from above.

A natural scenario that would explain these behaviors is that the localization peak (and an extremely  weak tail that likely extends below it to zero
energy \cite{Chernyshev02}) represents a limiting behavior of excitations that gradually become more localized as the energy is lowered; weakly dispersive
damped excitations that are more akin to diffusive, with a diffusion constant vanishing as the localization peak is approached. Most likely, the subsystem
that is strongly involved in these excitations consists predominantly of those spins that are effectively the least coupled to the bulk N\'eel order,
because of a small number of neighbors but also subtle details of their environment. These more easily excitable spins should be coupled to each other
through effective interactions mediated by the bulk degrees of freedom, which naturally leads to an energy scale below the bare Heisenberg coupling $J$.
Since the full diluted Heisenberg model produces these features and the tomographic maps $D_1$ and $D_2$ show that only a small subset of spins is involved
at these energies, it appears plausible that an effective sparse random-network model of spins at vacancies, with suitable interactions, should be able to
produce highly damped diffusive-like excitations that gradually involve fewer spins and localize at the lowest energy scale. Most likely, coupling of the
spins to the global N\'eel order also has to be taken into account by a weak external staggered field. Explicit construction of an effective model is
beyond the scope of the present work, but we are planning future investigations along these lines.

Before discussing the weight $D_3$ in the following window, it is more instructive to first look at $D_4$, from the window in Fig.~\ref{stomo} covering
the upper end of the main peak in $S_0(\omega)$ and also the maximum in the participation ratio. This window should be dominated by the higher-energy
magnons with energy close to the zone-boundary spin waves of the clean system, with only moderate contamination from multimagnon states at even higher
energy. We see in Fig.~\ref{tomo1234}(d) that these excitations generate spectral weight rather uniformly over the patches between the
vacancies, with typically much less weight on the spins immediately adjacent to the vacancies. This distribution supports our picture of high-energy spin
waves that can propagate over distances corresponding to several wave lengths before being scattered by the vacancies, though with very substantial damping.
The contrast with the spectral weight distribution $D_2$ of the random network excitations is striking and only partially captured by the participation ratio.

Returning to the energy window between the random network mode and the high-energy magnons, the map $D_3$ in Fig.~\ref{tomo1234}(c) essentially interpolates
between $D_2$ and $D_4$, with a large concentration of spectral weight on {\it all} the spins adjacent to vacancies, and essentially uniform smaller weight
on other spins. Thus, we conclude that the magnons, in the part of the BZ where the dispersion is anomalous [relatively close to ${\bf q}=(0,0)$ and
$(\pi,\pi)$] is affected by the fact that the spins adjacent to impurities have lower mean coupling to the rest of the system. It is also interesting to
note that the spins with highest spectral weight in $D_2$ often have lower weight than other spins adjacent to vacancies in $D_3$. As already discussed,
the low-energy magnons also to some extent affect the $D_2$ map, but the striking difference in spectral weight next to vacancies shows that the excitations
are different.

The high-energy magnons will of course also to some extent affect the $D_3$ map, and it is therefore likely that the low-energy magnons are even more
concentrated on the spins with less than four neighbors than what is suggested by the $D_3$ map. Thus, an effective model involving only this subset
of spins may be applicable here, and possibly such an effective model, if suitable interactions could be engineered, may reduce to the previously discussed
random network model (including only a subset of the spins with less than four neighbors) at the lowest energy scales.

To summarize what we learn from the real space maps---those shown above and in Appendix \ref{sec:l32maps}, as well as other cases not shown here---from low to
high energy the excitations produce spectral weight as follows: The localized mode ($D_1$) involves the smallest number of spins, most of them adjacent to
vacancies and mostly also corresponding to sites with high monomer density. The diffusive-like mode ($D_2$) involves a much larger subset of the spins
adjacent to vacancies, most of them not belonging to monomer regions. Both the localized and diffusive-like modes may originate from a sparse
random network including some of the spins close to vacancies. The low-energy magnons ($D_3$) occupy also all the other nearest neighbors of the
vacancies and to some extent also uniformly involve the spins in the patches between the vacancies---perhaps less than indicated by $D_3$, since it is
to some extent contaminated by the high-energy magnons. The high-energy magnons largely avoid the spins closest to the vacancies.

\section{Quantum Rotor Excitations}
\label{sec:rotor}

Here we discuss in more detail the remarkably sharp low-energy peak that we identify as the signature of the lowest state in the tower of Anderson
quantum rotor excitations. In a clean system, this state, which has total spin $S=1$ for a bipartite lattice with global sublattice
balance, is exclusively present in the dynamic structure factor at ${\bf q}=(\pi,\pi)$. However, as we have seen, in the diluted system there is some
rotor weight present for all ${\bf q} \not = (0,0)$ due to the loss of translational symmetry. $S({\bf q},\omega)$ is exactly zero at $q=0$ as a
consequence of the the conserved uniform magnetization. 

\subsection{The Anderson rotor tower}

Let us first briefly review the mechanism of the quantum rotor tower \cite{Anderson52}, following the elementary derivation in Ref.~\onlinecite{Sandvik10}.
In the N\'eel state, it is assumed that the spins on sublattices A and B effectively form two large spins, ${\bf S}_A$ and ${\bf S}_B$, with spin quantum numbers
$S_A=S_B \propto N/2$, $N$ being the number of spins. Reduction of the N\'eel order by quantum fluctuations can be taken into account by setting $S_A=S_B < N/2$,
but only the proportionality is important initially. Here we first assume that the system (a large cluster of spins) is sublattice balanced; otherwise
$S_A \not= S_B$.

The lowest-order rotationally invariant interactions between these spins is
\begin{equation}
H_{\rm eff} = J_{\rm eff} \boldsymbol{S}_A \cdot \boldsymbol{S}_B = \frac{J_{\rm eff}}{2}({\bf S}\cdot {\bf S}-S_A^2-S_B^2),
\end{equation}
where ${\bf S} = \boldsymbol{S}_A + \boldsymbol{S}_B$. Here the $S=0$ ground state for $J_{\rm eff} > 0$ corresponds the ordered N\'eel
state and the $S>0$ excited states form the Anderson tower of rotor states, with the terminology reflecting the similarity of the spectrum to that of an actual
quantum rotor. Furthermore, since the constant $-(S_A^2+S_B^2) \propto N^2$, we must take $J_{\rm eff} \propto 1/N$ for the ground state energy to be proportional
to $N$. We can therefore write the excitation energy as a function of the total spin $S$ as
\begin{equation}
E(S) \propto \frac{S(S+1)}{N},
\end{equation}
where, the constant of proportionality is the inverse of the transverse uniform susceptibility of the Heisenberg system for a complete match between
the effective model and the true eigenvalue spectrum of the model. This matching and numerical tests for the clean 2D Heisenberg model are discussed
in detail in Ref.~\onlinecite{Sandvik10}, but we will not need such a detailed matching here.

The $S^z_{\bf q}$ operator used in the dynamic spin structure factor can only excite the $S=1$ rotor mode from the $S=0$ ground state. However, the unbalanced
clusters typically present in large diluted systems (even when global sublattice balance is imposed) have $S > 0$ ground states, and the
rotor excitations then have spin $S+1$ and $S-1$, both of which have energy scaling as $1/N$. We confirmed this scaling of the disorder averaged
rotor contributions in Fig.~\ref{rotor_ens}.

\begin{figure}[t]
\includegraphics[width=55mm]{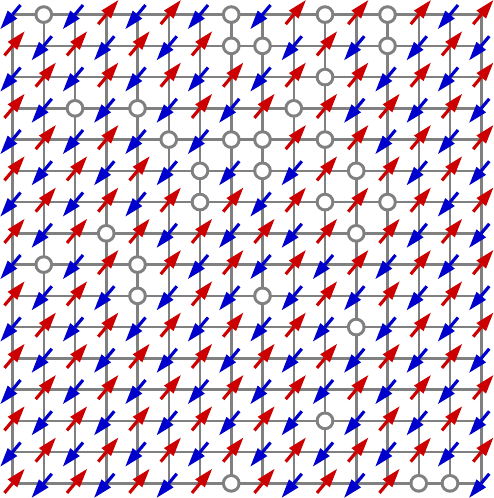}
\caption{Illustration of the classical rotor model for a system with vacancies (circles). The spins (arrows) are oriented
according to their sublattice, forming a degenerate ground state manifold under global $O(3)$ rotations.}
\label{sketchmap_cr}
\end{figure}

Here our interest is in the way in which the rotor peak at ${\bf q}=(\pi,\pi)$ in the clean system is spread out over the BZ in the diluted system.
We find interesting features, some of which can be explained by classical concepts related to sublattice imbalance and others arising from specific
quantum fluctuations that we further explore in Sec.~\ref{sec:jq} with the $J$-$Q$ model.
In Sec.~\ref{sub:classicalrotor} we first discuss the most basic aspect of inhomogeneity at $\omega=0$ in the N\'eel state (which should not be
exactly the same as the non-uniformity measured by equal-time correlations) and its rotor weight distribution within a classical model, before presenting
our results for the diluted Heisenberg model in Sec.~\ref{sub:hbergrotor}. In \ref{sub:sublattdimer} we show that some features not captured by the
classical rotor model can be naturally explained by the diluted dimer-monomer model. We discuss the distribution of the rotor weight in real space in
Sec.~\ref{sec:realrotor}.

\subsection{Classical rotor model}
\label{sub:classicalrotor}

The simplest way to model the distribution of the  rotor amplitude in the BZ
is with a ``classical rotor'' model, i.e., treating the spins as classical $O(3)$ vectors of length $S_c$ in a N\'eel
configuration, as illustrated in Fig.~\ref{sketchmap_cr}. The degeneracy under rotations can here be interpreted as a zero mode, and the ground states
are regarded as eigenstates of the spin operator  used in the dynamic spin structure factor. Thus, $S({\bf q},\omega)$ only consists of a $\delta$ function
at $\omega=0$. The amplitude of the $\delta$ function for given ${\bf q}$ is then also the total spectral weight, which equals the static structure factor
$S({\bf q}$), i.e., the Fourier transform of the equal-time spin-spin correlation function (with the appropriate normalization). Thus, we define the
classical rotor weight as
\begin{equation}
  W^0_c(\boldsymbol{q})=\frac{\pi}{N}\sum_{\boldsymbol{r},\boldsymbol{r}'}e^{-i\boldsymbol{q}\cdot (\boldsymbol{r-r}')}
  \left\langle {\bf S}_{c}(\boldsymbol{r})\cdot {\bf S}_{c}(\boldsymbol{r}')\right\rangle,
\label{w0class}
\end{equation}
where $S_{c}(\boldsymbol{r})\cdot S_{c}(\boldsymbol{r}') \in \{0,\pm S_c^2\}$, depending on the sublattices of ${\bf r}$ and ${\bf r}'$ and the presence of
vacancies [${\bf S}_{c}(\boldsymbol{r}=0$]. The factor $\pi$ comes from its conventional inclusion in the spectral function in Eq.~(\ref{sqwexact}).
The average over vacancy realizations can easily be computed analytically, with the result for $L \to \infty$ being
\begin{equation}
  W^0_{c}({\bf q}) = \left \lbrace
  \begin{array}{ll} \pi N(1-p)^2S_c^2, & {\rm for~} {\bf q}=(\pi,\pi), \\ \pi p(1-p)S_c^2, & {\rm for~} {\bf q}\not=(\pi,\pi), \end{array} \right.
\label{w0cform}
\end{equation}
Thus, not surprisingly, we find a divergence with the system size at ${\bf q}=(\pi,\pi)$ and a constant value for other ${\bf q}$.

We can take some of the quantum effects into account by renormalizing the classical spin length $S_c$ to the actual sublattice magnetization (per spin, not
per lattice site) of the diluted quantum model, which has previously been computed versus $p$ by SSE simulations \cite{Sandvik02}. A low-order fit valid
for small $p$ gave
\begin{equation}
S_c \approx  0.3072 - 0.134p-0.51p^2
\label{scform}
\end{equation}
in the thermodynamic limit. Finite size effects lead to larger values for smaller $L$ \cite{Sandvik02}, which we do not take into account here but will
observe in the results for the Heisenberg model. There is also a small size correction factor $N/(N-2)$ in the ${\bf q} \not=(\pi,\pi)$ weight that we
left out in Eq.~(\ref{w0cform}).

With the proper normalized value of $S_c$, $W^0_{c}(\pi,\pi)$ is a strict upper bound of the staggered (long-range)
rotor weight in the Heisenberg model in the thermodynamic
limit, since it represents the exact value of the full frequency integral of the spectral function, not just its quantum rotor contribution at
$\omega \propto 1/N$. This bound can be expected to be very accurate, because the quantum rotor weight almost exhausts the divergent static structure
factor at $(\pi,\pi)$, as can be seen for a small system in Fig.~\ref{sw_ed1}(c). At other momenta, where our main interest lays here, $W^0_{c}({\bf q})$
should just be taken as a zeroth-order estimate. We would expect larger true rotor weights typically, because the classical N\'eel state completely neglects
the spin fluctuations that give rise to non-trivial correlations for ${\bf q}\not=(\pi,\pi)$ and govern the total spectral weight when used
in Eq.~(\ref{w0cform}). The rotor weight is still of course only a small fraction of the total spectral weight for ${\bf q}\not=(\pi,\pi)$, and just
substituting the correct spin correlation function of the diluted Heisenberg model (computed with QMC) would not be of much use here.

\subsection{Rotor weight in the Heisenberg model}
\label{sub:hbergrotor}

As we saw in Sec.~\ref{sec:sqw}, the rotor peak for the larger system sizes is well separated from the spectral weight at higher energy. We can
therefore define the rotor weight as the integral over the sharp peak,
\begin{eqnarray}
W^0_{\delta}(\boldsymbol{q})=\left\langle\int^{\rm cut}_{0}S(\boldsymbol{q},\omega)d\omega\right\rangle,
\label{w0delta}
\end{eqnarray}
where the cut-off energy is easily identifiable in our data. Though not visible in the figures in Sec.~\ref{sec:sqw}, there is some very small spectral
weight also between the rotor peak and the localization peak or edge, since no completely sharp features can appear with unrestricted sampling in SAC
and there is also likely an actual thin low-energy tail of the localization peak. Therefore, a minimum exists that can be taken as the cut-off $\omega$ value in
the integral. The value of $S({\bf q},\omega$) at the cut-off is typically orders of magnitude below the weight at the localization edge, and
the results of the integral in Eq.~(\ref{w0delta}) is insensitive to the details of how the cut-off is chosen.

\begin{figure}[t]
\includegraphics[width=83mm]{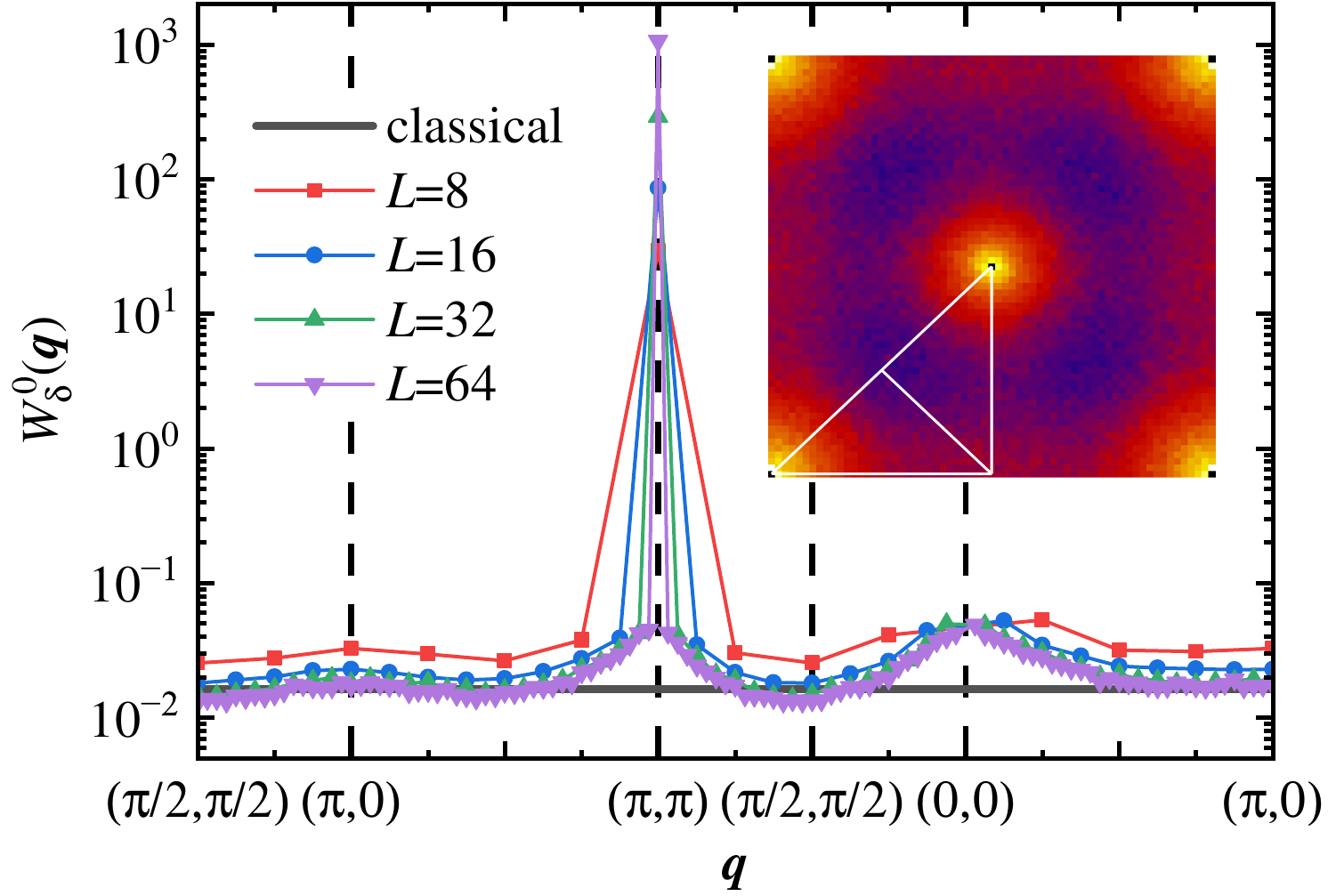}
\caption{Distribution of quantum rotor weight at $p=1/16$ along a path in the BZ for different system sizes. Note that the $L=32$ data are almost covered by
the $L=64$ points. The horizontal line shows the prediction Eq.~(\ref{w0cform}) of
the classical rotor model with the quantum renormalized spin $S_c=0.297$, i.e., the sublattice magnetization per spin according to
Eq.~(\ref{scform}) \cite{Sandvik02}. The inset shows the rotor distribution in the entire BZ, $q_x,q_y \in [0,2\pi]$, for $L=64$, with the
large weight at ${\bf q}=(\pi,\pi)$ left out. The path in the main graph is marked with lines in the BZ.}
\label{rotorweight}
\end{figure}

In Fig.~\ref{rotorweight} we graph the resulting rotor weight along our standard path in the BZ for different system sizes. Here we use a log scale
on the vertical axis in order to visualize both the divergent weight exactly at ${\bf q}=(\pi,\pi)$ and the distribution of $L \to \infty$ convergent
non-zero weight over the rest of the BZ. Overall, for ${\bf q}\not=(\pi,\pi)$  we observe slowly decreasing rotor weight with increasing $L$, with only minor
differences in the results for $L=32$ and $L=64$. Instead of the uniform weight in the classical model for ${\bf q} \not=(\pi,\pi)$, we see accumulation of weight
close to both $(\pi,\pi)$ and $(0,0)$, with a minimum close to (not exactly at) $(\pi/2,\pi/2)$. In this region the weight falls below the constant
value predicted using the classical rotor model, while at other momenta it is much higher. 

\begin{figure}[t]
\includegraphics[width=83mm]{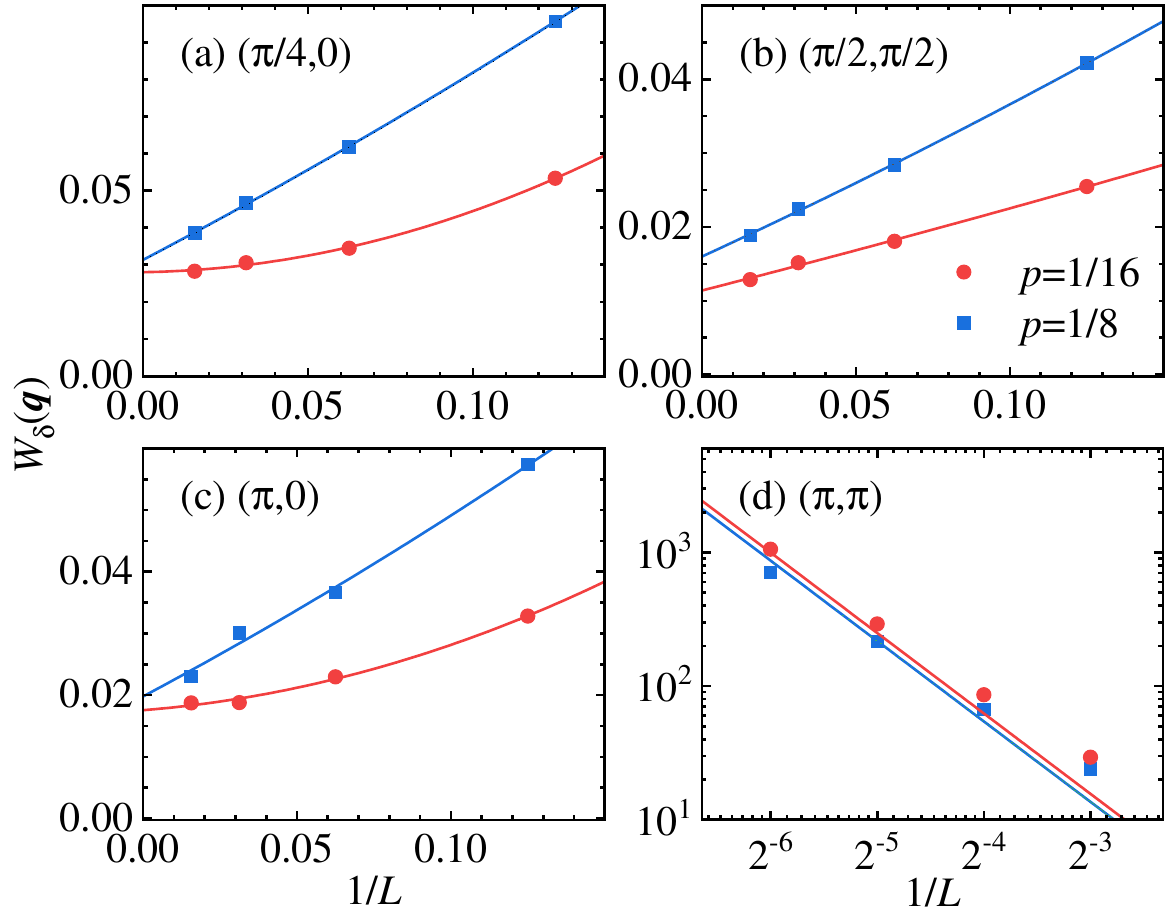}
\caption{Size dependence of rotor weight in $S({\bf q},\omega)$
at $p=1/16$ and $p=1/8$ for selected momenta: (a) $\boldsymbol{q}=(\pi/4,0)$, (b) $\boldsymbol{q}=(\pi/2,\pi/2)$,
(c) $\boldsymbol{q}=(\pi,0)$, and (d) $\boldsymbol{q}=(\pi,\pi)$. In (a)-(c) the curves are fitted second-order polynomials, while in (d) the line
corresponds to the predicted form Eq.~(\ref{w0cform}) with the effective spin given by Eq.~(\ref{scform}).}
\label{rw_size}
\end{figure}

In Fig.~\ref{rw_size} we analyze the size dependence of the rotor weight at selected momenta for both $p=1/16$ and $p=1/8$. In some cases the size correction
is very close to linear in $1/L$, while at generic points a quadratic fit works well. At $(\pi,\pi)$ we see the expected linear divergence in $N$, with
the exact form very close to the prediction based on Eq.~(\ref{w0cform}) when the values of $S_c$ corresponding to $p=1/16$ and $1/8$ are taken according
to Eq.~(\ref{scform}). For the other ${\bf q}$ points, we do not have any further insights in the size dependence expected, but it is nevertheless clear from
the results that the nonuniform rotor weight distribution persists in the thermodynamic limit.

In the inset of Fig.~\ref{rotorweight} we show a heat map of the weight in the entire BZ, where a broad minimum around $(\pi/2,\pi/2)$ (and the three other
symmetrically located points) is also apparent. Note again that there is no spectral weight at all at $(0,0)$, and in the heat map we have also left out
$(\pi,\pi)$ since the rotor weight there is orders of magnitude larger than at nearby points.

When translated to real space, the rotor weight distribution reflects a certain inhomogeneity in the breaking of the spin rotational symmetry in the N\'eel
state. It should be noted here again that the equal-time spin-spin correlation function, i.e., the static structure factor $S({\bf q})$, corresponds to the total
spectral weight for given ${\bf q}$, but the rotor weight is only the static, $\omega=0$ (when $L \to \infty$), contribution. At ${\bf q}=(\pi,\pi)$, the
equal-time and static structure factors are essentially the same, as we discussed above and illustrated in Fig.~\ref{rw_size}(d). There is also continuum
spectral weigh above the rotor peak in this case, as seen in Fig.~\ref{sw_p16}(d), but its relative weight vanishes as $L \to \infty$. At other momenta, the
continuum weight survives as a finite fraction of the static structure factor. It should  in principle be possible to detect the ${\bf q}$ dependent rotor
weight experimentally using elastic neutron scattering, in particular the broadened ${\bf q}=(\pi,\pi)$ Bragg peak and the new peak at ${\bf q}=(0,0)$,

We will go through several steps of trying to understand the origins of this $\omega=0$ spin texture, in the subsections below and further in Sec.~\ref{sec:jq},
there with the aid of the $J$-$Q$ model, where the suppression of weight around $(\pi/2,\pi/2)$ is amplified by the four-spin interaction $Q$.

\subsection{Dimer-monomer model}
\label{sub:sublattdimer}

We here explore the ability of the classical dimer-monomer model to generate features of the rotor weight related to local sublattice imbalance. We already
showed an example of enhanced local magnetic response matching regions of local sublattice imbalance in the dimer model in Fig.~\ref{monomer_locx}, following the
approach first used to model the 2D Heisenberg model when diluted close to the percolation point ($p_c\approx 0.407$) in Ref.~\onlinecite{Wang10}.

\renewcommand\arraystretch{1.2}
\begin{table}[t]
	\centering
	\begin{tabular}{c|c|c|c|c|c|c|c}
		\diagbox{L}{$R_m$}{$N_m$} &0 & 2 & 4 & 6 & 8 &$\cdots$ & $\overline{N}_m$  \\
		\hline 
		32 & 983 & 17 & 0 & 0 & 0 & 0 & 0.034  \\
		\hline
		40 & 957 & 38 & 5 & 0 & 0 & 0 & 0.096 \\
		\hline
		48 & 830 & 120 & 33 & 11 & 5 & 1 & 0.49\\
		\hline
		64 & 336 & 246 & 172 & 112 & 60 & 74 & 3.182 \\
		\hline
	\end{tabular}
	\caption{Number of realizations $R_m$ with $N_m$ monomers remaining in equilibrium in the dimer-monomer model with vacancy fraction $p=1/16$
          when a total of 1000 vacancy realizations were generated. Results are shown versus the system size, and in each case the mean value of
          monomers is shown in the rightmost column. The $\cdots$ column includes all $N_m>8$.}
	\label{tab1}
\end{table}

\begin{figure}[t]
\includegraphics[width=75mm]{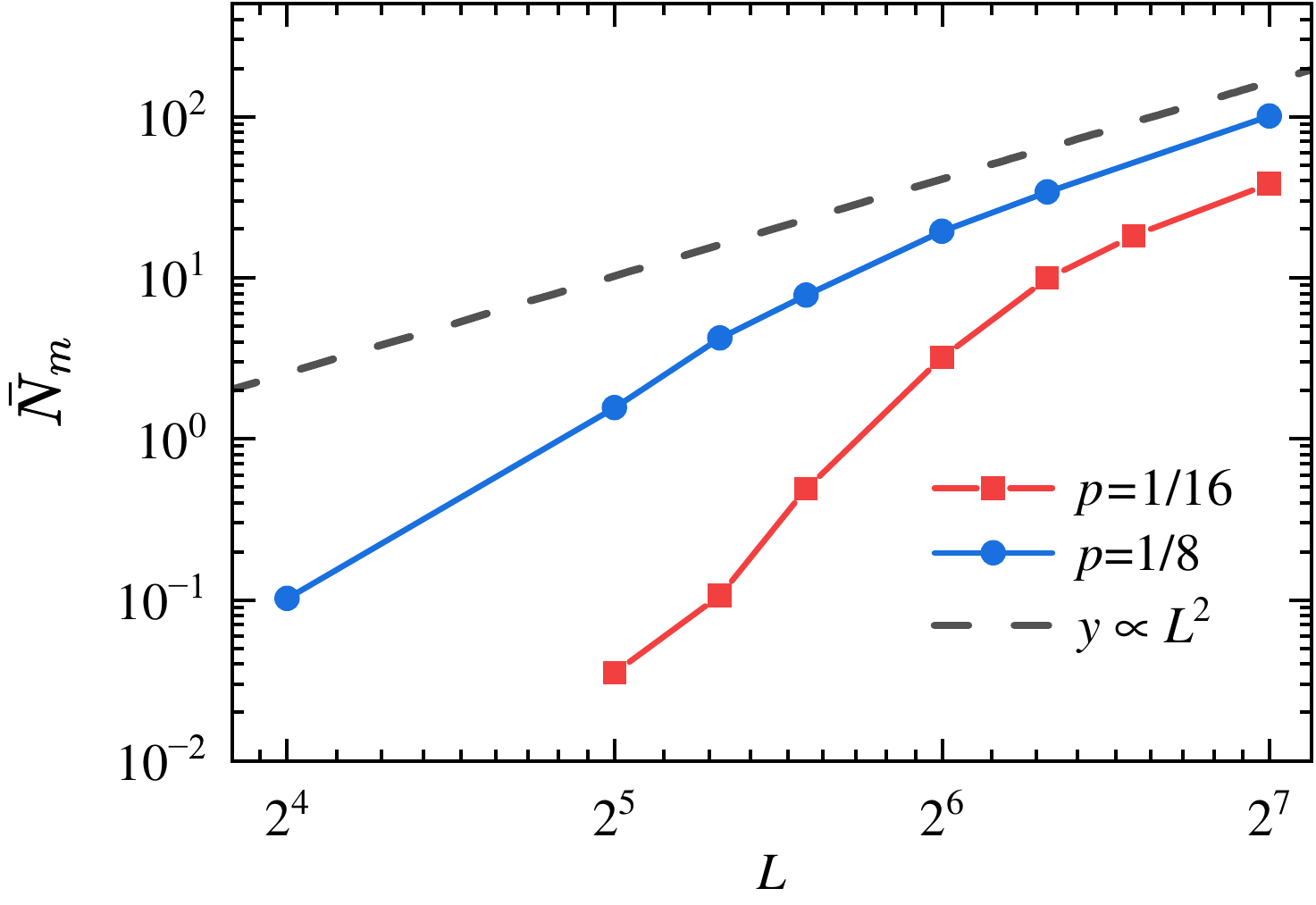}
\caption{Average monomer number vs system size for $p=1/16$ (red squares) and $p=1/8$ (blue circles). The dashed line shows qualitatively the
expected linear growth with the system volume.}
\label{nm_size}
\end{figure}

We consider configurations with the minimum number of monomers for each random realization, sampled with the algorithm explained in Fig.~\ref{dimer_update}.
To reduce statistical fluctuations for small system sizes, we again use a canonical ensemble for the vacancy fraction, i.e., for system size $N=L^2$, the number
of vacancies is exactly $pN$. For small systems, the number of remaining monomers
is often zero, and we generate a large number of configurations in order to obtain sensible averaged results. Table \ref{tab1} shows examples of the
distribution of the number of remaining monomers in system sizes up to $L=64$. It is only for system sizes larger than $L=64$ that the expected mean
number of monomers grows in proportion to the system volume, as seen in Fig.~\ref{nm_size}.

Regarding the monomers as spins $S_{\boldsymbol{r}}=\pm 1/2$ according their sublattices, as in Fig.~\ref{sketchmap_dm}, we again perform a Fourier transform
similar to Eq.~(\ref{w0class}) in the classical rotor model,
\begin{eqnarray}
F(\boldsymbol{q})= \frac{1}{\bar N_m}\left | \sum_{\boldsymbol{r}_m}S_{\boldsymbol{r_m}}{\rm e}^{-i\boldsymbol{q}\cdot \boldsymbol{r}_m}\right |^2
\label{fqdef}
\end{eqnarray}
where $\boldsymbol{r}_m$ are the positions of monomers.

The idea behind the dimer-monomer model in this context is that the sublattice imbalance described by the monomers enhance the staggered magnetic
order relative to the renormalized flat N\'eel background described by $S_c$ in Eq.~(\ref{scform}). Recall that we used a value for the classical spin
length $S_c < 1/2$ to account for the quantum fluctuations that reduce the magnitude of the N\'eel order parameter. In principle, the reduction of the
N\'eel order in the diluted system could be studied with a more complete quantum mechanical description by a variational state based on fluctuating
dimer states built from Eq.~(\ref{dimersinglet}) along with unpaired spins. In the classical dimer-monomer model studied here, we only use the monomers
to describe regions with locally enhanced N\'eel order, i.e., larger $S_c$, in (an unknown) proportion to the monomer density. The so generated nonuniform
N\'eel state will lead to deviations from the zeroth-order predictions in Eq.~(\ref{w0cform}). The results for the dimer-monomer model will only represent
these differences. Thus, we define the additional weight of the rotor state in momentum space as
\begin{equation}
W^0_{\rm dm}(\boldsymbol{q})=\langle F(\boldsymbol{q})\rangle,
\label{w0dmdef}
\end{equation}
with $F(\boldsymbol{q})$ defined in Eq.~(\ref{fqdef}) and averaged over large number of dimer-monomer configurations for each vacancy realization, and further
averaged over several thousand vacancy relations for small $L$ and about 1000 for the largest $L$ considered.

\begin{figure}[t]
\includegraphics[width=80mm]{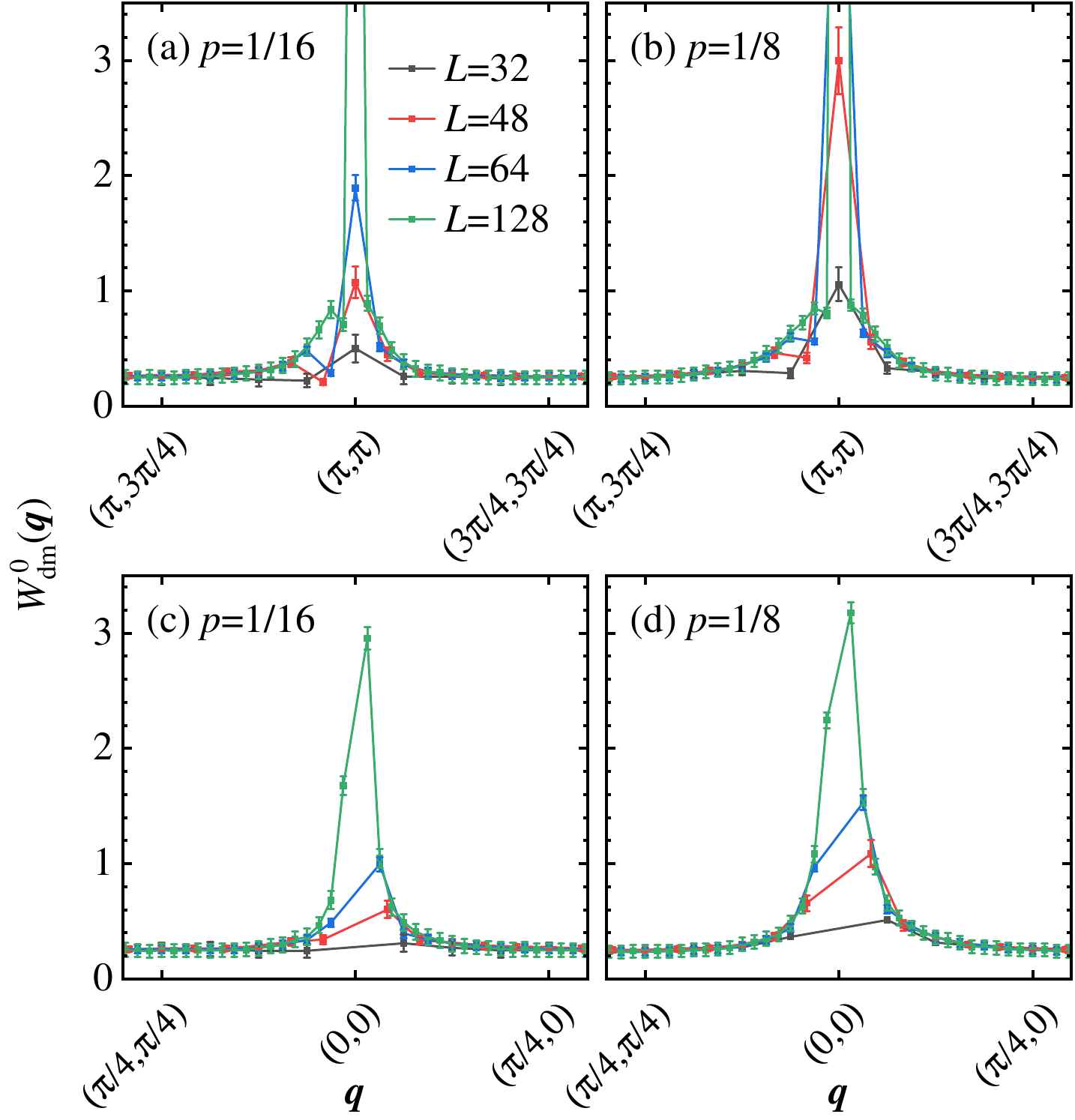}
\caption{Rotor weight distribution obtained with the dimer-monomer model near ${\bf q}=(\pi,\pi)$ and $(0,0)$.
Further away from these points the value does not change appreciably. Results for $p=1/16$ are shown in (a) and (c),
while (b) and (d) show analogous results for $p=1/8$.}
\label{rw_dimer}
\end{figure}

We show results for several system sizes at both $p=1/16$ and $p=1/8$ in Fig.~\ref{rw_dimer}. Here we observe two prominent features that are absent in the
classical rotor model: (i) weight spreads out from the $\delta$ function at ${\bf q}=(\pi,\pi)$ to form a broadened peak, and (ii) a second peak has emerged
at $(0,0)$, which is somewhat broader than the $(\pi,\pi)$ peak. The value exactly at $(0,0)$ is zero because of the imposed global sublattice balance.
These two peak features both agree qualitatively with the results for the rotor weight in the Heisenberg model in Fig.~\ref{rotorweight},
and we can therefore conclude that they indeed reflect the consequences of sublattice imbalance as modeled by the dimer-model model. Importantly, however,
the dimer-monomer model does not produce any other structure in the BZ beyond the two peaks; in particular, the differences between the zone-boundary points
${\bf q}=(\pi/2,\pi/2)$ and $(\pi,0)$ in Fig.~\ref{rotorweight} are not reproduced within the classical dimer-monomer model.

In attempts to explain this missing features, we have also carried out simulations with dimer-dimer interactions included, the motivation being that the
Heisenberg model already shows precursors to the formation of a columnar valence-bond solid state, one manifestation of which is the dip in magnon energy
close to ${\bf q}=(\pi,0)$ seen in Fig.~\ref{dispersion} \cite{Shao17}. As shown in Appendix \ref{sec:dimerinteractions}, the results in Fig.~\ref{rw_dimer}
are rather stable, however, and neither attractive nor repulsive columnar dimer interactions can generate the sought effect. Thus, we believe that the 
suppression of $(\pi/2,\pi/2)$ weight (where the minimum is actually not exactly at this ${\bf q}$ point) is a manifestly quantum dynamical effect, which
we explore in Sec.~\ref{sec:jq} with the $J$-$Q$ model.

\subsection{Rotor weight in real space}
\label{sec:realrotor}

We now return to the real-space distribution of spectral weight discussed in Sec.~\ref{sec:tomo}, considering the window including only the quantum rotor
excitation in Eq.~(\ref{stomo}). We here use the same $L=16$ vacancy realization as in Fig.~\ref{monomer_locx}, and in Appendix \ref{sec:l32maps} we show
supporting results for the $L=32$ realization also used in Fig.~\ref{tomo1234}. The ${\bf r}$ averaged spectral weight $S_0(\omega)$ of the $L=16$ sample looks
qualitatively similar to that for the $L=32$ sample in Fig.~\ref{tomography}, and we define the energy window $D_0({\bf r})$ in the same manner so that
it contains only the rotor peak.

\begin{figure}[t]
\includegraphics[width=55mm]{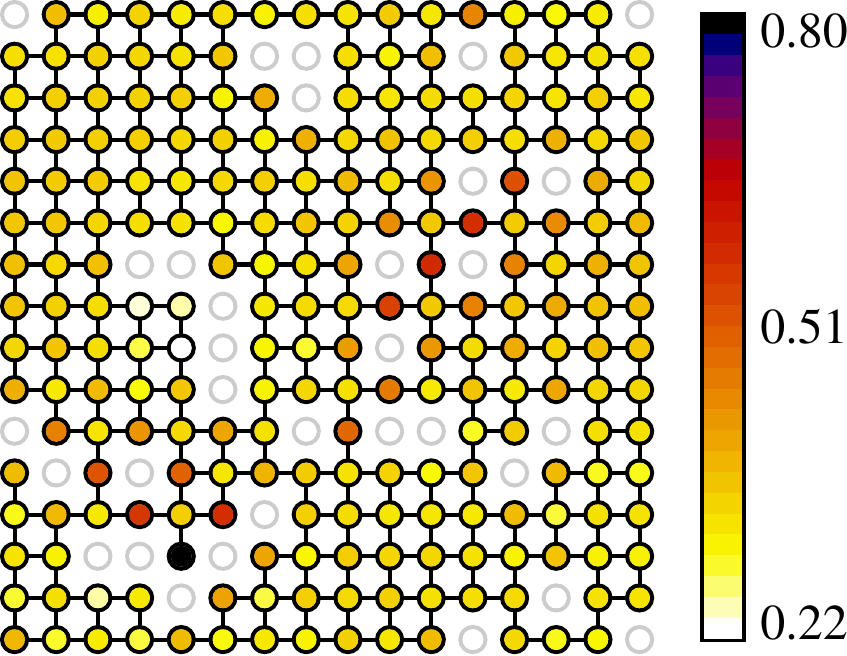}
\caption{Rotor weight distribution $D_0({\bf r})$ in real space in the same $L=16$ system as in Fig.~\ref{monomer_locx}, computed according to
Eq.~(\ref{stomo}) in an energy window similar to that indicated for an $L=32$ system in Fig.~\ref{tomography}.}
\label{tomo-s0}
\end{figure}

The real-space distribution $D_0$, shown in Fig.~\ref{tomo-s0}, looks very similar to the local susceptibility map in Fig.~\ref{monomer_locx}(a). This can
be explained by the well-known sum rule for the local susceptibility (see, e.g., Ref.~\cite{Wang10}),
\begin{equation}
\chi_{\bf r} = \frac{2}{\pi}\int_0^\infty \frac{S({\bf r},\omega)}{\omega}d\omega,
\end{equation}  
which implies that excitations at the lowest energies will dominate, provided that they carry significant spectral weight. A robust rotor peak indeed
always exists at energy far below the spectral weight in the continuum at higher energies. Thus, the rotor weight should be expected to contribute
significantly to $\chi_{\bf r}$.

Given the close match between the results in Fig.~\ref{tomo-s0} and the monomer distribution in Fig.~\ref{monomer_locx}(b), we conclude that the dimer-monomer
model is a good indicator for the spatial distribution of the quantum rotor state, which is also of course the prerequisite for using it to explain the rotor
weight distribution in ${\bf q}$ space above in Sec.~\ref{sub:sublattdimer}.

We show both the monomer and rotor weight distribution for the same $p=1/8$, $L=32$ vacancy realization as in Fig.~\ref{tomo1234} in Appendix \ref{sec:l32maps},
and also similar results for a case of $p=1/16$. Here again we find a strong correlation between monomer sites and rotor weight, with the rotor weight also being
large at all other sites (as is also the case in the $L=16$ case in Fig.~\ref{tomo-s0}), as expected because all spins participate in the N\'eel ordering and the
monomer sites only reflect elevated $\omega=0$ ordering. The localized excitations, corresponding to the energy window above the rotor peak, visualized in the
map in Fig.~\ref{tomo1234}(a), also have strong overlap with some of the monomer sites, but much lower spectral weight overall is distributed on the other spins.

\section{J-Q model}
\label{sec:jq}

As we have seen above, the difference in rotor weight at ${\bf q}=(\pi,0)$ and ${\bf q}=(\pi/2,\pi/2)$ cannot be accounted for simply by sublattice imbalance
as described by the classical rotor model supplemented by the dimer-monomer model.
This deficit of the semi-classical description of the N\'eel state, with suppression of the overall N\'eel
order by quantum fluctuations and local enhancement in regions of sublattice imbalance, is not surprising in itself. However, the particular effect of
suppression of the rotor weight close to ${\bf q}=(\pi/2,\pi/2)$ may be related to the same mechanism that increases the magnon energy at this momentum while
reducing it at $(\pi,0)$, as observed both in the clean and diluted systems in Fig.~\ref{relativedisp}.

This effect on the dispersion relation was shown
to originate in quantum fluctuations related to incipient spinon deconfinement close to ${\bf q}=(\pi,0)$ \cite{Shao17},
a conclusion reached for the clean system by monitoring the
effects the four-spin interaction $Q$ in the $J$-$Q$ model defined in Eq.~(\ref{jqham}). It was found that the magnon peak in $S({\bf q},\omega)$ at
${\bf q}=(\pi,0)$ vanishes or becomes extremely small already when $Q/J=1$, far below the quantum phase transition into a valence-bond solid state at
$Q/J \approx 22$ \cite{Sandvik07}. In contrast, the magnon peak remains robust at ${\bf q}=(\pi/2,\pi/2)$. Here we use the $J$-$Q$ model for the same
purpose to investigate the effects of enhanced correlated singlet fluctuations on the distribution of the rotor weight in the BZ.

\begin{figure}[t]
\includegraphics[width=80mm]{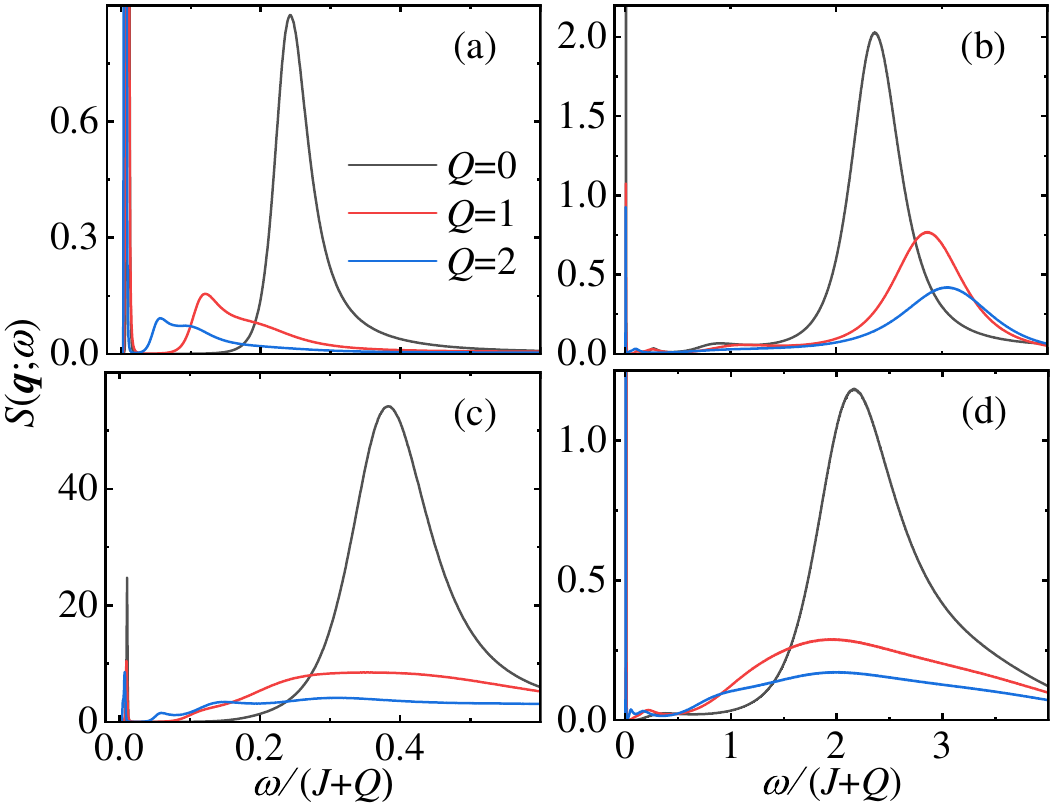}
\caption{$S(\boldsymbol{q},\omega)$ computed with the $J$-$Q$ model at $Q/J=0$ (black curves), $1$ (red),
and $2$ (blue) at representative $\boldsymbol{q}$ points for $p=1/16$, $L=32$.
(a) $\boldsymbol{q}=(2\pi/L,0)$, (b) $\boldsymbol{q}=(\pi/2,\pi/2)$, (c) $\boldsymbol{q}=(\pi-2\pi/L,\pi-2\pi/L)$, (d) $\boldsymbol{q}=(\pi,0)$.}
\label{sw_jq_p16}
\end{figure}

In Fig.~\ref{sw_jq_p16} we show the dynamic structure factor at $p=1/16$ in a system of size $L=32$ at selected momenta, comparing results for
$Q/J=0$, $1$, and $2$. In all cases except for ${\bf q}=(\pi/2,\pi/2)$, the location of the most prominent peak is shifted dramatically to lower energy
when $Q$ increases. A large quantum rotor peak persists at $Q>0$, as it should since the system is still strongly N\'eel ordered at these values
of $Q$. Much more continuum spectral weight extends to low energies, ending at a small peak far below the energy of the localization
edge for this system size at $Q=0$ (Fig.~\ref{sw0_8}). Thus, it appears that the $Q$ interaction further facilitates the formation of a low-energy localization
mode---perhaps in this case reflecting localized spinons. In contrast to the other momenta, at ${\bf q}=(\pi/2,\pi/2)$ in Fig.~\ref{sw_jq_p16}(b)
the high-energy mode remains with significant spectral weight and its energy increases, as in the clean system \cite{Shao17}.

\begin{figure}[t]
\includegraphics[width=75mm]{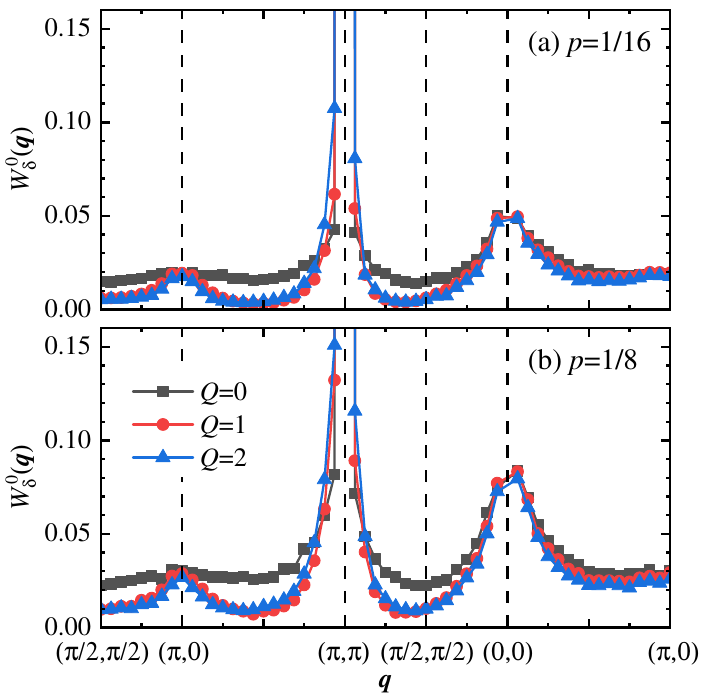}
\caption{Rotor weight distribution in the $J$-$Q$ model with $Q/J=0$ (black squares), $1$ (red circles), and $2$ (blue triangles), graphed along the
standard path in the BZ at vacancy fraction $p=1/16$ in (a) and $p=1/8$ in (b).}
\label{rw_jq}
\end{figure}

In Fig.~\ref{rw_jq} the weight of the quantum rotor peak is graphed along the standard path in the BZ for both $p=1/16$ and $p=1/8$ and for
the same three values of $Q/J$ as above. Here the depletion of the weight close to $(\pi/2,\pi/2)$ is obviously much more dramatic at $Q=1$ than at $Q=0$,
with not much further change when increasing $Q/J$ to $2$. An interesting feature here is that the suppression is actually stronger at the lower
vacancy fraction. This counter-intuitive result has its natural explanation in our proposed mechanism: The precursor fluctuations related to spinon
deconfinement, i.e., correlated singlets resonating in columnar patterns, which reflect the eventual transition to the columnar dimerized state at large $Q/J$.
At the larger vacancy fraction, these extended fluctuations of the bulk are to some extent already hindered by the presence of the vacancies, which
individually favor certain vortex-like patterns (at least for large $Q/J$) \cite{Kaul08} and collectively prohibit the formation of a uniform pattern
in the dimerized phase \cite{Liu18}.

\begin{figure}[t]
\includegraphics[width=60mm]{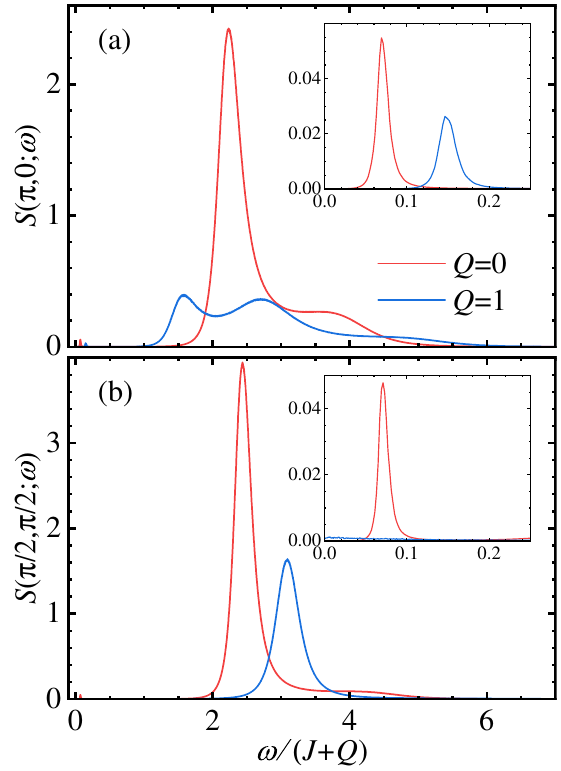}
\caption{Dynamic structure factor in the $L=16$ Heisenberg model ($Q=0$) and $J$-$Q$ model ($Q=1$), both with one vacancy and shown for $\boldsymbol{q}=(\pi,0)$
in (a) and $\boldsymbol{q}=(\pi/2,\pi/2)$ in (b). The insets focus on the quantum rotor peaks, which in (a) have weight $W^0_\delta \approx 0.0011$
at $Q=0$ and $W^0_\delta \approx 0.0015$ at $Q=1$. In (b) $W^0_\delta \approx 0.0008$ at $Q=0$, while at $Q=1$ a much smaller spectral weight is spread out
between $\omega=0$ and $\omega \approx 0.1$ and does not form a sharp peak.}
\label{jqholes1}
\end{figure}

\begin{figure}[t]
\centering
\includegraphics[width=60mm]{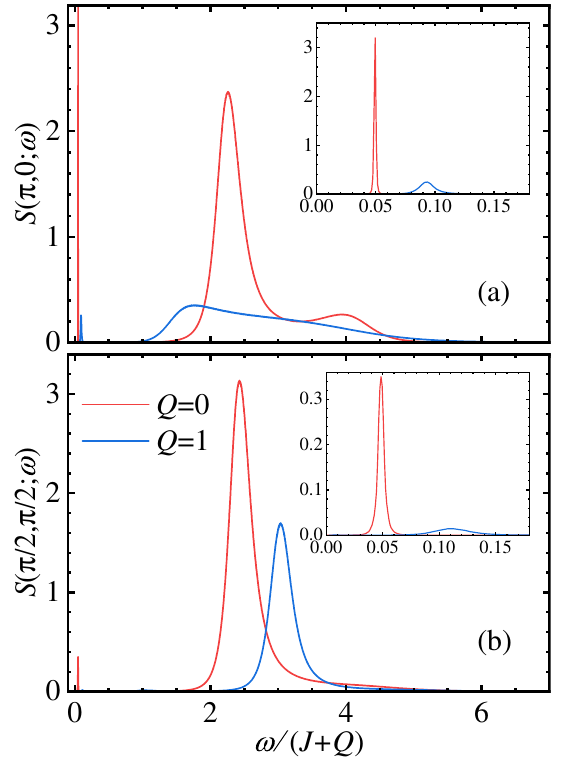}
\caption{Results as in Fig.~\ref{jqholes1} but for two vacancies at maximum separation. The rotor weight in (a) is
$W^0_\delta \approx 0.0066$  at $Q=0$ and $W^0_\delta \approx 0.0089$ at $Q=1$, while in (b) $W^0_\delta \approx 0.0025$ at $Q=0$ and 
$W^0_\delta \approx 0.0014$ at $Q=1$.}
\label{jqholes2}
\end{figure}

Our assertion that the $Q$-enhanced singlet fluctuations are responsible for the suppression of rotor weight at ${\bf q}=(\pi/2,\pi/2)$ is further
supported by results for systems with one and two vacancies, shown in Figs.~\ref{jqholes1} and \ref{jqholes2}, respectively. Comparing results at
$Q/J=0$ and $1$ for a small system, $L=16$, we again see completely different effects of the $Q$ interaction at ${\bf q}=(\pi,0)$ and $(\pi/2,\pi/2)$, 
with the magnon peak pushed up in energy by $Q$ in the latter case while the spectral weight instead broadens out to lower energies in the former
case. The rotor weight increases slightly with $Q$ at ${\bf q}=(\pi,0)$ but is diminished at ${\bf q}=(\pi/2,\pi/2)$. In most of the
cases, the rotor energy is higher when $Q/J=1$, which also explains the more significant broadening of the peak by the SAC. 
However, in the case of the single
vacancy, the $(\pi/2,\pi/2)$ rotor weight in Fig.~\ref{jqholes1}(b) is extremely small and spread out over a wide range of low energies, which may
just be an effect of the weight being too small to resolve properly.

The main message of all these results for the $J$-$Q$ model is that the precursor fluctuations to spinon deconfinement in the clean
Heisenberg model and $J$-$Q$ model with small $Q$ are also responsible for the depletion of the rotor weight at ${\bf q}=(\pi/2,\pi/2)$  in the
diluted system. This depletion, which reflects a nonuniform spin texture at $\omega=0$ in the N\'eel state, may possibly be experimentally detectable by
elastic neutron scattering in layered antiferromagnets doped by nonmagnetic impurities \cite{Vajk02}, though the effect is rather weak in the
Heisenberg model. The peak around ${\bf q}=(0,0)$ may be easier to detect but is not related to the spinon precursor.

\section{Conclusions and Discussion}
\label{sec:conc}

\subsection{Brief summary of main results}

By studying the disorder averaged dynamic structure factor $S({\bf q},\omega)$ using state-of-the-art QMC and SAC methods, we have obtained a wealth of
insights into the excitations of the 2D Heisenberg model diluted with vacancies. In total, we identified six different types of excitations forming a
multiscale structure factor. Some of our observations were previously predicted in spin wave theory with a $T$-matrix approach for the vacancies. Specifically,
localized excitation at a low energy \cite{Chernyshev02} and a high-energy magnon with dispersion relation close to that in the clean system, crossing
over to an anomalous dispersion relation at lower energies above the localization energy \cite{Brenig91,Chernyshev02}. However, there are both quantitative
and qualitative discrepancies. We fond a second previously missed peak between the localization peak and the low-energy magnons, which we propose originates
from a subset of spins close to vacancies, whose excitations are diffusive-like and gradually become localized with lowered energy. We also investigated
novel aspects of the quantum rotor excitation associated with the formation of the long-range N\'eel order in the thermodynamic limit. Finally, there is
a significant multimagnon continuum for all ${\bf q}$, extending as far as to $\omega \approx 5$ at the zone boundary.

The energy $\omega_{\rm loc} \propto {\rm e}^{-\pi/4p}$ of the localized mode was obtained in Ref.~\onlinecite{Chernyshev02} at small dilution fractions $p$.
At $p=1/8$, we were able to extrapolate the energy of the localization mode to infinite system size, giving $\omega_{\rm loc} \approx 0.1$, which is about
50 times larger than the predicted energy scale. At $p=1/16$, the localization energy is too low for a meaningful extrapolation with the system sizes
available. However, in this case we find that the localization peak shifts as $\omega_{\rm loc} \propto L^{-1}$, which is consistent with the prediction
$\omega_{\rm loc} \propto l^{-1}$, where $l$ is the localization length, when $L \ll l$.

As for the discrepancy at $p=1/8$, it should be noted that the prefactor of the exponential form of $\omega_{\rm loc}$ was not calculated analytically, and
the form also perhaps applies only at much smaller dilution fractions. It is also possible that the localized excitation found here is not exactly
of the same type as in the analytical calculation, because of the approximations made there.

The real-space resolved structure factor $S({\bf r},\omega)$ proved to be very useful in further characterizing the excitations by the manner in which
individual spins participate in production of spectral weight within different energy windows. In particular, using this tomographic approach, we conclude
that some of the spins located at sites in the immediate vicinity (nearest neighbors) of the vacancies form a subsystem---a random network of interacting
moments---that is mainly involved in the diffusive-like excitations that gradually localize as the energy is lowered. In contrast, the magnons
generate almost uniform spectral weight in the patches between vacancies, with much less weight on the spins adjacent to vacancies at high energy. At
lower energy, where the dispersion relation becomes anomalous, close to the predicted logarithmic form \cite{Chernyshev02}, the spectral weight is
larger on the spins adjacent to the vacancies.

Because of these qualitative differences between low- and high-energy magnons, we regard the spin waves as two different types of excitations within their
respective energy windows. As the energy is further lowered toward that of the random network mode, the weight between the vacancies diminishes, and also
on the majority of the spins next to vacancies, thus leaving behind the rather sparse random network of spins with large spectral weight. This energy window
is also to some extent influenced by the tail of the low-energy magnons, and the observed uniform weight within the patches may to a large extent arise
from these excitations, for which we find further support in the fact that the participation ratio decreases when $\omega$ is lowered.

The quantum rotor excitation is not included in spin wave calculations, where the $O(3)$ symmetry of the N\'eel order parameter is broken from the outset.
In a finite system, the symmetry is not broken and the tower of $S>0$ states that become degenerate in the thermodynamic limit (which can combine to form the
symmetry-broken state) is present in the excitation spectrum. An important aspect of our work was to investigate how the rotor weight in the dynamic
structure factor at $\omega \propto N^{-1}$ spreads out in momentum space from ${\bf q} = (\pi,\pi)$, where it is located in the uniform system. We found
broadened peaks close to both $(\pi,\pi)$ and $(0,0)$, which can be explained by local sublattice imbalance modeled by a simple classical dimer-monomer
model. Other structure, specifically a depletion of spectral weight close to ${\bf q}=(\pi/2,\pi/2)$, can be magnified by turning on the four spin interaction
of the $J$-$Q$ model, thus suggesting signatures of incipient spinon deconfinement at momenta close to ${\bf q}=(\pi,0)$, similar to the clean 2D Heisenberg
model \cite{Shao17}. We also found an enhanced tendency of localization in the $J$-$Q$ model.

\subsection{Contrasts with spin wave theory}

Spin waves with anomalous dispersion had been predicted early on \cite{Harris77,Brenig91,Wan93} but it was only the work of Chernyshev et al.~\cite{Chernyshev02}
that resolved the way in which this mode merges with localized state at low energy. Thus, very close to ${\bf q}=(0,0)$ and $(\pi,\pi)$, the use of a disorder
averaged medium, in which the spin waves propagate at higher energy, becomes invalid. However, the relationship between the anomalous spin waves and essentially
conventional magnons at higher energy (in reality highly damped) remained unsettled, as the interactions between the spin waves of the clean system and the
impurity modes (from the spins next to vacancies)
at energy $\omega \approx J$ was not sufficiently included to obtain reliable spectral functions at these energies. The high-energy mode then exists as an
apparently well defined separate excitation close to the magnetic zone boundary. In some parts of the BZ (where the magnon energy is of order $J$), they
both exist as individual excitations, producing two separate peaks in $S({\bf q},\omega)$. In our calculations, we always observe only one peak at these energy
scales. In the spin wave theory there is also no peak in the $({\bf q},\omega)$ regions where we identify the diffusive-like random-network excitations.

Our calculations differ in a fundamental manner from the $T$-matrix calculations, in that we perform the disorder average of the dynamic correlation function
(in imaginary time) and do not assume that the excitations are formed in an already averaged medium.  As already mentioned above we demonstrated that the
excitations below the anomalous magnon actually propagate only within a subset of the spins; a fraction of those in the closest vicinity of (distance one
lattice spacing from) the vacancies. We conjecture that these spins form a random network of moments coupled to each other by an effective interaction much
less than $J$, and that such a model (which we have not constructed but outlined the ingredients for) may host both diffusive-like and localized excitations.
It is likely that disorder averaging at the early stage in the $T$-matrix approach modifies crucial aspects of the diffusive-like excitations from the outset,
even though accounting for the localization peak. In particular, a featureless almost flat continuum was found above the localization peak \cite{Chernyshev02},
while we find a second peak below the anomalous spin wave mode. It is at present unclear exactly what the nature is of the states forming this peak, but a
natural hypothesis is weakly dispersive but predominantly diffusive-like excitations, which does not contradict the qualitative discussion in
Ref.~\onlinecite{Chernyshev02} of the excitations that eventually become fully localized.

There is an alternative to the disorder averaged medium with the $T$-matrix in spin wave theory: the linear spin wave Hamiltonian can be diagonalized
numerically in real space for specific disorder realizations and then averaged \cite{Mucciolo04}. This approach is closer to our work in terms of the way
the disorder averaging is carried out. However, it appears that the linear spin wave Hamiltonian is not sufficient to describe the different excitation
modes accurately, as the dynamic structure factor at low dilution (computed for system size $L=32$ \cite{Mucciolo04}) is essentially a double peak below
the zone boundary energy \cite{Mucciolo04}, with neither signs of a lower-energy diffusive mode (which in our calculation is far below the lower of
the two peaks in the spin wave calculation) nor a localization peak at low energy. Overall, the profile for momenta close to the zone boundary is much
more broadened by disorder than in the $T$-matrix approach, though still lacking the high-energy (multimagnon) tails that we have found here. Thus,
we conclude that the interactions neglected within linear spin wave theory (which also is the starting point of the $T$-matrix calculations) are crucial
in forming the true excitations of the diluted Heisenberg model.

The lack in the numerical spin wave calculation of features seen in the $T$-matrix calculation \cite{Chernyshev02} is surprising, given that the exact
diagonalization of the linear spin wave Hamiltonian in principle contains all the processes of the $T$-matrix approach, and also the multi-impurity scattering
processes neglected there. Perhaps the spectral weight of the localization feature only appears for larger system sizes within linear spin wave theory, while
in our fully interacting systems it appear sooner (though the localization energy decreases as $L^{-1}$ until $L$ exceeds the true localization length
$l$). As mentioned, it is also possible that the localization mode and states above it are somewhat different within the $T$-matrix approach because of the
impurity averaged medium, which differs from our calculations here as well as the exact spin wave diagonalization in Ref.~\onlinecite{Mucciolo04}.

\subsection{Relevance of the dimer-monomer model}

An important remaining question is how, or even if, the 2D monomer percolation transition at $p_c = 0$ (at $p_c>0$ in three dimensions) in the dimer-monomer
model \cite{Bhola22} is related to the localization mode of the Heisenberg model. On a superficial level, if the monomer regions correspond to spins involved
in the localization process, then the size of these regions diverges as $p \to 0$ and the excitation energy approaches $0$. The lowering of the localization
mode in the $T$-matrix analysis is exponentially rapid as $p$ is decreased, $\omega_{\rm loc} \propto {\rm e}^{-\pi/4p}$ \cite{Chernyshev02}, while the
percolation phenomenon is associated with critical behavior (with, e.g., an unusually large correlation length exponent, $\nu \approx 5$ \cite{Bhola22}).
This difference is not necessarily a contradiction, because more than one monomer region may be involved in a low-energy excitation. The localization length
scale is therefore not necessarily that of a single domain of spins, but may instead reflect a typical distance between far separated groups of small
numbers of spins.

From our real-space results in Sec.~\ref{sec:rotor}, it is clear that the monomer regions correlate strongly with the spectral weight distribution of
the rotor mode. However, all the spins carry substantial relative spectral weight of the rotor mode---which is natural since the entire system is affected by
the N\'eel order and participate in the symmetry breaking. There is only elevated weight on the monomer sites (and some other sites) reflecting a
real-space texture of the N\'eel state as observed at zero frequency (corresponding to inelastic neutron scattering). This texture is different from that
manifested in the equal-time spin correlation function, which corresponds to the entire spectral weight integrated over frequency.

The monomer regions also correlate strongly with the localized excitations, as shown in Sec.~\ref{sec:tomo} and Appendix \ref{sec:l32maps}, but in this case
there is significant spectral weight only on some of the monomer rich sites, and almost no spectral weight is carried by spins not belonging to monomer regions.
Since there are likely many large localized excitations in a given sample and our tomographic approach only reflects their collective impact on the spins, it
is not possible to draw any quantitative conclusions on the number of spins involved in a single localized state. Our system sizes are also too small to fully
contain even one typical localized state (though at $p=1/8$ the $L=64$ value of $\omega_{\rm loc}$ is already close to the extrapolated value). Nevertheless,
the fact that the monomer sites are strongly matched with the low-energy states above the quantum rotor suggests that the localized mode in some way is
dominated by a rather small number of weakly interacting spins belonging to regions of sublattice imbalance.

More broadly, our real-space tomographic analysis overall suggests that the excitations become gradually concentrated on the spins immediately adjacent to the
vacancies as the energy is lowered from the zone boundary energy. Some of these spins become ``deactivated'' as the energy approaches that of the low-energy
spin waves, and finally the localized states involves only a very small subset of those spins. Therefore, the most likely interpretation of the length scale
associated with the localized mode is one characterizing the distance between spins within that subset of effective moments in different parts of the lattice,
not the size of a single connected patch of a large number of spins. Thus, the classical dimer-monomer may contain the salient features of the localized mode,
but exactly how remains an open question.

A natural step in the future would be to construct an effective model containing only the spins adjacent to vacancies, with suitable interactions between
disconnected groups of spins and also couplings to a global staggered field. Such a model may possibly describe the anomalous magnons as well as the
diffusive-like excitations (where most of the spins have been frozen out) and also the localized mode, though our main goal with a very sparse effective
model would be to account for the cross-over from highly damped, almost diffusive excitations to localized excitations at the lowest energies. Effective
models would not only be useful both for computational studies on larger length scales, but also in analytical work to further elucidate the nature of
the excitations.

Another potentially useful further development in effective descriptions beyond the classical dimer-monomer model would be to study variational wave functions
based on the ``N\'eel biased'' singlets defined in Eq.~(\ref{dimersinglet}). Such wave functions can be simulated more easily than with full QMC calculations
of the Hamiltonian \cite{Sandvik10b}, though dynamical properties would still demand imaginary-time correlations and numerical analytic continuation..

Since a monomer percolation transition at $p_c > 0$ was predicted in three dimensions \cite{Bhola22}, it would be interesting to repeat the kind of
calculations we have reported here for the 3D diluted Heisenberg model. A direct connection between monomer regions and localized excitations would
then imply a transition at $p>0$ from a gapless spin conductor to a disordered spin insulator with gapped localized excitations at $T=0$.

\subsection{Experimental significance}

While there have been substantial past experimental studies of cuprate square-lattice antiferromagnets (parent compounds of high-T$_c$ superconductors)
with Cu substituted by Zn and Mg \cite{Cheong91,Ting92,Corti95,Carretta97,Vajk02,Papinutto05}, the focus was typically on the effects of vacancies on the
long-range N\'eel order and on dynamic and thermodynamic properties at rather high temperatures. The NQR relaxation experiments in Ref.~\onlinecite{Corti95}
are intriguing in our context here, because a low-energy peak was detected at low temperatures, shifting down with $p$, which was not compatible with the
energy expected for conventional spin waves and was instead interpreted as a ``freezing'' of local moments. In light of our results here and in
Refs.~\cite{Brenig91,Chernyshev02}, the peak could possibly reflect the damped propagating low-energy excitations of the random spin network.

The best experimental realization of the 2D Heisenberg model to date is Cu(DCOO)$_2\cdot$4D$_2$O \cite{Piazza15}, which could also possibly be doped by Zn or Mg
(replacing Cu as in the cuprates). Most likely, this would also correspond to vacancies in the Heisenberg model. It should be noted that such substitutions can
also lead to modifications of the exchange interactions around the nonmagnetic sites, which has detectable consequences in the N\'eel ordered moment versus $p$
\cite{Liu09,Liu13}. The weakly modified interactions would certainly also lead to quantitative shifts in the excitations discussed here, but we expect
all types of excitation modes to still be present. In Cu(DCOO)$_2\cdot$4D$_2$O the exchange constant is approximately 70 K, more than an order of magnitude
smaller than in the cuprates, thus the spin dynamics is more easily accessible in neutron scattering experiments over the full magnetic bandwidth. The cuprates
also are known to have more effects of ring-exchange and other higher-order interactions when the spin-only model is derived from the Hubbard model
\cite{Delannoy09a,Delannoy09b}. Thus, Cu(DCOO)$_2\cdot$4D$_2$O doped with Zn of Mg may be the best realization also of the site diluted Heisenberg model,
if such compounds are stable.

The peak formed in $S({\bf q},\omega)$ in the model at the low localization scale may in principle be observable in NMR experiments at low $p$ in the form
of the dependence of the spin-lattice relaxation rate $1/T_1$ on the nuclear resonance frequency $\omega_{\rm N}$, since $1/T_1 \propto S_0(\omega_{\rm N})$
(with idealized on-site only hyperfine couplings). However, it is not clear from our $T=0$ work how finite-temperature effects (and also magnetic field effects
in NMR experiments) will affect the likely fragile spectral weight structure at this very low energy scale. Overall, neutron scattering experiments at
dilution fractions $p \sim 10$-$20$\% may be more realistic, if sufficiently low temperature can be reached.

With the $T=0$ results for the diluted Heisenberg model at hand, to connect more closely to experiments it would be useful to also carry out $T>0$ QMC
calculations, of the spectral functions studied here and also of low-temperature thermodynamics. In addition to the single layer, the expected weak
interlayer couplings present in materials can also be taken into account, though of course the larger volume also makes the simulations more demanding of
computing resources---the calculations presented here already consumed about five million CPU hours.

A promising simplified way to include the effects of 3D N\'eel order at low temperature is to add a weak staggered magnetic field to the Hamiltonian
Eq.~(\ref{ham}), to break the spin-rotation symmetry of the order parameter. This symmetry breaking also enables the separation of the transversal and
longitudinal components of $S({\bf q},\omega)$ (which can be experimentally separated with polarized neutrons). This approach was taken in an early QMC
and numerical analytic continuation study of the 2D Heisenberg model \cite{Sandvik01} and can be repeated, now with more powerful analytic continuation
tools, for diluted systems at not much more computational effort than in the work on the rotationally symmetric model studied here.

\subsection{Technical aspects}

On a technical level, the spectral functions resolved here are unprecedented, as far as we know, in their rich details at energy scales ranging from
$\omega \approx 10^{-3}$ $J$ to $\omega \approx 5J$. Looking at the local spectral functions in Fig.~\ref{sw0_8}, there are five distinct peaks, each of which we
have explained in terms of the different types of excitations discussed above. The ability to resolve structure at low energy is related to the very
low temperatures used in the QMC simulations; typically $T = 1/4N$. With such low temperatures, and with excitations present down to a similar energy scale,
the imaginary-time correlations $G(\tau)$ remain significant within statistical errors up to $\tau$ of order $4000$ for the largest system size, thus
containing a large amount of information that can be converted to the frequency domain using the SAC method.

Our results should serve as an example of the possibility to reach these very low temperatures and resolve spectral functions in remarkable detail.
Many other disordered spin models, in particular, should have interesting spectral features amenable to the same methods used here, e.g., other types
of disordered 2D Heisenberg models \cite{Roscilde05,Yu05,Sandvik06,Laflorencie06,Yu06,Yu10} and the random $J$-$Q$ model \cite{Liu18}.

\begin{figure}[t]
\includegraphics[width=65mm]{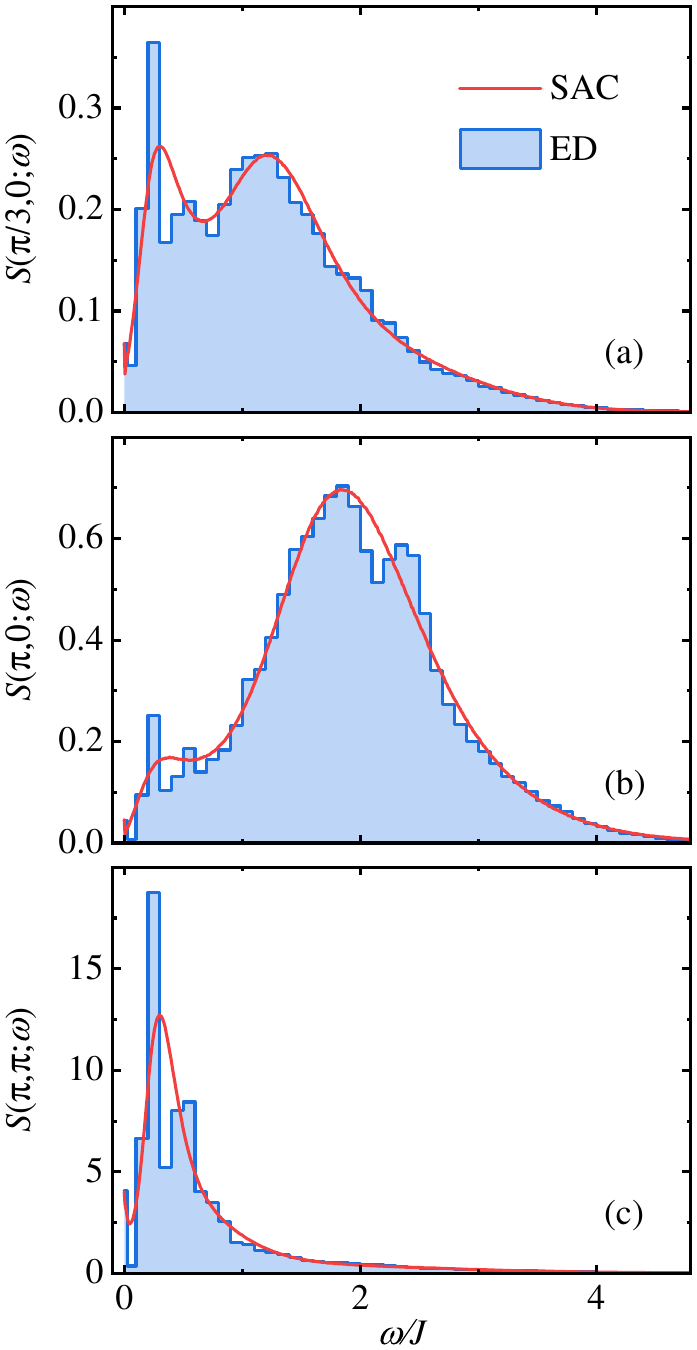}
\caption{Disorder averaged (1000 samples) $S(\boldsymbol{q},\omega)$ for system size $L=6$ diluted at $p=5/18$, with only average sublattice
  balance; $\langle N_A\rangle =\langle N_B\rangle=(1-p)N/2$. Results are shown at (a) ${\bf q}=(\pi/3,0)$, (b) ${\bf q}=(\pi,0)$, (c) ${\bf q}=(\pi,\pi)$.
  The histograms represent the distribution of $\delta$ functions in the exact expression Eq.~(\ref{sqwexact}) and the red curves are the results
  of the SAC method applied to SSE data.}
\label{sw_ed2}
\end{figure}

\begin{figure}[t]
\includegraphics[width=84mm]{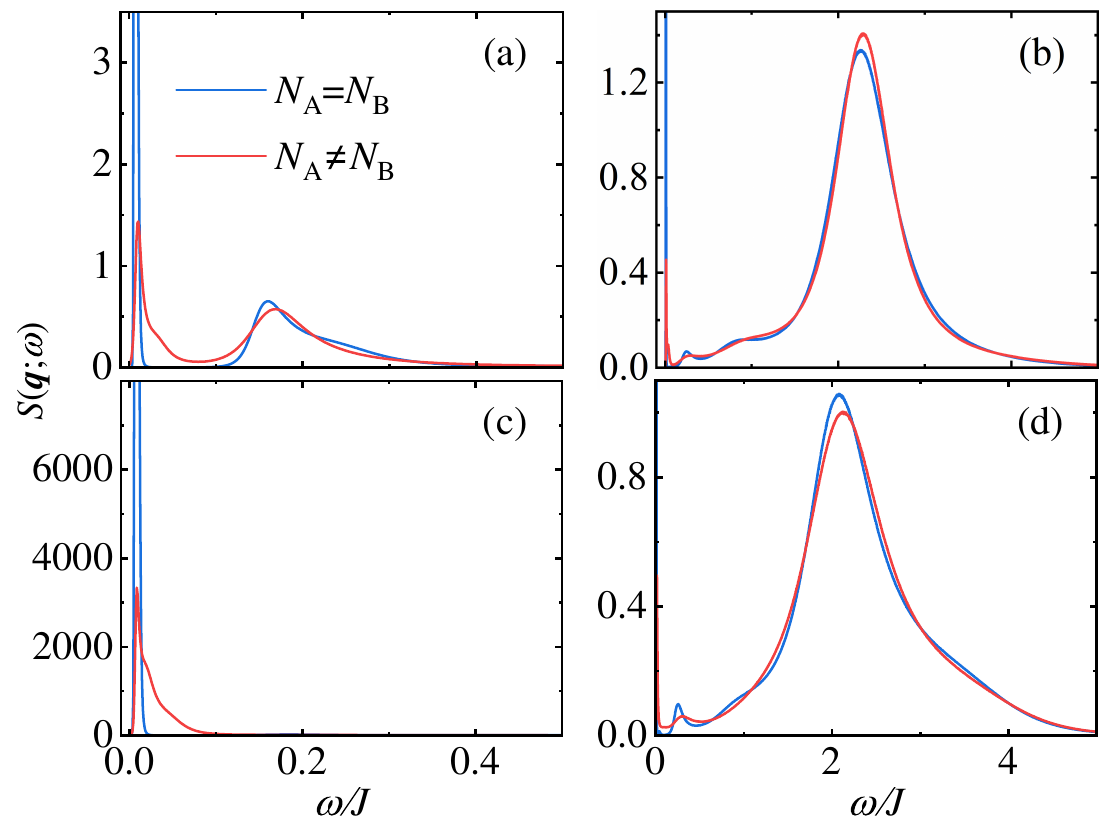}
\caption{$S(\boldsymbol{q},\omega)$ at representative $\boldsymbol{q}$ points for $p=1/8$, $L=32$. The blue and red curves show results in the
balanced and unrestricted vacancy ensembles, respectively. The momentum is $\boldsymbol{q}=(\pi/16,0)$ in (a), $\boldsymbol{q}=(\pi/2,\pi/2)$ in (b),
$\boldsymbol{q}=(\pi,\pi)$ in (c), and $\boldsymbol{q}=(\pi,0)$ in (d).}
\label{sw_unrest}
\end{figure}

\begin{acknowledgments}
We would like to thank Alexander Chernyshev, Wenan Guo, and Shiliang Li for valuable discussions, and Ling Wang for help with the simulations of the
dimer-monomer model. H.S.~was supported by the National Key Projects for
Research and Development of China under Grant No.~2021YFA1400400, the National Natural Science Foundation of China under Grant No.~12122502, and the Fundamental Research Funds for the Central Universities. A.W.S.~was supported by
the Simons Foundation under Grant No.~511064. Some of the numerical calculations were carried out on the Shared Computing Cluster managed by Boston University's
Research Computing Services.
\vskip3mm
\end{acknowledgments}

\appendix

\section{Sublattice imbalanced systems}
\label{sec:nanb}

In the main paper we have only shown results for systems with global sublattice balance; $N_A=N_B$. In a real 2D Heisenberg quantum magnet doped with nonmagnetic
impurities, this constraint would of course not be realistic requirement. Given that the individual clusters can be imbalanced even with global balance, the
$N_A=N_B$ and $\langle N_A\rangle =\langle N_B\rangle$ ensembles should not be expected to differ in the thermodynamic limit. Nevertheless, we here show some
results for systems where $N_A= N_B$ was not imposed for each sample but $\langle N_A\rangle =\langle N_B\rangle=(1-p)N/2$.

\begin{figure}[t]
\includegraphics[width=75mm]{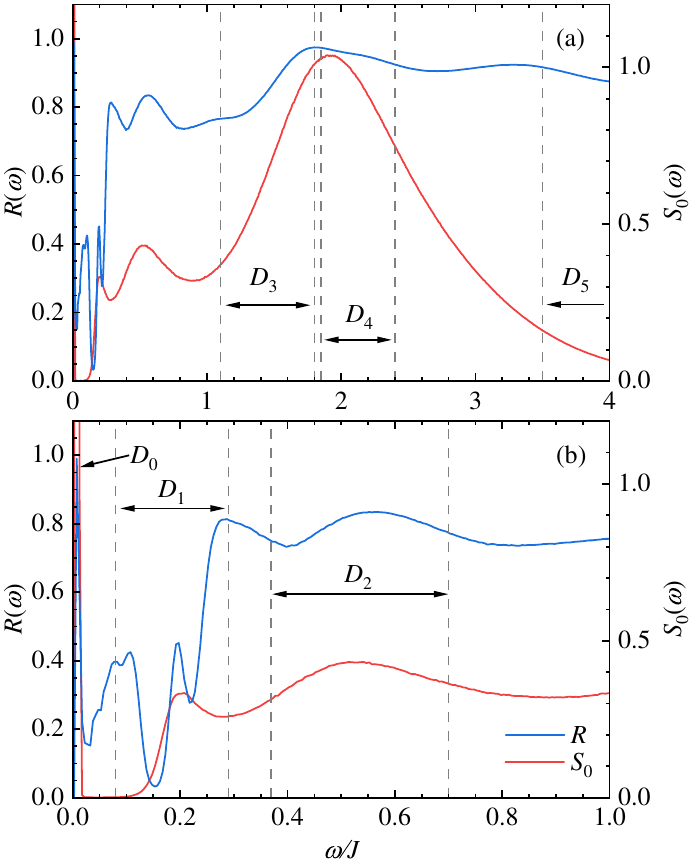}
\caption{Local spectral function for an individual $L=32$ vacancy realization at $p=1/16$ along with the frequency windows defining the maps $D_0,\ldots,D_5$
discussed in this Appendix. Corresponding results for the $p=1/8$ configuration used in Figs.~\ref{d0} and \ref{d5}
are shown in the main paper, Fig.~\ref{tomography}.}
\label{toms0_16}
\end{figure}

\begin{figure*}[t]
\includegraphics[width=140mm]{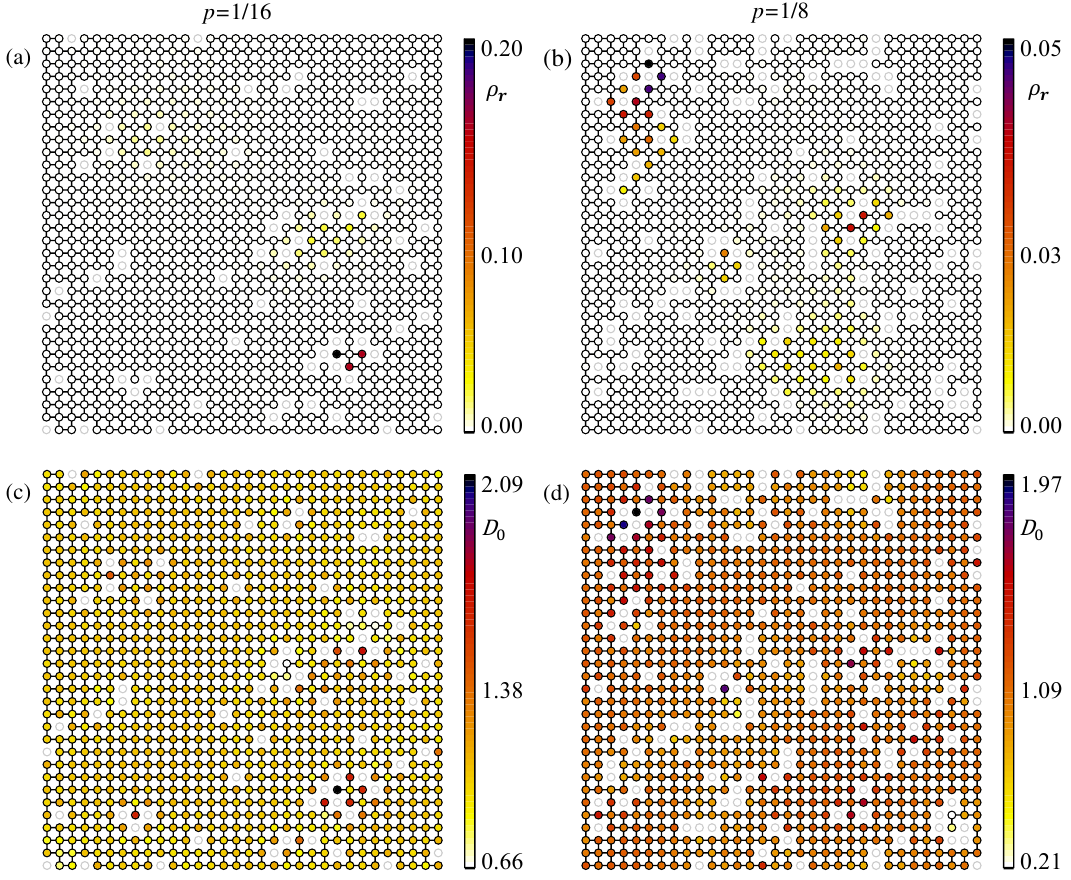}
\caption{Monomer distributions for $L=32$ systems with $p=1/8$ and $p=1/16$ are shown in (a) and (b) respectively. The tomographic quantum rotor weight
distributions for the same samples are shown in (c) and (d).}
\label{d0}
\end{figure*}

First, in Fig.~\ref{sw_ed2} we show results analogous to Fig.~\ref{sw_ed1}, comparing SAC and exact Lanczos results for $p=5/18$ averaged over 1000 disorder
samples. Here we observe that it is not possible to clearly distinguish the contributions from the rotor excitations. While there are low-energy peaks in the
Lanczos based histograms at the same energy for all three momenta shown, there is much more spectral weight on the right hand side of this peak than in
Fig.~\ref{sw_ed1}. The agreement between SAC and Lanczos results is also less impressive, though still acceptable, because of the difficulty in reproducing
the shape of the shallow minimum between the two peaks.

In Fig.~\ref{sw_unrest} we show SAC results for $p=1/8$, $L=32$, comparing the two ensembles directly. At the two momenta close to the uniform and
staggered wave-vectors, Fig.~\ref{sw_unrest}(a) and Fig.~\ref{sw_unrest}(c), the rotor peak is again much more spread out in the unrestricted
$\langle N_A\rangle =\langle N_B\rangle$ ensemble. At the other two momenta, Fig.~\ref{sw_unrest}(b) and Fig.~\ref{sw_unrest}(d), the differences are
rather small, but the unrestricted ensemble does not show the localization peak as clearly, again because of some spectral weight between it and the
(now rather sharp) rotor peak. We fully expect the agreement between the two ensembles to improve as $L$ is further increased.

\section{L=32 Real-space spectral maps}
\label{sec:l32maps}

\begin{figure*}[t]
\includegraphics[width=140mm]{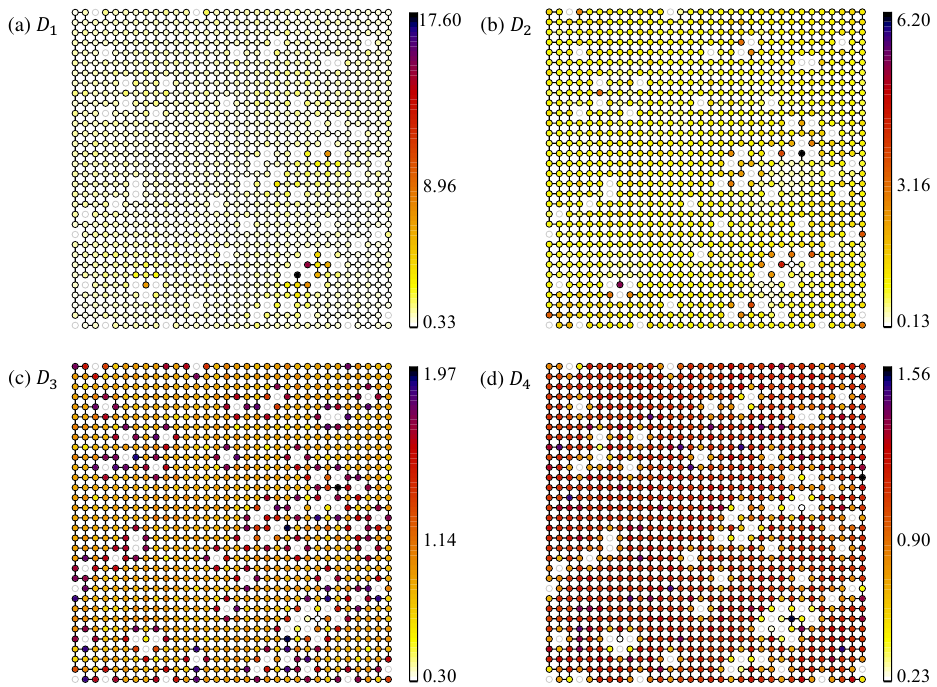}
\caption{Tomographic maps $D_1,\ldots,D_4$ for the same $p=1/16$ samples as in Figs.~\ref{d0}(a) and \ref{d0}(c). The maps for the same $p=1/8$ samples
as in Figs.~\ref{d0}(b) and \ref{d0}(d) are shown in Fig.~\ref{tomo1234}.}
\label{tomol32_p16}
\end{figure*}

\begin{figure*}[t]
\includegraphics[width=130mm]{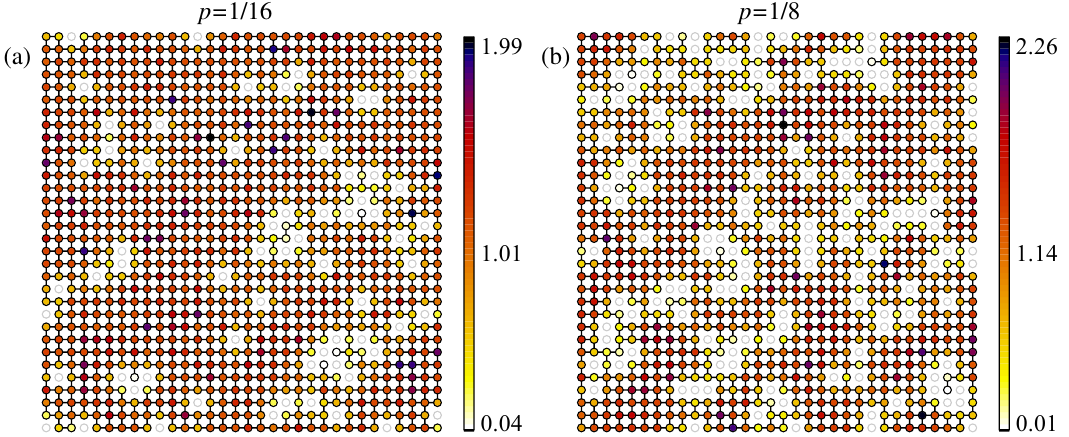}
\caption{Tomographic maps $D_5$ corresponding to the high energy tails of the same vacancy samples as in Fig.~\ref{d0}. The energy windows are
shown in Figs.~\ref{tomography} and \ref{toms0_16}.}
\label{d5}
\end{figure*}

\begin{figure}[t]
\includegraphics[width=55mm]{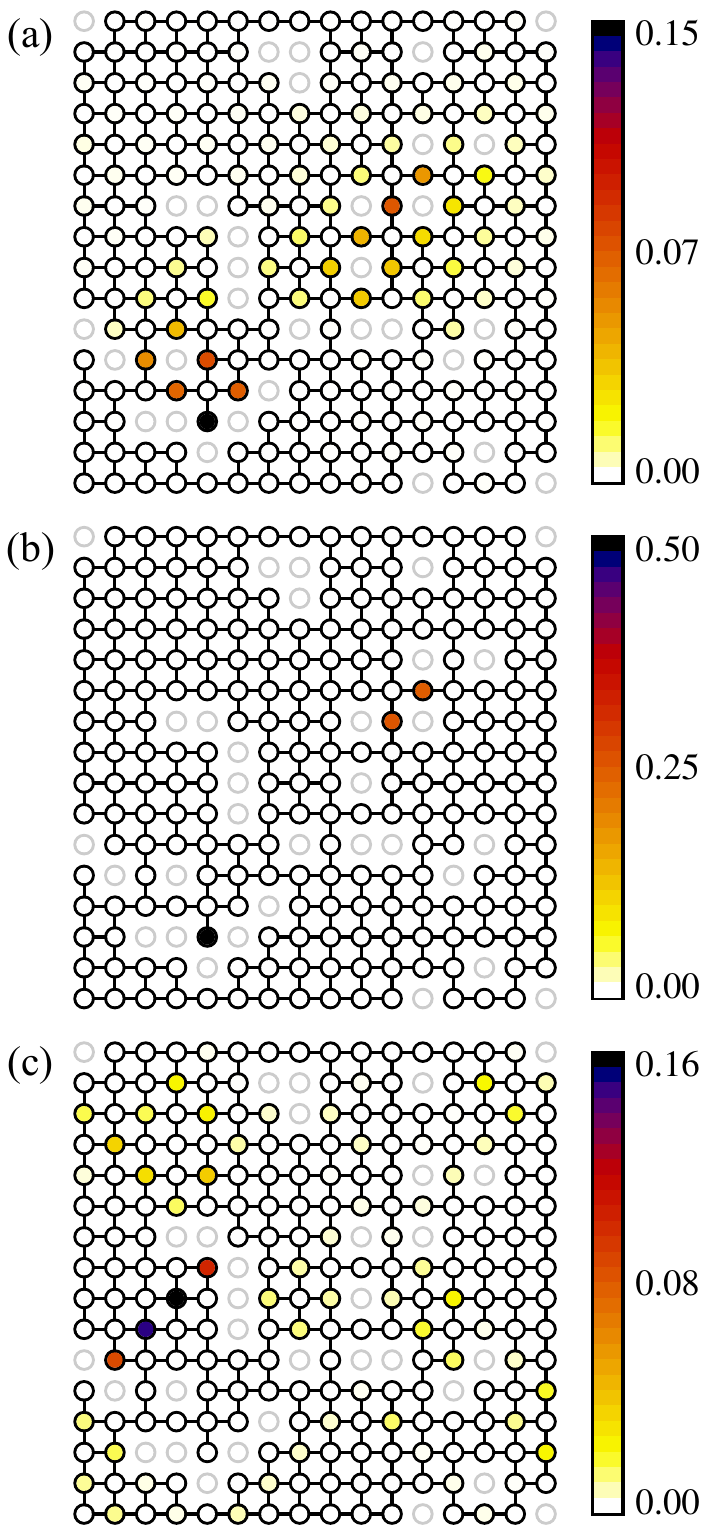}
\caption{Monomer distribution in the system previously considered with the noninteracting dimer-monomer model in Fig.~\ref{monomer_locx}(a), here repeated in
  (a). In (b) and (c), results are shown for the interacting models in Eq.~(\ref{zpm}); $Z_+$ (favoring parallel dimers) and $Z_-$ (disfavoring parallel dimers).}
\label{dimer_interaction}
\end{figure}

To complement the tomographic analysis of the real-space spectral weight in Secs.~\ref{sec:realrotor} and \ref{sub:hbergrotor}, we here present additional
results for $L=32$ vacancy realizations; one at $p=1/16$ and one at $p=1/8$. The ${\bf r}$ averaged quantity $S_0(\omega)$ for $p=1/16$ (the analogue of
the $p=1/8$ case in Fig.~\ref{tomography}) is shown in Fig.~\ref{toms0_16} along with the windows used to define the integrals for the $D_0,\ldots,D_5$
in Eq.~(\ref{stomo}).

In Sec.~\ref{sec:dimer}, we compared the monomer distribution with the rotor weight only for a system of size $L=16$ at $p=1/8$. In Fig.~\ref{d0} we show
results for both $p=1/16$ and $p=1/8$. In both cases, it is clear that sites with monomers correspond to elevated rotor weight, though there are also some
nonuniformities in the rotor weight also among spins with no corresponding monomer occupation, especially some of the spins next to vacancies.

In Fig.~\ref{tomol32_p16} we show the tomographic maps $D_1$-$D_4$ for the $L=32$ configuration at $p=1/16$, complementing the one for $p=1/8$ in
Fig.~\ref{tomo1234}. The energy windows are defined in Fig.~\ref{toms0_16}. The spectral weight distributions are qualitatively similar to the
$p=1/8$ case. Visually, there is a bit more variability in the spectral weight on the spins adjacent to vacancies in $D_3$, but this aspect of
the low-energy magnons depend to some extent on how the energy window is chosen. The trend of spectral weight moving from the patches between
vacancies to spins next to vacancies when the magnon energy is lowered from is a very stable feature for all the cases we have studied.

In the main paper we did not discuss the tomographic maps $D_5$ corresponding to the high-energy tails of the spectral weight, within the frequency windows
illustrated in Figs.~\ref{tomography} and \ref{toms0_16} (which in both cases extend beyond the figure frames up to the highest frequencies of the tails of the
distributions). Results for both $p=1/16$ and $p=1/8$ are shown in Fig.~\ref{d5}. Similar to the the high-energy magnon dominated $D_4$ map, the dominant
spectral weight in $D_5$ falls within the uniform patches between vacancies, with even less weight on the spins next to vacancies.

\begin{figure}[h]
\includegraphics[width=75mm]{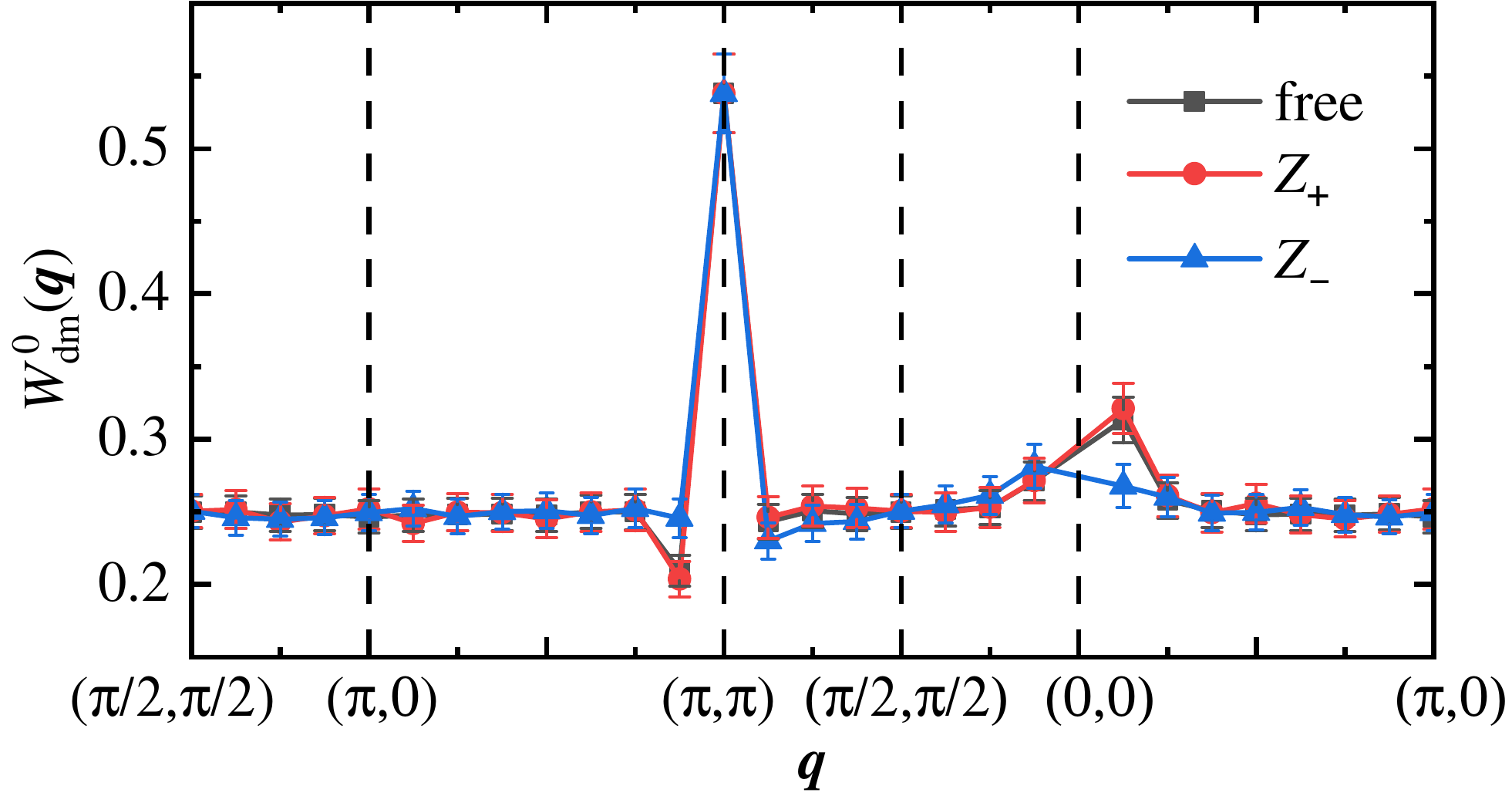}
\caption{Rotor weight distribution $W^0_{\rm dm}(\boldsymbol{q}$) along a cut in the BZ based on simulations of the noninteracting dimer model (black squares)
as well as the interacting models $Z_+$ (red circles) and $Z_-$ (blue triangles) for system size $L=16$ at $p=1/8$, averaged over 2000 vacancy samples, each
constrained to have two monomers.}
\label{rw_dm_interaction}
\end{figure}

\section{Interacting dimer model}
\label{sec:dimerinteractions}

In Sec.~\ref{sec:rotor} we found that the simple dimer-monomer model was able to qualitatively explain the spread of quantum rotor weight in the form
of peaks around ${\bf q}=(0,0)$ and $(\pi,\pi)$, but beyond these peaks the distribution is flat. It is then interesting to include interactions in
the dimer-monomer model, which could potentially change the nature of the sublattice imbalance and produce more structure in ${\bf q}$ space. Given our
proposal that correlated singlet fluctuations are responsible for the observed structure, we here consider dimer-dimer interactions either favoring or
disfavoring columnar order.

We use the partition functions
\begin{eqnarray}
Z_\pm=\sum_{\mathcal{C}}\exp(\pm \beta \cdot N_p(\mathcal{C})),
\label{zpm}
\end{eqnarray}
where $N_p(\mathcal{C})$ is the number of parallel dimers in configuration $\mathcal{C}$, i.e, the number of $2\times 2$ plaquttes with two dimers
either horizontally or vertically oriented. We take the inverse temperature $\beta$ to be large, so that the $Z_+$ and $Z_-$ system has the maximum
and minimum number of parallel dimers, respectively. We again sample the configurations with minimal number of monomers for given vacancy realization.

In Fig.~\ref{dimer_interaction} we compare the previous noninteracting results for the specific vacancy realization in Fig.~\ref{monomer_locx}(a), here
reproduced in Fig.~\ref{dimer_interaction}(a) for convenience, with results for $Z_+$ in Fig.~\ref{dimer_interaction}(b) and for $Z_-$ in
Fig.~\ref{dimer_interaction}(c). There are significant effects in the monomer distribution, with more parallel dimers leading to smaller areas of
sublattice imbalance, while the suppression of parallel dimers cause larger regions.

An example of the rotor weight distribution computed as before using Eq.~(\ref{w0dmdef}) is shown in Fig.~\ref{rw_dm_interaction}. Here we only see small
changes at the peaks relative to the noninteracting case for the repulsive model $Z_-$. For the attractive $Z_+$ model, there are no detectable changes
within statistical errors for this small system size. Thus, it seems unlikely that any classical dimer-monomer model would be able to describe the ${\bf q}$
dependence of the rotor weight beyond the peaks at $(0,0)$ and $(\pi,\pi)$. However, our results also show that those peaks are very robust, being insensitive
to the details of the model, even when the imbalanced regions themselves can vary strongly with the model parameters, as seen in Fig.~\ref{dimer_interaction}.
This result reinforces the notion that the peaks in the diluted Heisenberg model are indeed caused by local sublattice imbalance, even though the distribution
of the ``dangling spins'' in the quantum model must differ from that of the monomers in the classical models. Given the previous results in
Fig.~\ref{monomer_locx} and the results with interactions included in Fig.~\ref{dimer_interaction}, the non-interacting model is the best one in terms
of matching the monomer distribution to the local magnetic response of the Heisenberg model.

\end{document}